\documentclass[a4paper,11pt]{article}
\usepackage[left=2cm,right=2cm,top=2cm,bottom=2cm]{geometry}
\usepackage[usenames,dvipsnames,svgnames,table]{xcolor}
\usepackage{psfrag,amsmath,amsfonts,amssymb,latexsym,amsthm,verbatim,color,lscape,multirow,graphicx,subcaption}
\usepackage[figuresright]{rotating}
\usepackage[compress]{natbib}
\usepackage{url}
\usepackage{hyperref}
\hypersetup{colorlinks,
citecolor=black,
filecolor=black,
linkcolor=black,
urlcolor=black,
pdftex}

\usepackage{booktabs}
\usepackage{colortbl}
\definecolor{gray}{rgb}{0.9,0.9,0.9}

\usepackage{times}
\usepackage{blindtext}
\usepackage[font={footnotesize,it}]{caption}

\usepackage{sectsty}  
\subsectionfont{\normalsize\bf}
\sectionfont{\large\bf}

\usepackage[symbol]{footmisc}

\renewcommand{\thefootnote}{\fnsymbol{footnote}}

\def\p{\partial}

\def\Cbar{{\overline C}}
\def\C{\mathbf{C}}

\def\pp{\mathbf{p}}
\def\R{\mathbf{R}}
\def\y{\mathbf{y}}
\def\Y{\mathbf{Y}}

\def\b{\beta}
\def\de{\delta}

\def\gbf{\boldsymbol{\gamma}}
\def\pibf{\boldsymbol{\pi}}

\def\th{\theta}
\def\thbf{\boldsymbol{\theta}}
\def\pibf{\boldsymbol{\pi}}
\def\Xibf{\boldsymbol{\Xi}}

\def\Cbar{{\overline C}}
\def\C{\mathbf{C}}

\def\pp{\mathbf{p}}
\def\R{\mathbf{R}}
\def\y{\mathbf{y}}
\def\Y{\mathbf{Y}}

\def\de{\delta}
\def\Debf{\boldsymbol{\Delta}}

\def\th{\theta}
\def\thbf{\boldsymbol{\theta}}
\def\pibf{\boldsymbol{\pi}}
\def\Xibf{\boldsymbol{\Xi}}

\long\def\symbolfootnote[#1]#2{\begingroup
\def\thefootnote{\fnsymbol{footnote}}\footnote[#1]{#2}\endgroup}

\newcommand{\tr}{\mathrm{tr}} 

\usepackage[normalem]{ulem}

\let\proglang=\textsf
\newcommand{\pkg}[1]{{\fontseries{b}\selectfont #1}}

\begin{document}

\title{Factor copula models for mixed data}

\author{Sayed H. Kadhem\footnote{School of Computing Sciences, University of East Anglia, Norwich Research Park, Norwich NR4 7TJ, U.K.}  \and Aristidis~K.~Nikoloulopoulos \footnote{Correspondence to: \href{mailto:a.nikoloulopoulos@uea.ac.uk}{a.nikoloulopoulos@uea.ac.uk}, Aristidis K. Nikoloulopoulos, School of Computing Sciences, University of East Anglia, Norwich NR4 7TJ, U.K.} }
\date{}

\maketitle
\begin{abstract}
\baselineskip=25pt
 \noindent
We develop  factor copula models for analysing the dependence  among  mixed continuous and discrete responses. Factor copula models are canonical  vine copulas that involve both observed and latent variables, hence they allow 
 tail, asymmetric and non-linear dependence. 
They can be explained as conditional independence models with  latent variables that don't  necessarily have an additive latent structure. 
We focus on important issues that would interest the social data analyst, such as  model selection and goodness-of-fit.  
Our general methodology is demonstrated with an extensive simulation study and illustrated by re-analysing three mixed response  datasets. Our studies suggest that there can be a substantial improvement over the standard factor model for mixed data and makes the argument for moving to factor copula models.

\noindent \textbf{Key Words:} Canonical vines; Conditional independence; Goodness-of-fit; Latent variable models; Model selection; Tail dependence/asymmetry.
\end{abstract}

\maketitle

\baselineskip=24.5pt

\section{Introduction}
It is very common in social science, e.g., in surveys,  to deal with datasets that have mixed continuous and discrete responses. 
In the literature, two broad frameworks  have been considered  to model the dependence among such mixed continuous and discrete responses, namely the latent variable and copula framework. 

There are two approaches for modelling multivariate mixed data with latent variables: the underlying variable approach that treats all variables as continuous by assuming  the  discrete responses as a manifestation of underlying continuous variables that usually follow the normal distribution (e.g., \citealt{Muthen1984-PKA,Lee&Poon&Bentler1992,Quinn2004}), and the response function approach that postulates distributions on the observed variables conditional on the latent variables usually from the exponential family (e.g., \citealt{Moustaki1996,Moustaki&Knott2000-PKA,Wedel&Kamakura2001,
Huber-etal-2004-RSS,Moustaki&Victoria-Feser2006}). 
The former method almost invariably  assumes that
the underlying variables (linked to the observed variables via a threshold process to yield ordinal data and  an identity process to yield continuous data)   follow a multivariate normal (MVN) distribution,
while  the latter assumes that  the observed variables are conditionally  independent  usually given MVN distributed latent variables.  They  are equivalent  when in the underlying and the response function approach the MVN distribution has a factor and an independence correlation structure, respectively \citep{Takane1987-PKA}.

The underlying variable approach calls the  MVN distribution as a latent model for the discrete responses, and therefore  maximum likelihood (ML) estimation  requires multidimensional integrations \citep{nikoloulopoulos13b,nikoloulopoulos2015a}; their dimension is equal with the number of observed discrete variables. This is why alternative estimation methods such as the three stage weighted least squares and  composite likelihood have been proposed; see e.g.,
\cite{Katsikatsou-etal-2012-csda}. The  response function approach, with  
the dependence coming from $p$ latent (unobservable) variables/factors where $p<<d$  (the number of observed variables), requires $p$- rather than $d$- dimensional integration. Hence, ML  estimation is feasible, especially  when the number of latent variables is small.

Nevertheless, both approaches are restricted to  MVN assumption for the observed or the latent variables  
that is not valid if tail asymmetry or tail dependence exists in the mixed data which is a realistic scenario. \cite{Ma&Genton2010-RSS},  \cite{Montanari&Viroli10}, and \cite{Irincheeva-etal-2012-JABES} stress that 
the MVN assumption might not be adequate, and acknowledge that the effect of misspecifying the distribution of the latent variables could lead to biased model estimates and poor fit.
To this end, \cite{Irincheeva&Cantoni&Genton2012-SJS} proposed a more flexible response function approach by strategically multiplying the MVN density of the latent variables by a polynomial function to achieve departures from normality.

As we have discussed, the underlying variable approach exploits the use of the MVN assumption to model the joint distribution of mixed data. The univariate margins are transformed to normality and then the MVN distribution is fitted to the transformed data. This construction  is apparently the MVN copula applied to mixed data \citep{shen-Weissfeld06,Hoff2007-AAS,Song-etal-2009-Biometrics,Jingetal2012,Jiryaie-etal-2016}, but previous papers (e.g., \citealt{Quinn2004})  do not refer to copulas as the approach can be explained without copulas.

\cite{Smith&Khaled2012}, \cite{Stober-etal-2015} and \cite{zilko&Kurowicka2016} called vine copulas to model mixed data. Vine copulas  have two major advantages over the MVN copula as emphasized in \cite{Panagiotelis-etal-2017-CSDA}. The first is that the computational complexity of computing the joint probability distribution  function grows quadratically with $d$, whereas for the MVN copula the computational complexity grows exponentially with $d$. The second is that vine copulas are highly flexible through their specification  from bivariate parametric copulas with different  tail dependence or asymmetry properties. They  have as special case the MVN copula, if all the bivariate parametric copulas are bivariate normal (BVN).

In this paper, we extend the factor copula models in \cite{Krupskii&Joe-2013-JMVA} and \cite{Nikoloulopoulos2015-PKA} to the case of mixed continuous and discrete responses. 
Factor copula are vine copula models that involve both observed and latent variables. Hence, they are highly flexible through their specification  from bivariate parametric copulas with different  tail dependence or asymmetry properties. 
The underlying variable approach  where the MVN distribution has  a $p$-factor correlation structure or its equivalent, i.e., the response function approach where the MVN distribution has an independence correlation structure   is  a special case  of factor copula models when all the  bivariate parametric copulas are BVN (hereafter standard factor model). 
Factor copula  models are more interpretable and fit better than vine copula models, when dependence can be explained through latent variables. 
Furthermore, they are closed under margins, i.e.,  
lower order  marginals belong to the same parametric family  of copulas and  a different 
permutation of the observed variables has exactly the same distribution. This is not the case for vine copulas without latent variables, where a different permutation of the observed variables could lead to a different distribution.

We tackle issues that particularly interest the social data analyst as model selection and goodness-of-fit. 
Model selection in previous papers  on factor copula models \citep{Krupskii&Joe-2013-JMVA,Nikoloulopoulos2015-PKA} was mainly based on simple diagnostics.   
In addition to simple diagnostics based on semi-correlations, we propose  an heuristic method that automatically selects the bivariate parametric copula families. 
As regard as to the issue of goodness-of-fit testing, we propose a technique that is  based on the  $M_2$ goodness-of-fit statistic \citep{Maydeu-Olivares&Joe2006} in multidimensional contingency tables to overcome the shortage of goodness-of-fit statistics for mixed continuous and discrete response data (e.g.,  \citealt{Moustaki&Knott2000-PKA}).

The remainder of the paper proceeds as follows. Section \ref{model-sec} introduces the factor copula  models for mixed data and provides choices of parametric bivariate copulas with latent variables. Estimation techniques and computational details are provided in Section \ref{Est-section}. Sections \ref{modSel-section} and  \ref{gof-section}  propose methods for model selection and   goodness-of-fit, respectively. Section  \ref{sec-application} presents applications of our methodology to three mixed response data sets. 
Section \ref{sec:simulations} contains an extensive simulation study to gauge the small-sample efficiency  of the proposed estimation,  investigate the misspecification of the  bivariate copulas,  and examine the reliability of the model selection and goodness-of-fit techniques. 
We conclude with some discussion in Section \ref{sec-discussion}, followed by a brief section with software details.

\section{\label{model-sec}The factor copula model for mixed responses}
Although the  factor copula models can be explained as truncated canonical vines rooted at the latent variables, we derive the models as conditional independence models,  i.e.,  a response function approach with dependence coming from latent (unobservable) variables/factors. 
The $p$-factor model assumes that the mixed continuous and discrete responses  $\Y=(Y_1,\ldots,Y_d)$ are conditionally
independent given $p$ latent variables $X_1,\ldots,X_p$. In line with  \cite{Krupskii&Joe-2013-JMVA} and \cite{Nikoloulopoulos2015-PKA}, we use a general copula construction, based on a set of bivariate copulas that link observed to latent  variables, to specify the factor copula models for mixed continuous and  discrete variables.  
The idea in the derivation of this $p$-factor model will be shown below for the 1-factor and 2-factor case. It can be extended to
$p\ge 3$ factors or latent variables in a similar manner. The  evaluation of  a $p$-dimensional integral   can be successively performed as we strategically assume that the factors or latent variables are independent. 

For the 1-factor model,
let $X_1$ be a latent variable, which we assume to be standard uniform
(without loss of generality).
From \cite{sklar1959}, there is a
bivariate copula $C_{X_1j}$
such that $\Pr(X_1\le x, Y_j\le y)=C_{X_1j}\bigl(x,F_j(y)\bigr)$ for $0\leq x\leq 1$ where $F_j$ is the
cumulative distribution function (cdf) of $Y_j$. Then it follows that
\begin{equation}
  F_{j|X_1}(y|x):=\Pr(Y_j\le y|X_1=x) = {\p C_{X_1j}\bigl(x,F_j(y)\bigr)\over\p x}.
  \label{eq-condjL1}
\end{equation}
Letting $C_{j|X_1}(F_j(y)|x)=\p C_{X_1j}(x,F_j(y))/\p x$ for shorthand notation and $\y=(y_1,\ldots,y_d)$  be  realizations of $\Y$,
the density\footnote[3]{We mean the density of  $\Y$ w.r.t. the product measure on the respective supports of the marginal variables. For discrete margins with integer values this is the counting measure on the set of possible outcomes, for continuous margins we consider the Lebesgue measure in $\mathbb{R}$.} of the observed data in  the 1-factor  model case is
\begin{equation}
\label{1-pdf}
f_\Y(\mathbf{y})=\int_0^1\prod_{j=1}^d f_{j|X_1}(y_j|x) \,dx, 
\end{equation}
where 
$$f_{j|X_1}(y|x)=\left\{\begin{array}{ccc}
C_{j|X_1}\bigl(F_j(y)|x\bigr) -  C_{j|X_1}\bigl(F_j(y-1)|x\bigr) &\mbox{if} & Y_j \quad \mbox{is discrete};\\
c_{X_1j}\bigl(x,F_j(y)\bigr)f_j(y)&\mbox{if} & Y_j \quad \mbox{is continuous},
\end{array}\right.
$$
is the  density of $Y_j=y$ conditional on $X_1=x$; $c_{X_1j}$ is the bivariate copula density of $X_1$ and $Y_j$  and $f_j$ is the univariate  density of $Y_j$.

For the 2-factor model, consider two latent variables $X_1,X_2$ that are, without loss of generality, independent uniform $U(0,1)$ random variables.
Let $C_{X_1j}$ be defined as in the 1-factor model,
and let $C_{X_2j}$ be a bivariate copula such that
  $$\Pr(X_2\le x_2,Y_j\le y|X_1=x_1)
  =C_{X_2j}\bigl(x_2,F_{j|X_1}(y|x_1)\bigr),$$
where $F_{j|X_1}$ is given in (\ref{eq-condjL1}).
Then for $0\leq x_1,x_2\leq1$,
  $$\Pr(Y_j\le y|X_1=x_1,X_2= x_2)
  = {\p\over \p x_2} \Pr(X_2\le x_2,Y_j\le y|X_1=x_1)$$
  $$= {\p\over \p x_2} C_{X_2j}\bigl(x_2,F_{j|X_1}(y|x_1)\bigr)=
   C_{j|X_2}\bigl(F_{j|X_1}(y|x_1)|x_2\bigr).$$

The density of the observed data in   the 2-factor  model case is 
\begin{equation}
\label{2-pdf}
f_\Y(\mathbf{y})=
\int_0^1\int_0^1\prod_{j=1}^d
f_{X_2j|X_1}\bigl(x_2,y_j|x_1\bigr)\,dx_1 dx_2,
\end{equation}
where
$$f_{X_2j|X_1}(x_2,y|x_1)=\left\{\begin{array}{ccc}
C_{j|X_2}\bigl(F_{j|X_1}(y|x_1)|x_2\bigr)-C_{j|X_2}\bigl(F_{j|X_1}(y-1|x_1)|x_2\bigr) &\mbox{if} & Y_j \quad\mbox{is discrete};\\
c_{jX_2;X_1}\bigl(F_{j|X_1}(y|x_1),x_2\bigr)c_{X_1j}\bigl(x_1,F_j(y)\bigr)f_j(y)&\mbox{if} & Y_j \quad \mbox{is continuous}.
\end{array}\right.$$

Note that the copula $C_{X_1j}$ links the $j$th response to the
first latent variable $X_1$, and the copula $C_{X_2j}$ links the $j$th
response to the second latent variable $X_2$ conditional on $X_1$. 
In our  general statistical model 
there are no constraints in the choices of the parametric marginal $F_j$ or copula $\{C_{X_1j},C_{X_2j}\}$  distributions.

\subsection{\label{bivcop}Choices of bivariate copulas with latent variables}

We provide choices of parametric bivariate copulas  that can be used to link the latent to the observed variables. 
We will consider copula families that have different tail dependence \citep{Joe1993-JMVA} or tail order \citep{Hua-joe-11}.

A bivariate copula $C$ is {\it reflection symmetric}
if its density 
 satisfies $c(u_1,u_2)=c(1-u_1,1-u_2)$ for all $0\leq u_1,u_2\leq 1$.
Otherwise, it is reflection asymmetric often with more probability in the
joint upper tail or joint lower tail. {\it Upper tail dependence} means
that $c(1-u,1-u)=O(u^{-1})$ as $u\to 0$ and {\it lower tail dependence}
means that $c(u,u)=O(u^{-1})$ as $u\to 0$.
If $(U_1,U_2)\sim C$ for a bivariate copula $C$, then $(1-U_1,1-U_2)\sim
\widehat C$,  
where $\widehat C(u_1,u_2)=u_1+u_2-1+C(1-u_1,1-u_2)$   
is the survival or reflected 
copula of $C$; this ``reflection"
of each uniform $U(0,1)$ random variable about $1/2$ changes the direction
of tail asymmetry. Under some regularity conditions (e.g., existing finite density in the interior of the unit square, ultimately monotone in the tail), if there exists  $\kappa_L(C)>0$ and some $L(u)$ that is slowly varying at $0^+$ (i.e., $
 \frac{L(ut)}{L(u)} \sim 1,$ as $u\to 0^+$ for all  $t>0$), then  $\kappa_L(C)$  is the   \textit{lower tail order} of $C$. The \textit{upper tail order} $\kappa_U(C)$ can be defined by the reflection of $(U_1,U_2)$, i.e., $\Cbar(1-u,1-u) \sim u^{\kappa_U(C)} L^*(u)$ as $u\to0^+$, where $\Cbar$ is the survival function of the  copula and $L^*(u)$ is a slowly varying function. 
With $\kappa=\kappa_L$ or $\kappa_U$,  
 a bivariate copula has \textit{intermediate tail dependence} if $\kappa \in (1,2)$, \textit{tail dependence} if $\kappa=1$, and  \textit{tail quadrant independence} if $\kappa=2$ with $L(u)$ being asymptomatically a constant.

After briefly providing definitions of  tail dependence and  tail order we provide below a list of bivariate parametric copulas with varying tail behaviour: 
\begin{itemize}
\itemsep=0pt

\item Reflection symmetric copulas with intermediate tail dependence such as the BVN copula with $\kappa_L=\kappa_U=2/(1+ \th)$, where $\theta$ is the copula (correlation) parameter. 
\item Reflection symmetric copulas with tail quadrant independence ($\kappa_L=\kappa_U= 2$), such as the Frank copula.

\item Reflection asymmetric copulas with upper tail dependence only such as 

\begin{itemize}
\itemsep=0pt
\item  the Gumbel copula with $\kappa_L=2^{1/\theta}$ and $\kappa_U=1$, where $\theta$ is the copula parameter;  

\item  the Joe copula with $\kappa_L=2$ and $\kappa_U=1$.
\end{itemize}
\item Reflection symmetric copulas with tail dependence, such as the
$t_\nu$ copula with $\nu$ the degrees of freedom and $\kappa_L=\kappa_U=1$.

\item Reflection asymmetric copulas with upper and lower tail dependence that can range independently from 0 to 1,  such as the BB1 and BB7 copulas with $\kappa_L=1$ and $\kappa_U=1$.

\item Reflection asymmetric copulas with  tail quadrant independence, 
such as the 
 the BB8 and BB10 copulas.

\end{itemize}

The 
BVN, Frank, and $t_\nu$  are comprehensive copulas, i.e., they interpolate between countermonotonicity  (perfect negative dependence) to comonotonicity (perfect positive dependence). 
The other aforementioned parametric families of copulas, namely Gumbel, Joe, BB1, BB7, BB8 and BB10  interpolate between independence and perfect positive dependence.
Nevertheless, negative dependence can be obtained from these copulas by considering reflection of one of the uniform random variables on (0, 1). If $(U_1,U_2)\sim C$ for a bivariate copula $C$ with positive dependence, then
\begin{itemize}
\item $(1-U_1,U_2)\sim
\widehat{C}^{(1)}$,   
where $\widehat{C}^{(1)}(u_1,u_2)=u_2-C(1-u_1,u_2)$  
is the 1-reflected 
copula of $C$ with negative lower-upper tail dependence;
\item $(U_1,1-U_2)\sim
\widehat{C}^{(2)}$,   
where $\widehat{C}^{(2)}(u_1,u_2)=u_1-C(u_1,1-u_2)$  
is the 2-reflected  
copula of $C$ with negative upper-lower dependence.
\end{itemize}
{\it Negative upper-lower tail dependence} means that $c(1-u,u)=O(u^{-1})$ as $u\to 0^+$ and {\it negative lower-upper tail dependence} means that $c(u,1-u)=O(u^{-1})$ as $u\to 0^+$ \citep{joe2010b}.

\begin{figure}[!h]
\begin{center}
\caption{\label{contours}Contour plots of bivariate copulas with standard normal margins and dependence parameters corresponding to Kendall's $\tau$ value of 0.5 on absolute value.} 
\begin{tabular}{|ccc|}
\hline
BVN& Frank&$t_3$\\\hline
\includegraphics[width=0.3\textwidth]{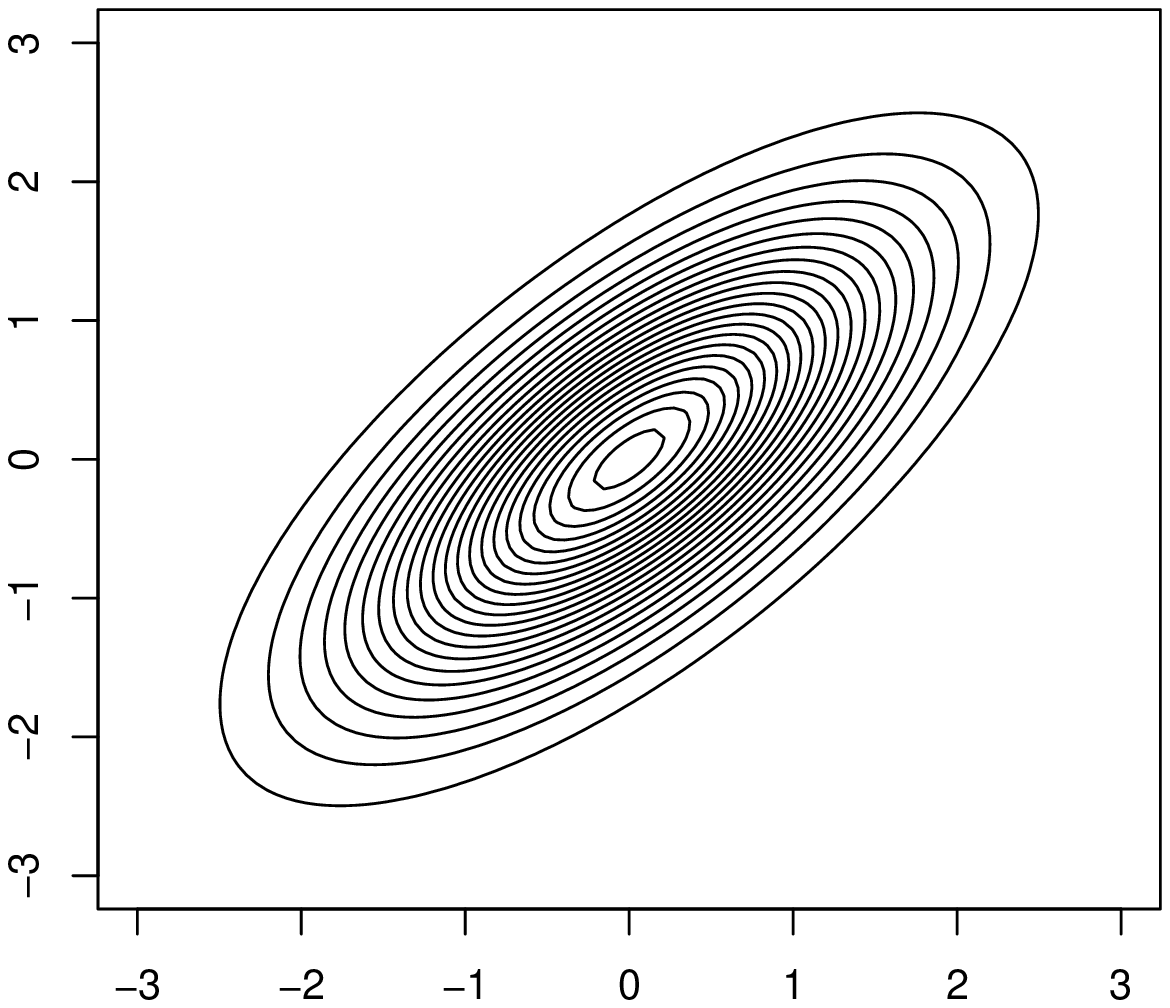}
&
\includegraphics[width=0.3\textwidth]{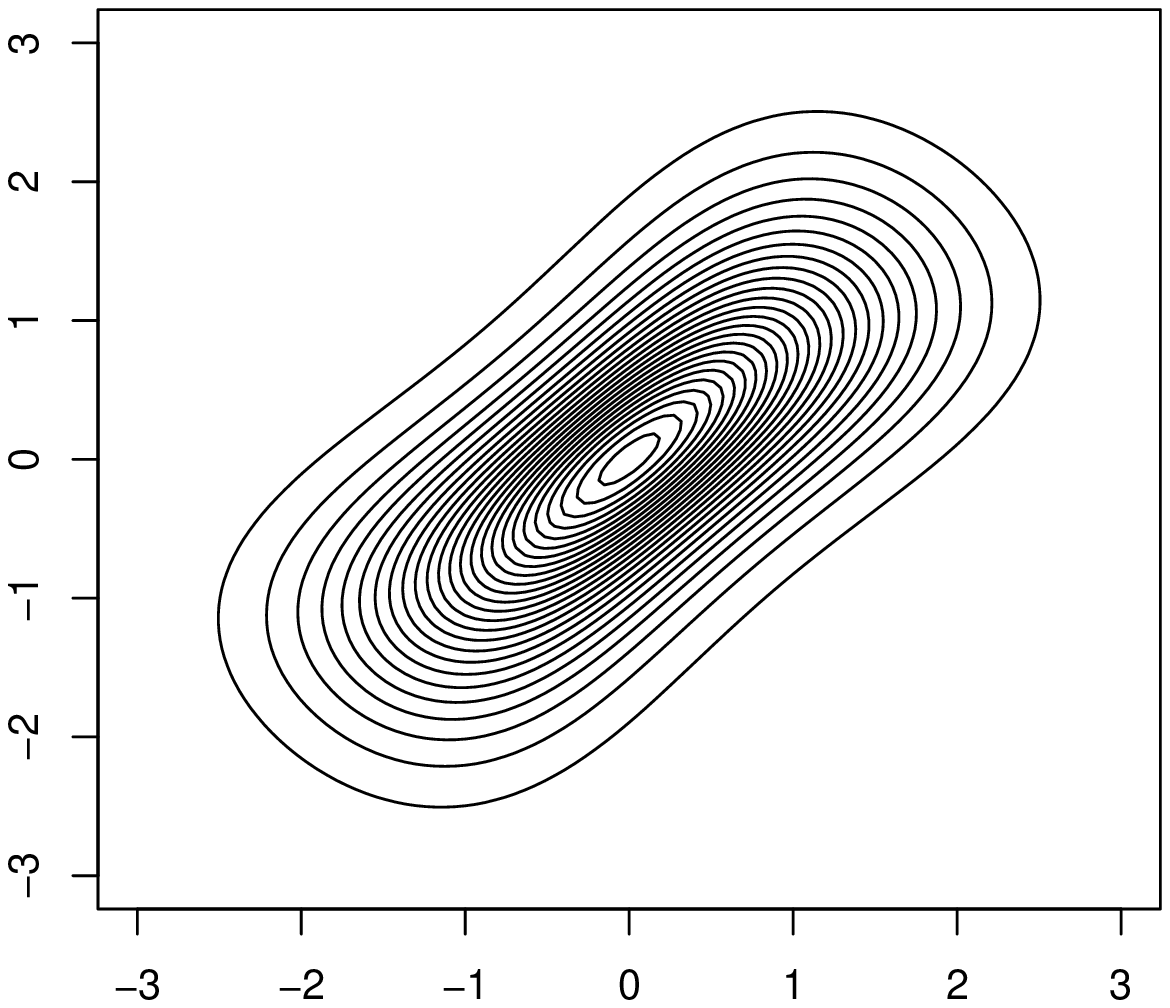}&
\includegraphics[width=0.3\textwidth]{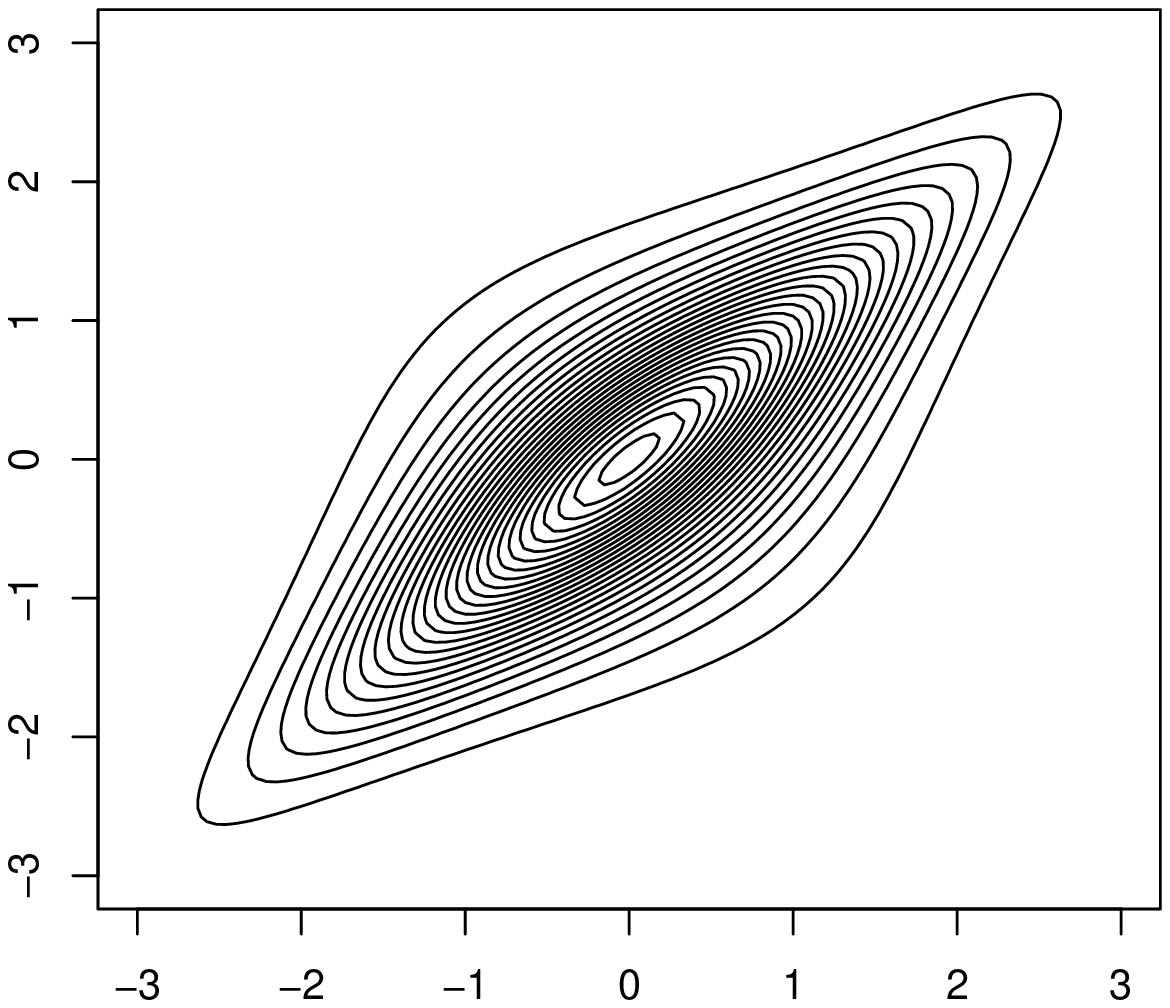}\\\hline

Joe& reflected Joe
&1-reflected Joe
\\\hline
\includegraphics[width=0.3\textwidth]{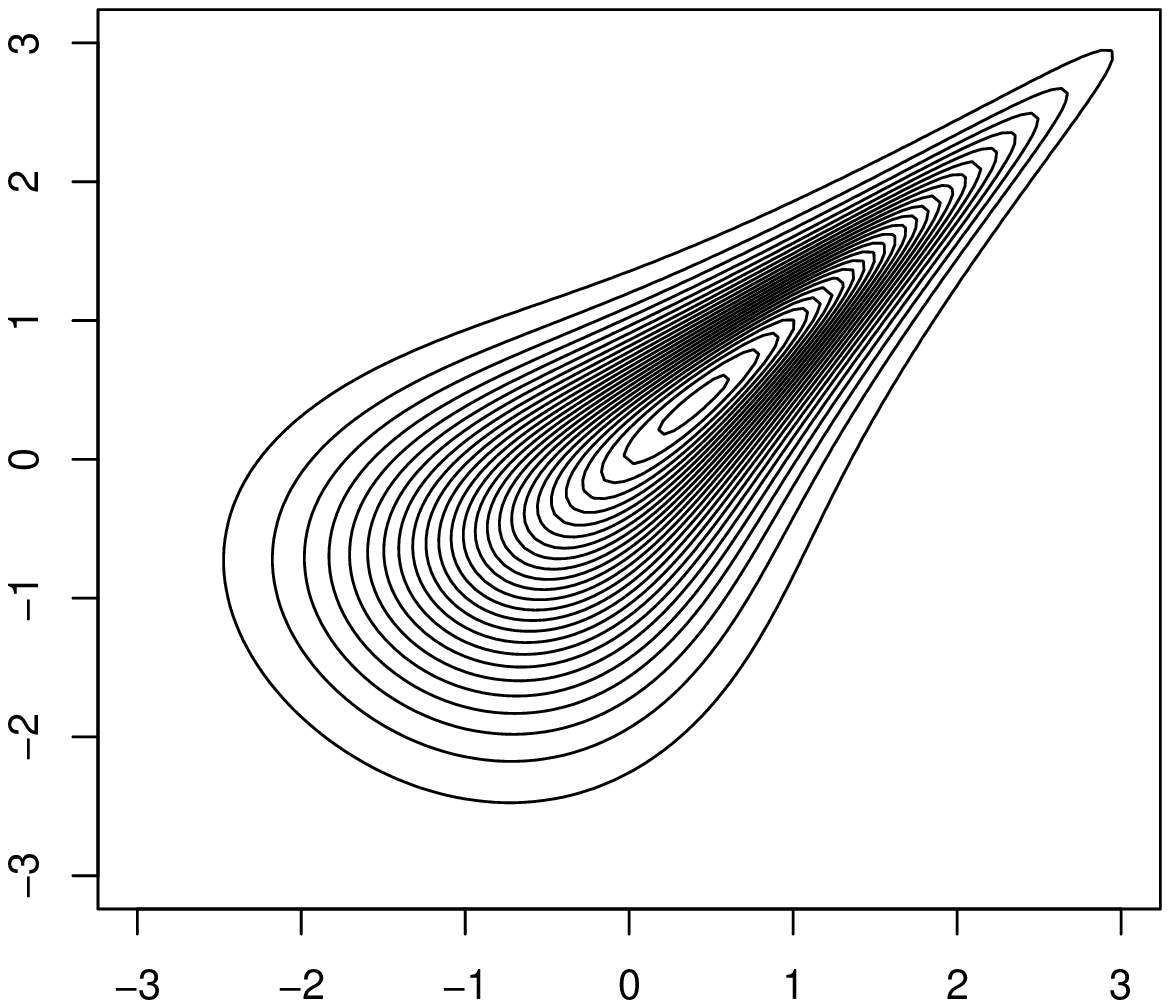}
&
\includegraphics[width=0.3\textwidth]{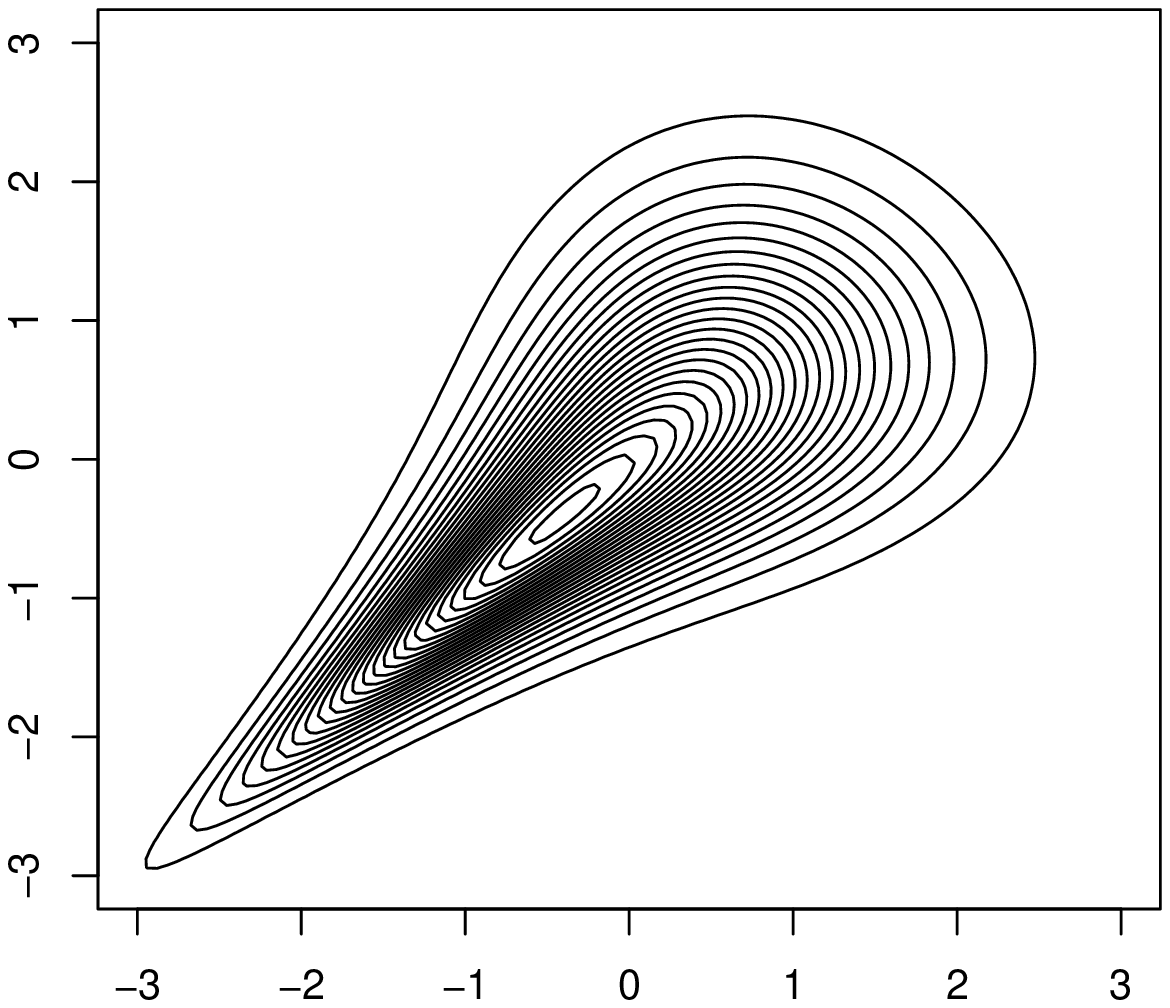}&
\includegraphics[width=0.3\textwidth]{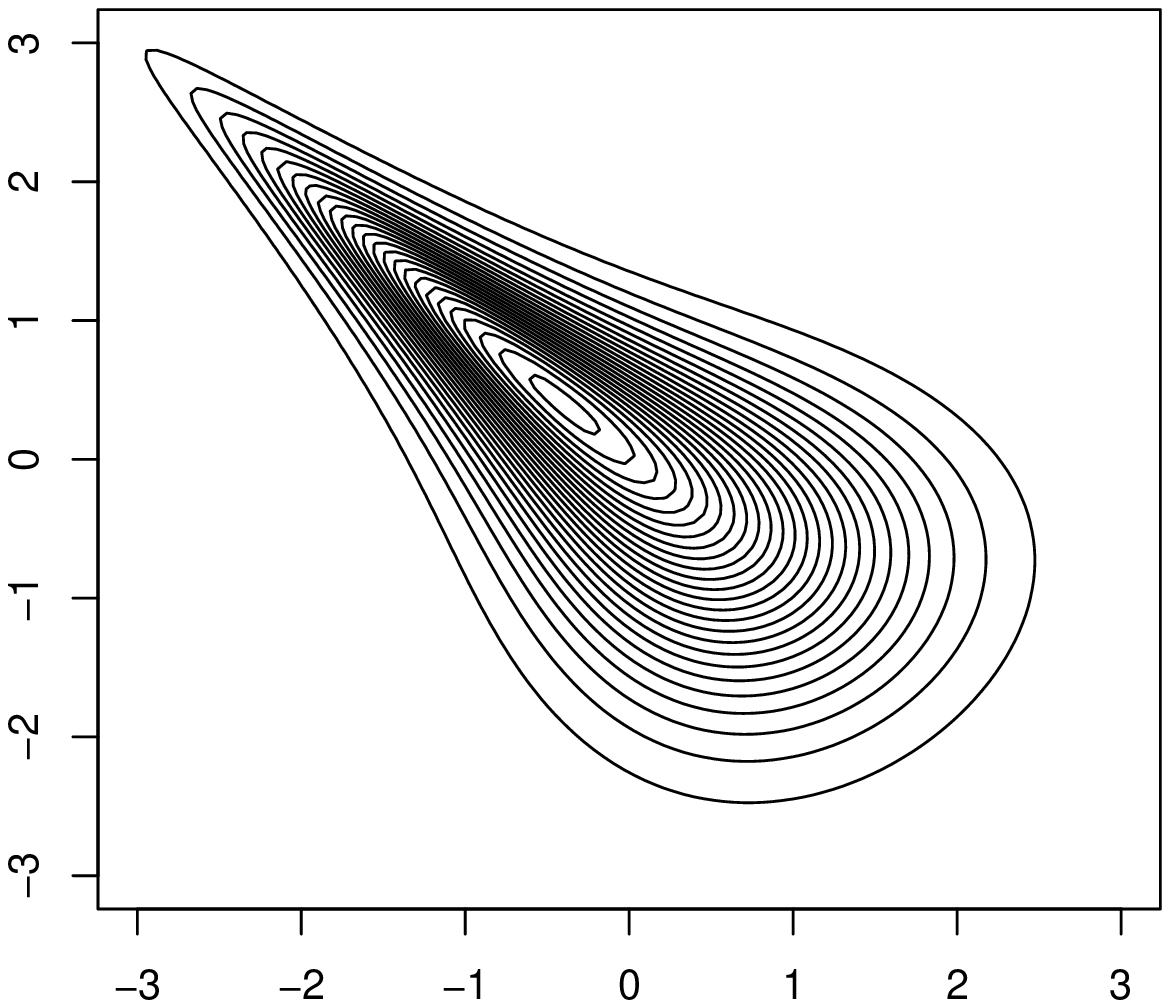}
\\
\hline
BB10& reflected BB10
&2-reflected BB10
\\\hline
\includegraphics[width=0.3\textwidth]{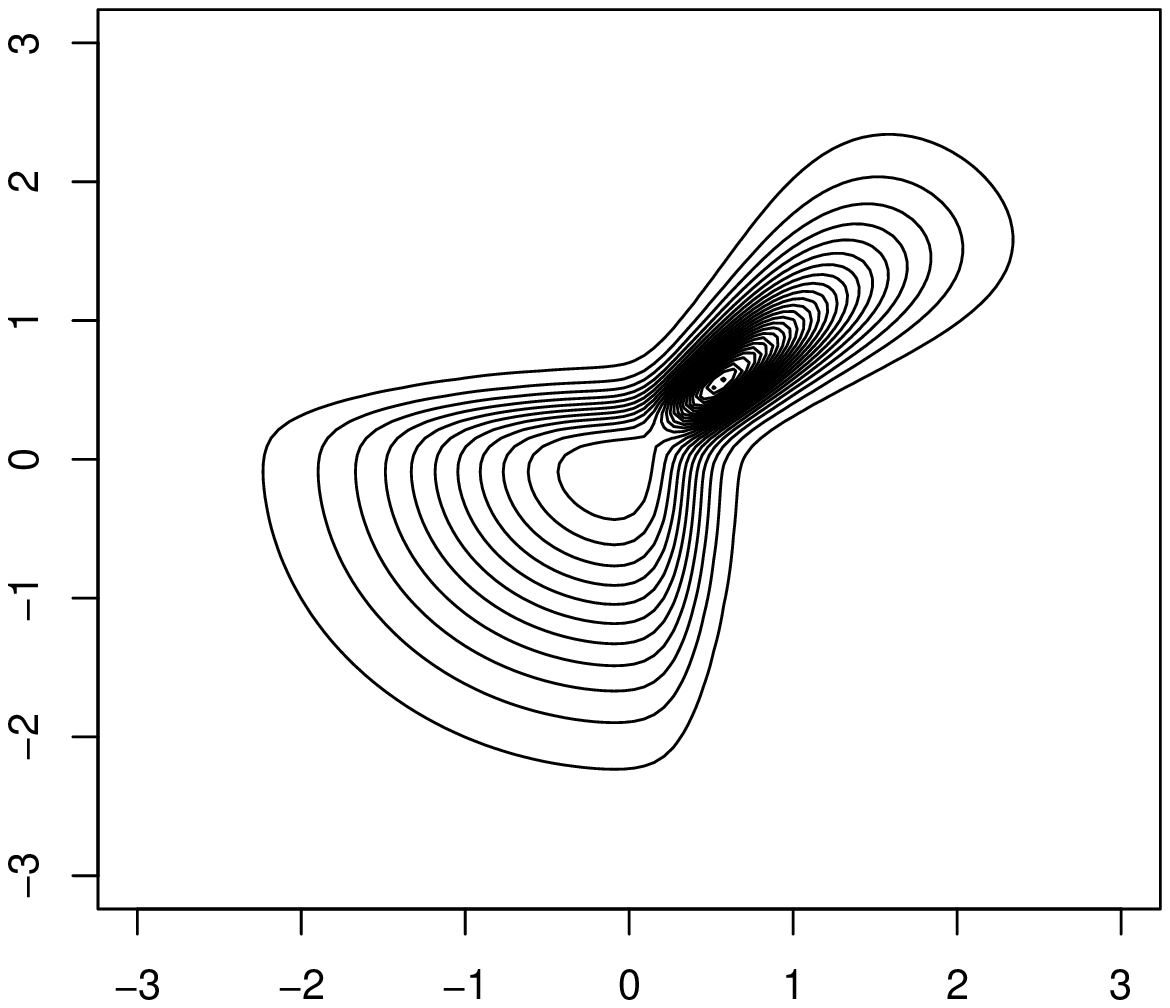}
&
\includegraphics[width=0.3\textwidth]{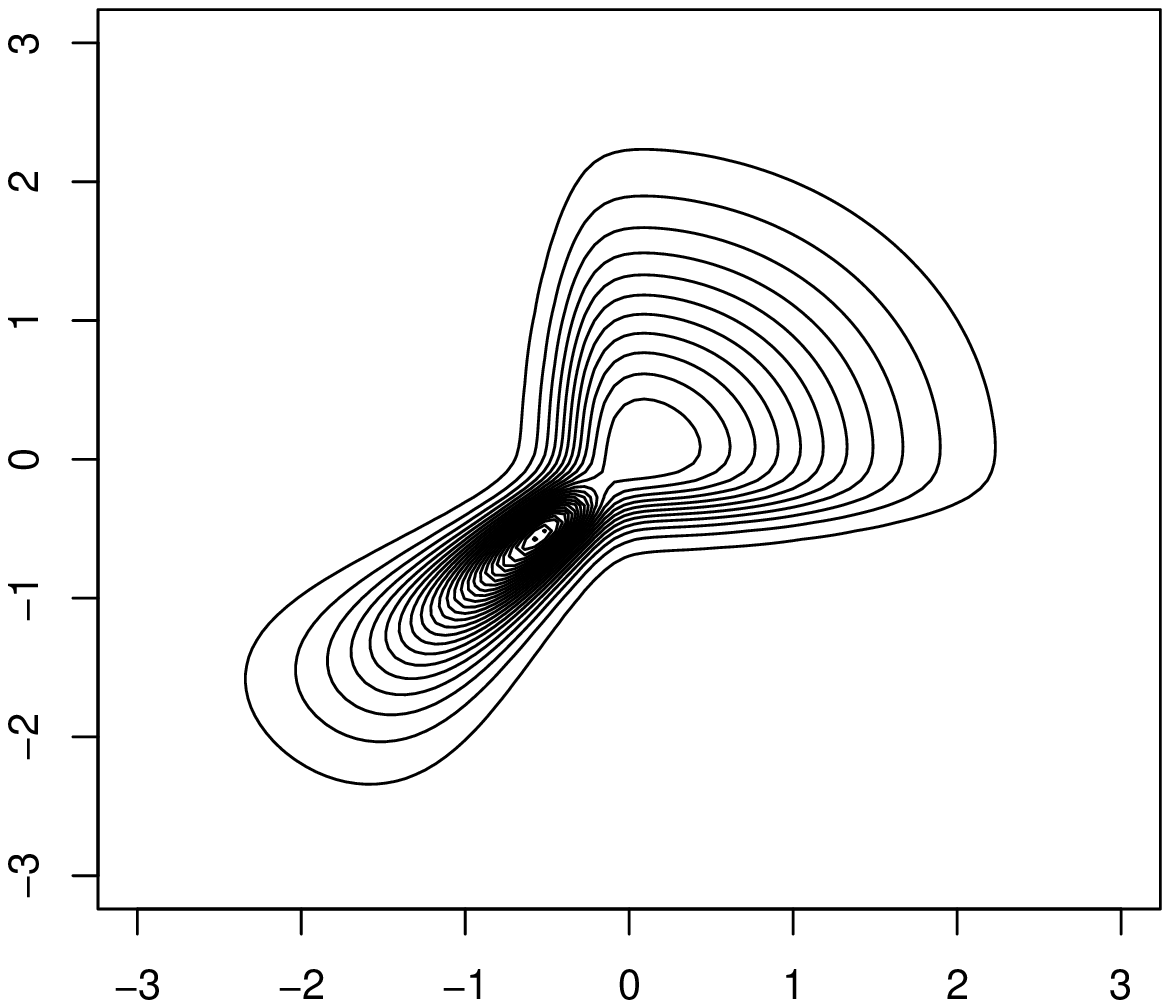}&
\includegraphics[width=0.3\textwidth]{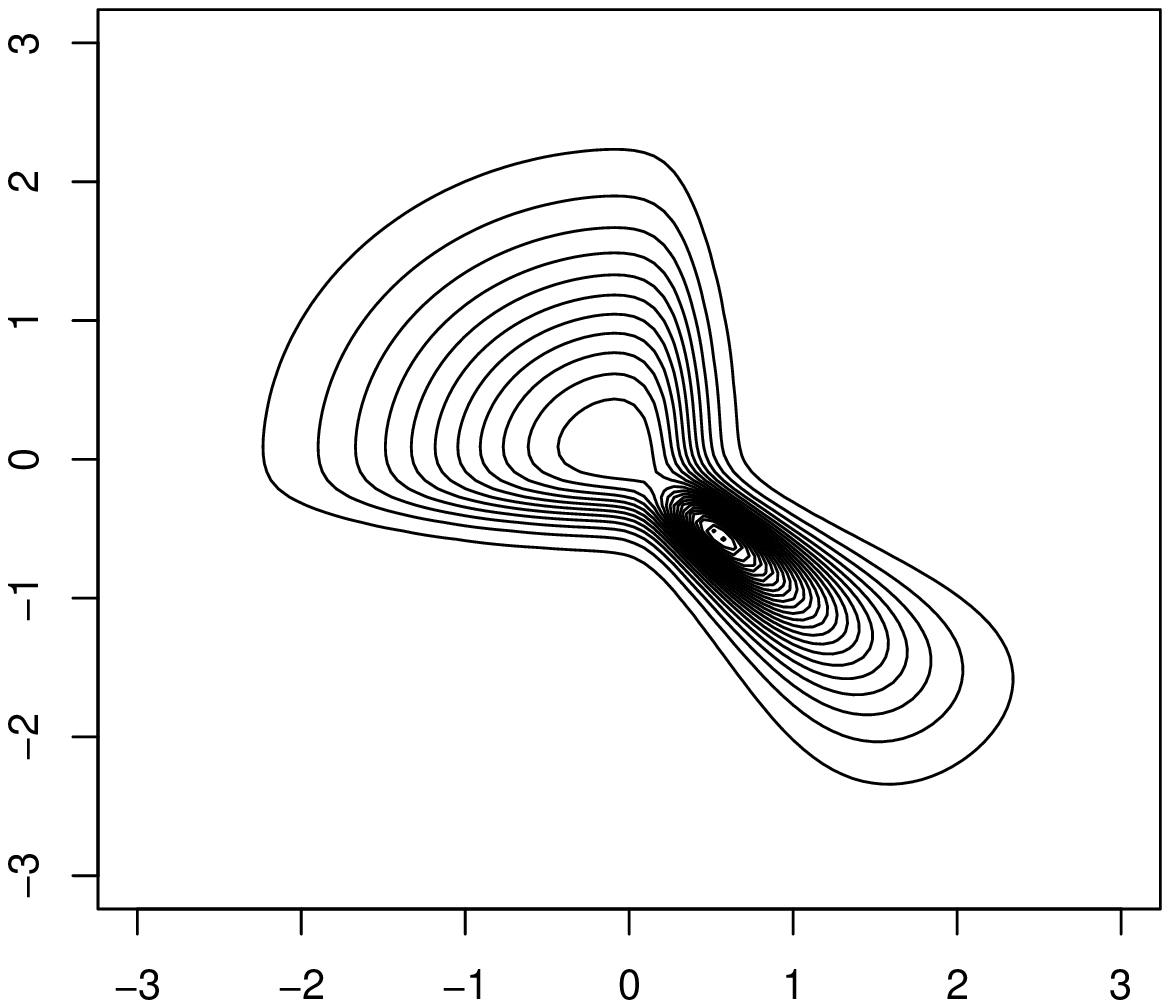}\\\hline
\end{tabular}
\end{center}
\end{figure}

In  Figure \ref{contours}, to depict the concepts of refection symmetric or asymmetric tail dependence or quadrant tail independence, we show contour plots of the corresponding copula densities with standard normal margins and dependence parameters corresponding to Kendall's $\tau$ value of 0.5. Sharper corners (relative to ellipse) indicate tail dependence.

\subsection{Semi-correlations to detect  tail dependence or tail asymmetry }
Choices of copulas with upper or lower tail dependence are better if the observed variables  have more probability in joint upper or lower tail than would be expected with the standard factor model. 
This can be shown with  summaries of  
correlations in the upper joint tail and lower joint tail.

For continuous variables, although copula theory uses transforms to standard uniform margins $U_j=F_j(Y_j)$, we convert to normal scores  $Z_j=\Phi^{-1}(U_j)$ to check deviations from the elliptical shape that would be expected with the BVN copula \citep{Nikoloulopoulos&Joe&Li2012-CSDA}.  
The correlations of normal scores in the upper and lower tail, hereafter semi-correlations, are defined as \citep[page 71]{Joe2014-CH}: 
\begin{eqnarray*}
\rho_N^+&=&\mbox{Cor}\Bigl(Z_{j_1},Z_{j_2}|Z_{j_1}>0,Z_{j_2}>0\Bigr)\\
&=&\frac{\int_{0}^{\infty}\int_{0}^{\infty} z_1z_2\phi(z_1)\phi(z_2)c\bigl(\Phi(z_1),\Phi(z_2)\bigr)dz_1dz_2-{\biggr(\int_0^\infty z\phi(z)\Bigr(1- C_{2|1}\bigr(0.5|\Phi(z)\bigl)\Bigr)dz\biggr)^2}/{C(0.5,0.5)}}
{\int_0^\infty z^2\phi(z)\Bigr(1- C_{2|1}\bigr(0.5|\Phi(z)\bigl)\Bigr)dz-{\biggr(\int_0^\infty z\phi(z)\Bigr(1- C_{2|1}\bigr(0.5|\Phi(z)\bigl)\Bigr)dz\biggr)^2}/{C(0.5,0.5)}};\\
\rho_N^-&=&\mbox{Cor}\Bigl(Z_{j_1},Z_{j_2}|Z_{j_1}<0,Z_{j_2}<0\Bigr)\\
&=&\frac{\int_{-\infty}^{0}\int_{-\infty}^{0} z_1z_2\phi(z_1)\phi(z_2)c\bigl(\Phi(z_1),\Phi(z_2)\bigr)dz_1dz_2-{\biggr(\int_{-\infty}^0 z\phi(z)C_{2|1}\bigr(0.5|\Phi(z)\bigl)dz\biggr)^2}/{C(0.5,0.5)}}
{\int_{-\infty}^0 z^2\phi(z) C_{2|1}\bigr(0.5|\Phi(z)\bigl)dz-{\biggr(\int_{-\infty}^0 z\phi(z)C_{2|1}\bigr(0.5|\Phi(z)\bigl)dz\biggr)^2}/{C(0.5,0.5)}}.
\end{eqnarray*}
Note in passing that for the BVN copula $\rho_N^{+}=\rho_N^{-}$ and has a closed form; see \citep[page 71]{Joe2014-CH}. 

From the above expressions, it  is apparent that  the normal scores 
semi-correlations depend only on the copula $C$ of $(U_{j_1},U_{j_2})$. Table \ref{tab:semicor} has semi-correlations for all the aforementioned bivariate parametric copulas with $\tau=\{0.3, 0.5, 0.7\}$. 
From the table we can see that $\rho_N^{+}=\rho_N^{-}$ for any reflection symmetric copula, while they are different for any reflection asymmetric one. If there is stronger upper (lower) tail dependence than with the BVN, then the upper (lower) semi-correlation is larger.

\begin{table}[!h]
  \centering
  \caption{ \label{tab:semicor} Lower semi-correlations $\rho_N^{-}$,  upper semi-correlations $\rho_N^{+}$, lower tail dependence $\lambda_{L}$, and  upper tail dependence $\lambda_{U}$, with $\tau = \{0.3, 0.5, 0.7\}$  for 1-parameter and 2-parameter bivariate copulas. }
  
  	\setlength{\tabcolsep}{12pt}	
	\renewcommand{\arraystretch}{0.85}			
	
    \begin{tabular}{ccccccccccc}
    \toprule
    Bivariate copula  & $\tau$ &       & {$\theta$}&$\delta$ &       & \multicolumn{1}{l}{  $\rho_N^{-}$ } &  $\rho_N^{+}$ &       & $\lambda_{L}$ & $\lambda_{U}$\\
    \midrule
    \multirow{3}[1]{*}{BVN} & 0.3  &       & 0.45  &       &       & 0.23  & 0.23  &       & 0.00  & 0.00 \\
          & 0.5  &       & 0.71  &       &       & 0.47  & 0.47  &       & 0.00  & 0.00 \\
          & 0.7  &       & 0.89  &       &       & 0.75  & 0.75  &       & 0.00  & 0.00 \\
          &       &       &       &       &       &       &       &       &       &  \\
    \multirow{3}[0]{*}{$t_3$} & 0.3  &       & 0.45  &       &       & 0.45  & 0.45  &       & 0.29  & 0.29 \\
          & 0.5  &       & 0.71  &       &       & 0.61  & 0.61  &       & 0.45  & 0.45 \\
          & 0.7 &       & 0.89  &       &       & 0.80  & 0.80  &       & 0.66  & 0.66 \\
          &       &       &       &       &       &       &       &       &       &  \\
    \multirow{3}[0]{*}{Frank} & 0.3  &       & 2.92  &       &       & 0.15  & 0.15  &       & 0.00  & 0.00 \\
          & 0.5  &       & 5.74  &       &       & 0.32  & 0.32  &       & 0.00  & 0.00 \\
          & 0.7  &       & 11.41 &       &       & 0.60  & 0.60  &       & 0.00  & 0.00 \\
          &       &       &       &       &       &       &       &       &       &  \\
    \multirow{3}[0]{*}{Joe} & 0.3  &       & 1.77  &       &       & 0.05  & 0.58  &       & 0.00  & 0.52 \\
          & 0.5  &       & 2.86  &       &       & 0.14  & 0.78  &       & 0.00  & 0.73 \\
          & 0.7  &       & 5.46  &       &       & 0.37  & 0.92  &       & 0.00  & 0.86 \\
          &       &       &       &       &       &       &       &       &       &  \\
    \multirow{3}[0]{*}{Gumbel} & 0.3  &       & 1.43  &       &       & 0.16  & 0.46  &       & 0.00  & 0.38 \\
          & 0.5  &       & 2.00  &       &       & 0.36  & 0.67  &       & 0.00  & 0.59 \\
          & 0.7  &       & 3.33  &       &       & 0.64  & 0.85  &       & 0.00  & 0.77 \\
          &       &       &       &       &       &       &       &       &       &  \\
    \multirow{3}[0]{*}{BB1} & 0.3  &       & 0.50  & 1.14  &       & 0.43  & 0.25  &       & 0.30  & 0.17 \\
          & 0.5  &       & 0.35  & 1.71  &       & 0.52  & 0.59  &       & 0.31  & 0.50 \\
          & 0.7  &       & 1.33  & 2.00  &       & 0.85  & 0.72  &       & 0.77  & 0.59 \\
          &       &       &       &       &       &       &       &       &       &  \\
    \multirow{3}[0]{*}{BB7} & 0.3  &       & 1.40  & 0.40  &       & 0.28  & 0.37  &       & 0.18  & 0.36 \\
          & 0.5  &       & 1.50  & 1.57  &       & 0.66  & 0.42  &       & 0.64  & 0.41 \\
          & 0.7  &       & 4.00  & 2.00  &       & 0.73  & 0.85  &       & 0.71  & 0.81 \\
          &       &       &       &       &       &       &       &       &       &  \\
    \multirow{3}[0]{*}{BB8} & 0.3  &       & 3.92  & 0.60  &       & 0.10  & 0.22  &       & 0.00  & 0.00 \\
          & 0.5  &       & 4.51  & 0.80  &       & 0.20  & 0.52  &       & 0.00  & 0.00 \\
          & 0.7  &       & 6.89  & 0.90  &       & 0.41  & 0.84  &       & 0.00  & 0.00 \\
          &       &       &       &       &       &       &       &       &       &  \\
    \multirow{3}[1]{*}{BB10} & 0.3  &       & 1.60  & 0.83  &       & 0.18  & 0.09  &       & 0.00  & 0.00 \\
          & 0.5  &       & 2.50  & 0.98  &       & 0.43  & 0.19  &       & 0.00  & 0.00 \\
          & 0.7  &       & 10.00 & 1.00  &       & 0.25  & 0.66  &       & 0.00  & 0.00 \\
    \bottomrule
    \end{tabular}%
\end{table}%

The population versions $\rho_N^{+},\rho_N^{-}$    also apply when the variables $Y_j$ are ordinal. Under the univariate probit model \citep[Section 3.3.2]{agresti2010} $Z_j$ are standard normal underlying latent variables,  such that 
\begin{equation}\label{underlying}
Y_j=y_j \quad \mbox{if} \quad 
\alpha_{y_j-1,j}\leq Z_j\leq  \alpha_{y_jj},\,y_j=1,\ldots,K_j,
\end{equation}
where $K_j$ is the number of categories of $Y_j$ and $a_{1j},\ldots,a_{K_j-1,j}$ are the univariate cutpoints
(without loss of generality, we assume $\alpha_{0j}=-\infty$ and  $\alpha_{K_jj}=\infty$). Note in passing that for  binary variables ($K_j=2$) the calculation of the semi-correlations is meaningless as the binary variables have  no tail asymmetries.

The sample versions of $\rho_N^{+},\rho_N^{-}$  are sample linear  (when both variables are continuous),  polychoric (when both variables are ordinal),  and polyserial (when one variable is continuous and the
other is ordinal) correlations  in the joint lower and upper  quadrants of the two variables.  The sample polychoric and polyserial correlation is defined as
$$\hat\rho_N=\mbox{argmax}_\rho\sum_{i=1}^n\log\Bigl(\Phi_2(\alpha_{y_{i1}},\alpha_{y_{i2}})-\Phi_2(\alpha_{y_{i1}-1},\alpha_{y_{i2}})-\Phi_2(\alpha_{y_{i1}},\alpha_{y_{i2}-1})+\Phi_2(\alpha_{y_{i1}-1},\alpha_{y_{i2}-1})\Bigr)$$
and 
$$\hat\rho_N=\mbox{argmax}_\rho\sum_{i=1}^n\log\biggl\{\phi(z_{i1})\biggl(\Phi\Bigl(\frac{\alpha_{y_{i2}}-\rho z_{i1}}{(1-\rho^2)^{1/2}}\Bigr)-\Phi\Bigl(\frac{\alpha_{y_{i2}-1}-\rho z_{i1}}{(1-\rho^2)^{1/2}}\Bigr)\biggr)\biggr\}$$ with $z_{ij}=\Phi\Bigl(\frac{1}{n+1}\sum_{i=1}^n\mathbf{1}(Y_{ij}\leq y_{ij})\Bigr)$, respectively.

\baselineskip=25pt

 \section{\label{Est-section}Estimation}
 
We use a two-stage copula modelling approach  toward the estimation of a multivariate model that borrows  the strengths  of the semi-parametric and IFM approach in   \cite{Genest-etal-1995} and  \cite{Joe2005-JMVA}, respectively. Suppose that the data are $y_{ij},\, j = 1,\ldots,d,\, i = 1,\ldots,n$, where $i$ is an index for individuals or clusters and $j$ is an index for the within-cluster measurements. For $i=1,\ldots,n$ we start from a $d$-variate sample $y_{i1},\ldots,y_{id}$  from which $d$ estimators  $F_1(y_{i1}),\ldots,F_d(y_{id})$ can be obtained. We use these   to transform the $y_{i1},\ldots,y_{id}$ sample to a uniform sample $u_{i1}=F_1(y_{i1}),\ldots,u_{id}=F_d(y_{id})$ on $[0, 1]^d$ and then  fit the factor copula model at the second step.  For continuous and discrete data $y_{ij}$, we  use   non-parametric and parametric univariate distributions, respectively, to transform the data $y_{ij}$ to copula data $u_{ij}=F_j(y_{ij})$, i.e., data on the uniform scale. Hence our proposed approach, in line with  the approaches in \cite{Genest-etal-1995} and \cite{Joe2005-JMVA},  can be  regarded as a two-step approach on the original data or simply as the standard one-step  ML method on the transformed (copula) data.

\subsection{Univariate modelling}
For continuous  random variables,  we estimate each marginal distribution non-parametrically by the empirical distribution function of $Y_j$, viz.  
$$F_j(y_{ij})=\frac{1}{n+1}\sum_{i=1}^n\mathbf{1}(Y_{ij}\leq y_{ij})=R_{ij}/(n+1),$$
where $R_{ij}$ denotes the rank of
$Y_{ij}$ as in the semi-parametric estimation of  \cite{Genest-etal-1995} and \cite{Shih&Louis1995}. Hence we  allow the distribution of the continuous margins to be quite free and not restricted by parametric families. 

Nevertheless, rank-based methods cannot be used for discrete variables with copulas \citep{genest&neslehova07}. Hence,  
for both  ordinal and count variables we have  chosen realistic parametric models: 

\begin{itemize}

\item For an ordinal response variable $Y_j$  we use the  univariate probit model in (\ref{underlying}).
The ordinal  response $Y_j$ is assumed to have density
$$f_j(y_j;\gbf_j)=\Phi(\alpha_{y_jj})-\Phi(\alpha_{y_j-1,j}),$$
where   $\gbf_j=(a_{1j},\ldots,a_{K_j-1,j})$ is the vector of the univariate cutpoints.

\item For a count response variable $Y_j$ we use the negative binomial distribution \citep{lawless87}. It allows for over-dispersion
and its probability mass function
is
  $$ f_j(y_j;\gbf_j)=\frac{\Gamma(\xi_j^{-1}+y_j)}{\Gamma(\xi_j^{-1})\; y_j!}
  \frac{\mu^y_j\xi^y_j}{(1+\xi_j^{-1})^{\xi_j^{-1} + y_j}},\quad
  y_j=0,1,2,\ldots,\quad \mu_j>0,\; \xi_j>0,$$
where $\gbf_j=\{\mu_j,\xi_j\}$ is the vector with the mean and dispersion parameters. In the limit  $\xi\to 0$ the negative binomial reduces to Poisson, which belongs to the
exponential family of distributions and it is the only distribution for count data that    existing latent variable models for mixed data can accommodate.   
\end{itemize}
To this end, for a discrete random variable $Y_j$, we approach estimation  by maximizing the univariate log-likelihoods 
$$\ell_j(\gbf_j)=\sum_{i=1}^n \log f_j(y_{ij};\gbf_j)$$
over the vector of the univariate parameters $\gbf_j$.  
That is equivalent with the first step of the IFM method in \cite{Joe1997-CH,Joe2005-JMVA}. 

 In line with the IFM method, if one uses a misspecified univariate model for the discrete responses at the first step, then the estimation of the copula parameters at the second step deteriorates as demostrated in \cite{Kim&Silvapulle07}. 
Nevertheless, there is no ``correct specification" of the margins or copula for
data analysis. If one does a proper analysis of the univariate margins
for goodness-of-fit, then the proposed two-stage  (or the  IFM) method should be fine. 
\cite{Kim&Silvapulle07} have ``true univariate distributions for simulations"
and ``specified univariate distributions for estimation" that were
very far apart and unrealistic, because the difference of the two
is easily detected without too much data.

\subsection{Copula modelling}
After estimating  the univariate marginal distributions we proceed to estimation of the dependence parameters.  For the 1-factor and 2-factor models, we let $C_{X_1j}$ and 
 $C_{X_2j}$ be parametric bivariate copulas, say with dependence parameters $\theta_j$ and $\delta_j$, respectively.  Let also  $\thbf=\{\gbf_{j}, \theta_j: j=1,\ldots,d\}$ and $\thbf=\{\gbf_{j},\theta_j,\delta_j: j=1,\ldots,d\}$ to denote the set of all parameters 
for the 1- and 2-factor model, respectively.  Estimation can be achieved by maximizing  the joint log-likelihood 
\begin{equation}\label{joint-loglik}
\ell_\Y(\thbf)=\sum_{i=1}^n\log f_\Y(y_{i1},\ldots,y_{id};\thbf).
\end{equation}
  over the copula parameters $\theta_j$ or $\delta_j,\, j=1,\ldots,d$ with the univariate parameters/distributions fixed as estimated at the first step of the proposed two-step estimation approach. 
 The estimated parameters can be obtained by 
using a quasi-Newton \cite{Nash1990} method applied to the logarithm of the joint likelihood.  
This numerical  method requires only the objective
function, i.e.,  the logarithm of the joint likelihood, while the gradients
are computed numerically and the Hessian matrix of the second
order derivatives is updated in each iteration. The standard errors (SE) of the  estimates can be obtained via the gradients and the Hessian computed numerically during the maximization process. These SEs are adequate to assess the flatness of the log-likelihood. Proper SEs that account for the estimation of univariate parameters can be obtained by maximizing the joint likelihood in (\ref{joint-loglik}) at one step over $\thbf$. 

For factor copula  models  numerical evaluation of the joint density $f_\Y(\mathbf{y;\thbf})$ can be  easily done using  Gauss-Legendre  quadrature  \citep{Stroud&Secrest1966}.
To compute one-dimensional integrals for the
1-factor model, we use the following approximation:

$$f_\Y(\mathbf{y})=\int_0^1\prod_{j=1}^d f_{j|X_1}(y_j|x) \,dx\approx\sum_{q=1}^{n_q}w_q\prod_{j=1}^df_{j|X_1}(y_j|x_q),$$
where $\{x_q: q=1,\ldots,n_q\}$ are the quadrature points
and $\{w_q: q=1,\ldots,n_q\}$ are the quadrature weights.
To compute two-dimensional integrals for the
2-factor model, the approximation  uses
Gauss-Legendre quadrature points in a double sum:
\begin{eqnarray*}f_\Y(\mathbf{y})&=&\int_0^1\int_0^1\prod_{j=1}^d
f_{X_2j|X_1}\bigl(x_2,y_j|x_1\bigr)\,dx_1 dx_2\\
&\approx&
\sum_{q_1=1}^{n_q}\sum_{q_2=1}^{n_q}w_{q_1}w_{q_2}\prod_{j=1}^df_{X_2j|X_1}\bigl(x_{q_2},y_j|x_{q_1}\bigr).
\end{eqnarray*}

With Gauss-Legendre quadrature, the same nodes and weights
are used for different functions;
this helps in yielding smooth numerical derivatives for numerical optimization via quasi-Newton \citep{Nash1990}.
Our comparisons show that $n_q=25$ is adequate with good precision.

\section{Model selection} \label{modSel-section}
In this section we propose an  heuristic method that automatically selects  the bivariate parametric copula families that link the observed to the latent variables.  This is very useful when the direction to the tail asymmetry based on semi-correlations is not consistent or clear. 
For multivariate mixed data, it is infeasible to estimate all possible combinations of bivariate parametric copula families, and compare them on the basis of information criteria. 
We develop an algorithm that can quickly select a factor  copula model that accurately captures  the (tail) dependence features in the data on hand. 
The linking copulas at each factor are selected  with a sequential algorithm under the initial assumption that linking  copulas are Frank, and  then sequentially  copulas with non-tail quadrant independence are assigned to any of pairs where necessary to account for tail asymmetry (discrete data) or tail dependence (continuous data).

For the 1-factor model, the proposed model selection algorithm  is summarized in the following steps:

\begin{enumerate}
\itemsep=10pt
\item For $j=1,\ldots,d$ estimate the marginal distributions $F_j(y)$.

\item Fit the 1-factor copula model with Frank  copulas  to link each of the  $d$ observed variables with the latent variable, i.e., maximise the log-likelihood function of the factor copula model in (\ref{joint-loglik}) over the vector of copula parameters $(\th_1,\ldots,\th_d)$.

\item If the $j$th linking copula has $\hat{\th}_j > 0$, then  select a set of copula candidates with ability to interpolate between independence and comonotonicity, otherwise 
 select a set of copula candidates with ability to interpolate between countermonotonicity and independence.

\item For $j= 1,\ldots,d$, 
\begin{enumerate}

\item Fit all the possible  1-factor copula models iterating over all the  copula candidates for the $j$th variable. 

\item Select the copula family that corresponds to the lowest information criterion, say the Akaike, that is  $\text{AIC}= -2 \times \ell +2 \times \#\text{copula parameters}$.

\item Fix the selected linking copula family for the $j$th variable.
\item Iterate through step (a) -- (c) to select the copulas that link all the observed variables to the 1st factor. 
\end{enumerate}
\end{enumerate}
For more than one factor  we can select the appropriate linking copulas accordingly. We first select copula families in the first factor, and then we proceed to the next factor and apply exactly the same algorithm.  

\section{\label{gof-section}Techniques for parametric  model comparison and goodness-of-fit}

Factor copula models with different bivariate linking copulas can be compared via the
log-likelihood or AIC at the maximum likelihood estimate. In addition, we will use the Vuong's test \citep{Vuong1989-Econometrica}
 to show if  a factor copula model  provides better fit than the standard factor  model with a latent additive structure, that is a factor copula model with BVN bivariate linking copulas \citep{Krupskii&Joe-2013-JMVA,Nikoloulopoulos2015-PKA}. 
The Vuong's test is the sample version of the difference in Kullback-Leibler divergence between two models and can be used to differentiate two  parametric models which could be non-nested. This test has been used extensively in the copula literature to compare vine copula models (e.g., \citealt{Brechmann-Czado-Aas-2012,Joe2014-CH,Nikoloulopoulos2015c}). We provide specific details in Section \ref{vuong}.

Furthermore, to assess the overall goodness-of-fit of the  factor copula models for mixed data, we will use appropriately the limited information $M_2$ statistic \citep{Maydeu-Olivares&Joe2006}. The $M_2$ statistic has been developed for goodness-of-fit testing in multidimensional contingency tables. 
\cite{Nikoloulopoulos2015-PKA} has used the $M_2$ statistic to assess the goodness-of-fit of factor copula models for ordinal data. We build on the aforementioned papers and propose a methodology to assess the overall goodness-of-fit of factor copula models for mixed continuous and discrete responses. We provide the specifics for the $M_2$ statistic in Section \ref{m2gof}.

\subsection{\label{vuong}Vuong's test for parametric  model comparison} 
In this subsection, we summarize the Vuong's test for comparing parametric models \citep{Vuong1989-Econometrica}.
Assume that we have Models 1 and 2 with parametric densities $f^{(1)}_\Y$ and $f^{(2)}_\Y$, respectively. We can
compare

\begin{align*}
& \Delta_{1f_\Y} = n^{-1} \Big[ \sum_{i=1}^{n}  \Bigl \{ E_{f_\Y} \log f_\Y(\y_i)-  E_{f_\Y}  \log f^{(1)}_\Y(\y_i;\thbf_1) \Bigr\} \Bigr],\\
& \Delta_{2f_\Y} = n^{-1} \Bigl[ \sum_{i=1}^{n}   \Bigl \{ E_{f_\Y}  \log f_\Y(\y_i)-  E_{f_\Y}  \log f^{(2)}_\Y(\y_i;\thbf_2)  \Bigr \} \Bigr].
\end{align*}
where $\thbf_1,\thbf_2$ are the parameters in Models 1 and 2, respectively, that lead to the closest Kullback-Leibler divergence to the true $f_\Y$; equivalently, they are the limits in probability of the MLEs based on Models 1 and 2, respectively.

Model 1 is closer to the true $f_\Y$, i.e., is the better fitting model if $\Delta=\Delta_{1f_\Y}-\Delta_{2f_\Y}<0$, and Model 2 is the better fitting model if $\Delta>0$. The sample version of $\Delta$ with MLEs $\widehat\thbf_1,\widehat\thbf_2$ is
$$\bar D=\sum_{i=1}^n D_i/n,$$
where $D_i=\log\left[\frac{f^{(2)}_\Y(\y_i;\widehat\thbf_2)}{f^{(1)}_\Y(\y_i;\widehat\thbf_1)}\right]$. 
 \cite{Vuong1989-Econometrica} 
has shown that asymptotically  that
$$\sqrt{n}\bar D/s\sim N(0,1),$$
where $s^2=\frac{1}{n-1}\sum_{i=1}^n(D_i-\bar D)^2$. 
Hence, its $95 \%$ confidence interval (CI) is $\bar{D}  \pm 1.96 \times \frac{1}{\sqrt{n}} \sigma$.

\subsection{\label{m2gof}M$_2$ goodness-of-fit statistic}

Since the $M_2$ statistic has been developed for multivariate ordinal data \citep{Maydeu-Olivares&Joe2006}, we propose to first transform the continuous and count variables to ordinal and then calculate the $M_2$ statistic at the maximum likelihood estimate before transformation.
 
Continuous variables 
can be transformed to ordinal with 
 categories that are meaningful  both practically and scientifically.
If this is not the case, we propose an unsupervised strategy of transforming a continuous  into an ordinal variable:

\begin{enumerate}
\item Set the number of ordinal categories $K_j$.
\item Transform $Y_j$  to a standard uniform random variable $U_j$ using its  empirical 
distribution function.  

\item Set the ordinal cutpoints on the uniform scale by generating  a regular sequence from 1 to $K_j-1$ and then dividing over $K_j$.   

\item Divide the range of $U_j$ into intervals with breaks the ordinal cutpoints. 

\item Transform $U_j$ to an ordinal variable $Y_j$ according to which interval its values fall.
\end{enumerate}

Count variables that  contain very high counts or very low counts, 
can be  treated as ordinal where  the first or the last category contains all the low or high counts, respectively, and their other values remain as they are. 
We further propose an unsupervised  strategy of categorising a count  into an ordinal variable:

\begin{enumerate}
\item Set the number of ordinal categories $K_j$.

\item Divide the range of $Y_j$ into intervals with breaks a regular sequence of length $K_j+1$ from $\min(Y_j)$ to  $\max(Y_j)$. 

\item Transform $Y_j$ to an ordinal variable  according to which interval its values fall.
\end{enumerate}

After applying the transformations as above for each continuous  or count variable,  we have   $d$ ordinal variables $Y_1,\ldots,Y_d$ (both the original and the transformed ones) where the $j$th $(1\leq j\leq d)$ variable consists of $K_j\geq 2$ categories labelled as   $ 0, 1, \ldots, K_j-1$. 
Consider the set of univariate and bivariate residuals that do not include category 0.  This is a residual vector of dimension
\begin{align*}
s = \sum_{j=1}^{d} (K_j-1)  + \sum_{ 1 \leq j_1 < j_2 \leq d }(K_{j_1} - 1)(K_{j_2} - 1).
\end{align*}

For a factor copula  model with parameter vector $\thbf$ of dimension $q$,
let $\pibf_2(\thbf)=\bigl(\dot{\pibf}_1(\thbf)^\top,\dot{\pibf}_2(\thbf)^\top\bigr)^\top$ be the column vector of the model-based marginal probabilities with  $\dot{\pibf}_1(\thbf)$
 the 
 vector of univariate marginal probabilities,
and $\dot{\pibf}_2(\thbf)$  the 
vector of bivariate marginal probabilities.
Also, let $\pp_2=(\dot{\pp}_1^\top,\dot{\pp}_2^\top)^\top$ be the vector of the observed sample proportions, with  $\dot{\pp}_1$
 the  
 vector of  univariate marginal proportions,
and $\dot{\pp}_2$  the  
vector of the bivariate marginal proportions.

With a sample size $n$,
the limited information statistic $M_2$ is given by
\begin{equation}\label{M_2}
M_2=M_2(\hat\thbf)=n\bigl(\pp_2-\pibf_2(\hat\thbf)\bigr)^\top \C_2(\hat\thbf)\bigl(\pp_2-\pibf_2\bigl(\hat\thbf)\bigr),
\end{equation}
with
\begin{equation}\label{C_2}
\C_2(\thbf)=\Xibf_2^{-1}-\Xibf_2^{-1}\Debf_2(\Debf_2^\top\Xibf_2^{-1}\Debf_2)^{-1}\Debf_2^\top\Xibf_2^{-1}
=\Debf_2^{(c)}\bigl([\Debf_2^{(c)}]^\top\Xibf_2\Debf_2^{(c)}\bigr)^{-1}[\Debf_2^{(c)}]^\top,
\end{equation}
where $\Debf_2=\partial\pibf_2(\thbf)/\partial\thbf^\top$ is an $s\times q$ matrix with the derivatives of all the  univariate and bivariate marginal probabilities with respect to the model parameters, $\Debf_2^{(c)}$ is an $s\times(s-q)$ orthogonal complement to $\Debf_2$,
such that $[\Debf_2^{(c)}]^\top\Debf_2=\mathbf{0}$, and $\Xibf_2=\mbox{diag}(\pibf_2(\thbf))-\pibf_2(\thbf)\pibf_2(\thbf)^\top$ is the $s\times s$ covariance matrix of all the univariate and bivariate marginal sample proportions, excluding category 0. Due to equality in (\ref{C_2}), $\C_2$ is invariant to the choice of orthogonal complement.
The limited information statistic $M_2$ has a null asymptotic
distribution that is $\chi^2$ with $s-q$ degrees of freedom
when the estimate $\hat \thbf$ is $\sqrt{n}$-consistent.
For details on  the computation of $\Xibf_2$ and  $\Debf_2$ for factor copula models we refer the interested reader to  \cite{Nikoloulopoulos2015-PKA}.

\section{Applications}\label{sec-application}
In this section we illustrate the proposed methodology by re-analysing three mixed response datasets.

Initially, we use the diagnostic method in \citet[pages 245-246]{Joe2014-CH}  to show that each dataset (or more precisely the 
correlation matrix of  the observed variables for each dataset)   has  a factor  structure based on linear factor analysis. The  correlation matrix $\R_{\mathrm{observed}}$  has been obtained based on the sample correlations  from the bivariate pairs of the observed variables. 
These are the linear  (when both variables are continuous),  polychoric (when both variables are discrete),  and polyserial (when one variable is continuous and the other is discrete) sample correlations among the observed variables. 
The resulting $\R_{\mathrm{observed}}$  is generally positive definite  if the sample size is not small enough; if not one has to convert it to positive definite. We calculate various measures of discrepancy between $\R_{\mathrm{observed}}$  and $\R_{\mathrm{model}}$ (the resulting correlation matrix of linear factor analysis), such as  the maximum absolute correlation
difference $D_1=\max|\R_{\mathrm{model}} - \R_{\mathrm{observed}}|$, the average absolute correlation
difference $D_2=\mathrm{avg}| \R_{\mathrm{model}} - \R_{\mathrm{observed}}|$, and  the correlation matrix discrepancy measure $D_3=\log\bigl( \det(\R_{\mathrm{model}}) \bigr) - \log\bigl( \det(\R_{\mathrm{observed}})\bigr) + \tr( \R^{-1}_{\mathrm{model}} \R_{\mathrm{observed}} ) - d$.

After confirming that a factor model with a parsimonious correlation structure is reasonable,  we calculate the semi-correlations   for  each  pair of observed variables  to check if there is tail asymmetry.  This will be a useful information for choosing potential parametric bivariate copulas other than the BVN copulas that lead to the standard factor model. Note that when the variables are negatively associated  we   calculate the sample semi-correlations in the lower-upper and upper-lower quadrant.

After 
motivating  why more flexible dependencies are needed in cases of mixed data and how those dependencies in the data can be captured by  suitable  bivariate copulas,  we proceed with factor copula models and   construct a plausible factor copula model, to capture any type of   
 reflection asymmetric  dependence, by using the proposed algorithm in Section 
 \ref{modSel-section}.  
For a baseline comparison, we first  fit the factor copula models with the  comprehensive bivariate parametric copula families that allow for  reflection symmetric dependence;  these  are the BVN, Frank, and $t_{\nu}$ copulas. For $t_{\nu}$ copulas, we summarize the choice of integer $\nu$ with the largest log-likelihood. For the standard 2-factor model, to obtain a unique solution we must impose sufficient constraints.  One
parameter for the second factor can be set to zero and the likelihood can be maximized with respect to
other $2d-1$ parameters. We report the varimax transform \citep{Kaiser1958}
of the loadings (a reparametrization of $2d$ parameters), converted to
factor copula parameters   via the relations  
\begin{equation}\label{2-fact-bvn-param}
\th_j=\b_{j1}, \quad  \de_j=\frac{\b_{j2}}{(1-\b_{j1}^2)^{1/2}},
\end{equation}
where $\b_{j1}$ and $\b_{j2}$ are the loadings at the first and second factor, respectively \citep{Krupskii&Joe-2013-JMVA, Nikoloulopoulos2015-PKA}. 

If the number of parameters is not the same between the models, we use the AIC as a rough diagnostic measure for goodness-of-fit between the models, otherwise we use  the likelihood at the maximum likelihood estimates. We further compute the Vuong's tests with Model 1 being the factor copula  model with BVN copulas, that is the standard factor  model, to reveal if any other factor copula  model provides better  fit than the standard factor model.  
To make it easier to compare strengths of dependence, we convert the estimated parameters to Kendall's $\tau$'s in $(-1, 1)$ via  the relations in \citet[Chapter 4]{Joe2014-CH};  SEs are also converted via the delta method. For the model that provides the best fit, we provide the estimates and SEs that are  obtained by maximizing the joint likelihood in (\ref{joint-loglik}) at one step over $\thbf$. Although, the two-stage estimation approach in Section \ref{Est-section} is a convenient way to quickly compare candidate factor copula models, the  full likelihood is applied for the best fitting factor copula model. 
The overall fit of the factor copula models is evaluated using the $M_2$ statistic. 
Note that the $M_2$ statistic in the case with $2d-1$ copulas (one set to
independence for the second
factor) is computed with $\Debf_2$ having one less column.

\subsection{\label{PE-sec}Political-economic dataset} 
 
\cite{Quinn2004} considered measuring  the (latent) political-economic risk  of 62 countries, for the year 1987. The political-economic risk is defined as the country's risk in manipulating economic rules for its own and constituents' advantages (see e.g.,  \citealt{North&Weingast1989}). \cite{Quinn2004} used 5 mixed variables, namely the continuous variable black-market  premium  in each country (used as a proxy for illegal economic activity),  the  continuous variable productivity as measured by real gross domestic product  per worker in 1985 international prices, the binary variable  independence of the national judiciary (1 if the judiciary is judged to be independent and 0 otherwise), and  the  ordinal variables measuring the lack of expropriation risk and lack of corruption. The dataset  and its complete description  can be found in \cite{Quinn2004}  or in the {\tt R} package {\tt MCMCpack} \citep{Martin-etal-2011-JSS}. Note that since the continuous variable black-market  premium is negatively associated with the remaining variables
(from the context), we  re-orient it   leading to positive dependence among all the observed variables. 

\begin{table}[!h]
  \centering
   \caption{\label{tab:correlationPE} The   sample correlation $ \rho_N$, lower semi-correlation $\rho_N^{-}$, and upper semi-correlation $\rho_N^{+}$ for each pair of variables, along with  the measures of discrepancy between the sample and the resulting correlation matrix of linear factor analysis with  1 and 2 factors for the political-economic risk data. }

    \setlength{\tabcolsep}{37pt}  

    \begin{tabular}{llccc}
    \toprule
    \multicolumn{2}{c}{pairs of variables}  & $\rho_N$ & $\rho_N^{-}$ & $\rho_N^{+}$ \\
    \midrule
    BM & GDP & 0.53 & -0.04 & 0.57 \\
    BM & IJ & 0.61 & - & - \\
    BM & XPR  & 0.67 & 0.88 & 0.63 \\
    BM & CRP  & 0.62 & 0.16 & 0.55 \\
    GDP & IJ & 0.78 & - & - \\
    GDP & XPR  & 0.55 & 0.11 & 0.75 \\
    GDP & CRP  & 0.77 & 0.24 & 0.63 \\
    IJ & XPR  & 0.91 & - & - \\
    IJ & CRP  & 0.87 & - & - \\
    XPR  & CRP  & 0.76 & 0.71 & 0.71 \\ 
 
    \bottomrule
    \end{tabular}%

       \setlength{\tabcolsep}{47pt}  
    \begin{tabular}{cccc}
   
    $\#$ factors    & $D_1$& $D_2$ & $D_3$ \\
    \hline
    1& 0.16  & 0.04  & 0.91 \\
    2 & 0.06  & 0.01  & 0.22 \\
    \bottomrule
    \end{tabular}%
   \begin{flushleft}
\begin{footnotesize}
BM: black-market premium; GDP: gross domestic product; IJ: independent judiciary; XPR: lack of expropriation risk; CPR: lack of corruption.
\end{footnotesize}  
\end{flushleft}
\end{table}%

Table \ref{tab:correlationPE} shows that the  sample correlation matrix   of the mixed responses has an 1-factor structure based
on linear factor analysis (large  $D_3$ is due to the small sample size as demonstrated using simulated data in Section \ref{sec:simulations}). The sample semi-correlations  in Table \ref{tab:correlationPE} show that there is more probability in the upper tail or lower tail compared with a discretized MVN, suggesting that a factor model with bivariate parametric copulas with upper  or lower tail  dependence might provide a better fit.
Table \ref{onefPE} gives the estimated parameters, their standard errors (SE) in Kendall’s $\tau$ scale, joint log-likelihoods, the $95\%$ CIs of Vuong's tests, and the $M_2$ statistics for the 1-factor copula models.
Table \ref{onefPE} also indicates the parametric copula family chosen for each pair using the proposed heuristic algorithm. Copulas with asymmetric  dependence are selected for all the copulas that link the  latent variable to each of the observed variables. Hence, it is  revealed that there are features in the data such as tail dependence and asymmetry  which cannot be captured by copulas with reflection symmetric dependence such as BVN, Frank and $t_\nu$ copulas.

\begin{table}[!h]
\setlength{\tabcolsep}{8pt}
  \centering
  \caption{\label{onefPE}Estimated parameters, their standard errors (SE) in Kendall's $\tau$ scale, joint log-likelihoods, the $95\%$ CIs of Vuong's statistics, and the $M_2$ statistics for the one-factor copula models for the political-economic risk data.}
    \begin{tabular}{lccccccccclcc}
    \toprule
1-factor          & \multicolumn{2}{c}{BVN$^\P$} &       & \multicolumn{2}{c}{$t_5$} &       & \multicolumn{2}{c}{Frank} &       & \multicolumn{3}{l}{Selected } \\
\cmidrule{2-3}\cmidrule{5-6}\cmidrule{8-9}\cmidrule{12-13}         & $\hat{\tau}$  & SE    &       & $\hat{\tau}$  & SE    &       & $\hat{\tau}$  & SE    &       & copulas & $\hat{\tau}$  & SE \\
    \midrule
    BM & 0.50 & 0.06  &       & 0.51 & 0.07  &       & 0.49 & 0.06  &       &  
    Joe 
     & 0.51 &{0.05} \\
   GDP & 0.57  & 0.05  &       & 0.57  & 0.06  &       & 0.58  & 0.06  &       & Joe     & 0.58  & {0.05} \\
   IJ & 0.80  & 0.09  &       & 0.81  & 0.09  &       & 0.75  & 0.09  &       & reflected Joe   
   & 0.80  & {0.07} \\
    XPR & 0.66  & 0.06  &       & 0.68  & 0.07  &       & 0.66  & 0.06  &       & Joe    & {0.69}  & 0.06 \\
    CRP & 0.71  & 0.06  &       & 0.70  & 0.06  &       & 0.72  & 0.06  &       & Gumbel     & 0.74  & 0.06 \\
 \midrule
    $\ell$ & \multicolumn{2}{c}{-165.15} &       & \multicolumn{2}{c}{-166.25} &       & \multicolumn{2}{c}{-164.89} &       & \multicolumn{3}{c}{-151.98} \\
 Vuong $95\%$CI &       &       &       & \multicolumn{2}{c}{{(-0.051,0.015)}} &       & \multicolumn{2}{c}{ {(-0.077,0.085)} } &       & \multicolumn{3}{c}{ {(0.073,0.352)}  } \\
    $M_2$    & \multicolumn{2}{c}{179.2} &       & \multicolumn{2}{c}{187.4} &       & \multicolumn{2}{c}{177.6} &       & \multicolumn{3}{c}{129.2} \\
    df    & \multicolumn{2}{c}{134} &       & \multicolumn{2}{c}{134} &       & \multicolumn{2}{c}{134} &       & \multicolumn{3}{c}{134} \\
    $p$-value & \multicolumn{2}{c}{$<0.01$} &       & \multicolumn{2}{c}{$<0.01$} &       & \multicolumn{2}{c}{$<0.01$} &       & \multicolumn{3}{c}{0.60} \\
    \bottomrule
\end{tabular}
\begin{flushleft}
\begin{footnotesize}
$^\P$: The resulting model is the same as the standard factor model; BM: black-market premium; GDP: gross domestic product; IJ: independent judiciary; XPR: lack of expropriation risk; CPR: lack of corruption.
\end{footnotesize}  
\end{flushleft}
\end{table}

In all the fitted models the estimated Kendall's $\tau$'s are similar.  
Kendall's $\tau$ only accounts for the dependence dominated by the middle of the data, and it is expected to be similar amongst different families of copulas. However, the tail dependence and tail order vary, as explained in Section \ref{bivcop}, and they are  properties to consider when choosing amongst different families of copulas \citep{Nikoloulopoulos&Karlis2008-CSDA}. 

The table shows that the selected model using the proposed algorithm  provides the best fit and there is a  substantial improvement over the standard factor model as indicated by the Vuong's  and  $M_2$ statistics. To compute the $M_2$ statistics we transformed the continuous variables to ordinal with 5 categories using the unsupervised strategy in Section \ref{m2gof}; similar inference  was drawn, when we transformed them to ordinal with 3, 4,  or 6 categories. 
The factor copula parameter of $0.51$ on negative  black market premium indicates a negative association between the illegal economic activity and the latent variable.  All the other estimated factor copula parameters 
 indicate a positive association between each of the other  observed variables   (independent judiciary, productivity, lack of expropriation, and  lack of corruption) with the latent variable. Hence, we can interpret  the latent variable to be the political economical certainty.

\subsection{\label{GSS-section}General social survey}

\cite{Hoff2007-AAS} analysed seven demographic variables of 464 male respondents to the 1994 General Social Survey. 
Of these seven, two were continuous (income and age of the respondents), three were ordinal with $5$ categories (highest degree of the survey respondent, income and highest degree of  respondent's parents),  and two were count variables (number of children of the survey respondent and respondent's parents). The data are available in \citet[Supplemental materials]{Hoff2007-AAS}.

\begin{table}[!h]
  \centering
      \caption{\label{tab:correlationGSS} The   sample correlation $ \rho_N$, lower semi-correlation $\rho_N^{-}$, and upper semi-correlation $\rho_N^{+}$ for each pair of variables, along with  the measures of discrepancy between the sample and the resulting correlation matrix of linear factor analysis with  1, 2 and 3 factors for the general social survey dataset. }
   \setlength{\tabcolsep}{34pt}  
             
    \begin{tabular}{lcccc}
    \toprule
    \multicolumn{2}{c}{pairs of variables} & $\rho_N$ & $\rho_N^{-}$ & $\rho_N^{+}$ \\
    \midrule
    
    income  & age  & 0.29 & 0.48 & 0.23 \\
    income  & degree  & 0.52 & 0.24 & 0.33 \\
    income  & pincome  & 0.14 & 0.02 & 0.28 \\
    income  & pdegree  & 0.24 & 0.04 & 0.08 \\
    income  & child  & 0.22 & 0.23 & 0.01 \\
    income  & pchild  & -0.09 & 0.06 & 0.00 \\
    age  & degree  & 0.06 & 0.22 & -0.04 \\
    age  & pincome  & -0.11 & -0.02 & 0.12 \\
    age  & pdegree  & -0.14 & -0.42 & 0.44 \\
    age  & child  & 0.58 & 0.36 & 0.26 \\
    age  & pchild  & 0.12 & 0.18 & 0.07 \\
    degree  & pincome  & 0.21 & 0.17 & -0.05 \\
    degree  & pdegree  & 0.46 & 0.46 & 0.41 \\
    degree  & child  & -0.11 & -0.10 & -0.09 \\
    degree  & pchild  & -0.25 & -0.14 & -0.30 \\
    pincome  & pdegree  & 0.44 & 0.44 & 0.34 \\
    pincome  & child  & -0.16 & -0.15 & 0.11 \\
    pincome  & pchild  & -0.23 & 0.13 & -0.30 \\
    pdegree  & child  & -0.21 & 0.08 & 0.10 \\
    pdegree  & pchild  & -0.34 & 0.19 & -0.32 \\
    child  & pchild  & 0.20 & -0.11 & -0.06 \\

    \bottomrule
    \end{tabular}%

       \setlength{\tabcolsep}{48pt}  
    \begin{tabular}{cccc}
    $\#$ factors    & $D_1$& $D_2$ & $D_3$ \\
    \hline
    1 & 0.55  & 0.09  & 0.82 \\
    2 & 0.15  & 0.03  & 0.13 \\
    3 & 0.02  & 0.00  & 0.00 \\
    \bottomrule
    \end{tabular}%

\end{table}%

Table \ref{tab:correlationGSS}   
shows that  the sample correlation matrix of the mixed responses has a 2- or even a 3-factor structure based
on linear factor analysis. The direction to the tail asymmetry based on  sample semi-correlations  in Table \ref{tab:correlationGSS}    is not consistent, and this emerges the usefulness of the proposed model selection technique.
Table \ref{res-GSS}  gives the estimated parameters, their standard errors (SE) in Kendall’s $\tau$ scale,  the joint log-likelihoods, the $95\%$ CIs of Vuong's tests,  and the $M_2$ statistics for the 1-factor and 2-factor copula models. 
The best fit for the 1-factor model is based on the bivariate  copulas selected by the proposed algorithm, where there is  improvement over the factor copula model with BVN copulas according to the Vuong's statistic. However, assessing the overall goodness-of-fit via the  $M_2$ statistic, it is revealed that one latent variable is not adequate to explain the dependencies among the mixed responses. 
To apply the $M_2$ statistic, age  and income were transformed to ordinal  with $4$  (18--24, 25--44, 45--64, and $65+$) and $5$  (0--10, 11--19, 20--29, 30--40, and $41+$) 
categories, respectively, and   number of children of the survey respondent and respondent's parents were treated as ordinal where the 4th (more than 3 children) and  8th (more than 7 children) category, respectively, contained all the high counts. 

\begin{table}[!h]
  \centering
 \caption{ \label{res-GSS}Estimated parameters, their standard errors (SE) in Kendall's $\tau$ scale, joint log-likelihoods, the $95\%$ CIs of Vuong's statistics, and the $M_2$ statistics for the 1- and 2-factor copula models for the general social survey dataset.}
  \setlength{\tabcolsep}{5.9pt}
     \begin{tabular}{lccccccccclcc}
    \toprule
    
1-factor          & \multicolumn{2}{c}{BVN$^\P$} &       & \multicolumn{2}{c}{$t_9$} &       & \multicolumn{2}{c}{Frank} &       & \multicolumn{3}{l}{{Selected}} \\
\cmidrule{2-3}\cmidrule{5-6}\cmidrule{8-9}\cmidrule{12-13}     & $\hat{\tau}$  & SE    &       & $\hat{\tau}$  & SE    &       & $\hat{\tau}$  & SE    &       & {copulas} & $\hat{\tau}$  & SE \\
    \midrule
     income     & 0.20  & 0.04  &       & 0.20  & 0.04  &       & 0.20  & 0.04  &       & Joe    & 0.29  & 0.04 \\
     age            & -0.14 & 0.04  &       & -0.14 & 0.04  &       & -0.14 & 0.04  &       & 2-reflected Joe 
     & -0.14 & 0.03 \\
     degree    & 0.40  & 0.04  &       & 0.39  & 0.04  &       & 0.38  & 0.04  &       & $t_3$    & 0.45  & {0.04}\\
     pincome  & 0.33  & 0.03  &       & 0.34  & 0.04  &       & 0.35  & 0.04  &       & $t_3$    & 0.33  & 0.05 \\
     pdegree   & 0.62  & 0.05  &       & 0.65  & 0.05  &       & 0.68  & 0.06  &       & reflected  Gumbel 
     & 0.56  & 0.05 \\
     child        & -0.20 & 0.04  &       & -0.19 & 0.04  &       & -0.19 & 0.04  &       & 2-reflected Joe 
     & -0.14 & 0.03 \\
     pchild      & -0.32 & 0.03  &       & -0.31 & 0.04  &       & -0.32 & 0.04  &       & 2-reflected Gumbel 
     & -0.27 & 0.03 \\
 \midrule
    $\ell$& \multicolumn{2}{c}{-3425.39} &       & \multicolumn{2}{c}{-3420.56} &       & \multicolumn{2}{c}{-3433.83} &       & \multicolumn{3}{c}{-3397.79} \\
    Vuong $95\%$CI &       &       &       & \multicolumn{2}{c}{(-0.005,-0.025)} &       & \multicolumn{2}{c}{(-0.037,0.001)} &       & \multicolumn{3}{c}{ {(0.022,0.097)} }  \\
    $M_2$    & \multicolumn{2}{c}{743.74} &       & \multicolumn{2}{c}{715.45} &       & \multicolumn{2}{c}{738.76} &       & \multicolumn{3}{c}{660.47} \\
    df    & \multicolumn{2}{c}{348} &       & \multicolumn{2}{c}{348} &       & \multicolumn{2}{c}{348} &       & \multicolumn{3}{c}{348} \\
    $p$-value & \multicolumn{2}{c}{$<0.001$} &       & \multicolumn{2}{c}{$<0.001$} &       & \multicolumn{2}{c}{$<0.001$} &       & \multicolumn{3}{c}{$<0.001$} \\\midrule
 \end{tabular}
 \setlength{\tabcolsep}{5.5pt}
    \begin{tabular}{lccccccccclcc}
2-factor          & \multicolumn{2}{c}{BVN$^\P$} &       & \multicolumn{2}{c}{{$t_9$}} &       & \multicolumn{2}{c}{{Frank}} &           & \multicolumn{3}{l}{Selected} \\
\cmidrule{2-3}\cmidrule{5-6}\cmidrule{8-9}\cmidrule{12-13}    & $\hat{\tau}$  &      &  & $\hat{\tau}$  &     &       & $\hat{\tau}$  & SE    &         & copulas & $\hat{\tau}$  & SE \\
    \midrule
     \multicolumn{13}{l}{   1st factor }    \\
 
    \hline
        income   & 0.36  &       && {0.35}  &  &       & 0.13  & 0.04  &         				& reflected Gumbel 
        & 0.34  & 0.03 \\
       age           &{-0.05} &       && {-0.06} &   &       & 0.50  & 0.05  &                	& reflected Joe
       & 0.49 & 0.03 \\
       degree  & 0.55  &       && {0.53}  &    &       & -0.12  & 0.04  &             				& BVN   & 0.18  & 0.04 \\
       pincome  & 0.27  &       && {0.28}  &   &       & -0.21  & 0.04  &       						&1-reflected  Joe 
        & -0.13  & 0.04 \\
       pdegree  & 0.48  &       && {0.50}  &   &       & -0.31  & 0.05  &       	  					&  1-reflected Joe 
       & -0.13  & 0.04 \\
       child    &   -0.13 &      & & {-0.14} &   &       &  0.52  & 0.05  &       	        	& reflected Joe
        & 0.44 & 0.04 \\
       pchild  & -0.28 &       && {-0.28} & &       & 0.23 & 0.04  &       		     	& Gumbel  & 0.11 & 0.03 \\
 
	\hline	           
     \multicolumn{13}{l}{   2nd factor }    \\ 
    \hline
     income      		&   0.38  &      & & {0.41}  &  &       & 0.50  & 0.06  &       						& Gumbel    & {0.40}  & 0.04 \\
     age            		& 0.54  &       && {0.55}  &   &       & 0.21  & 0.04  &       						& 2-reflected Joe  
      & {-0.14}  & 0.03 \\
     degree 			& 0.14  &       && {0.17}  &   &       & 0.57 & 0.07  &       						& reflected Joe 
        & 0.65  & 0.06 \\
     pincome 		& -0.09 &     &  & {-0.08} &   &       & 0.23 & 0.04  &       						& Gumbel   & {0.30} & 0.04 \\
     pdegree 			& -0.16 &     &  & {-0.14} &  &       & 0.44 & 0.05  &       		   & $t_5$   & {0.49} & 0.04 \\
     child    			& 0.53  &      & & {0.53}  &   &       & 0.08  & 0.04  &       						& BVN    & -0.24  & 0.04 \\
     pchild 			& 0.13  &       && {0.10}  &    &       & -0.24  & 0.04  &       						& 2-reflected Gumbel    
       & -0.26  & 0.03 \\
       \midrule
    $\ell$ & -3286.80 &      & & {-3278.88} &  &      & \multicolumn{2}{c}{-3300.07} &             & \multicolumn{3}{c}{-3235.86} \\    Vuong $95\%$CI &       &&       & {(-0.004,-0.038)}  &  &      & \multicolumn{2}{c}{ {(-0.058,0.001)} }&             & \multicolumn{3}{c}{(0.061,0.159)} \\
    $M_2$    & {471.47} &       && {{461.70}} & &       & \multicolumn{2}{c}{492.37} &             & \multicolumn{3}{c}{370.61} \\
    df    & 342   &       && {342} &   &     & \multicolumn{2}{c}{341} &            & \multicolumn{3}{c}{341} \\
    $p$-value & $<0.001$ &      & & $<0.001$ &    &    & \multicolumn{2}{c}{$<0.001$} &      & \multicolumn{3}{c}{0.13} \\
    \bottomrule
\end{tabular}
\begin{flushleft}
\begin{footnotesize}
$^\P$: The resulting model is the same as the standard factor model; p{demographic}: demographic variable of respondent's parents.
\end{footnotesize}  
\end{flushleft}
\end{table}

The 2-factor copula models with  BVN, $t_{\nu}$, and Frank copulas provide some improvement over the 1-factor copula models but according to the $M_2$ statistic they still  have a poor fit.  Note that the factor copula model with $t_9$ copulas was  not identifiable  (large SEs) in line with \cite{Nikoloulopoulos2015-PKA}, hence one  
parameter for the second factor was  set to zero and the likelihood was maximized with respect to the
remaining parameters. We report the varimax transform \citep{Kaiser1958}
of the loadings, converted to
factor copula parameters   via the relations in (\ref{2-fact-bvn-param}).

The selected 2-factor copula model using the algorithm in Section \ref{modSel-section} shows improvement over the standard factor model according to the Vuong's statistic and  better fit according to the $M_2$ statistic; it changes a $p$-value $<0.001$ to one $>0.10$.
For the 2-factor model based on the proposed algorithm for model selection, note that, without the need for a varimax rotation, the unique loading parameters ($\hat{\tau}$'s converted to normal copula parameters $\hat\th_j$'s and  $\hat\de_j$'s and then to loadings using the  relations in (\ref{2-fact-bvn-param})) show that one factor is loaded only on the demographic variables of the respondent's parents.

\subsection{Swiss consumption survey}

\cite{Irincheeva&Cantoni&Genton2012-SJS} considered measuring the latent variable `financial wealth of the household' in its different realizations by analysing  seven household variables of $n=9960$ respondents to the Swiss consumption survey. 
Out of these seven, three were 
continuous (food, clothing and leisure expenses), three were binary  (dishwasher, car, and motorcycle), and one was count variable (the number of bicycles in possession of the household).

\begin{figure}[!h]
\begin{center}
\caption{\label{normal-scores} Bivariate normal scores plots, along with correlations and semi-correlations for the continuous data from the Swiss consumption survey.}
\begin{tabular}{cc}
\includegraphics[width=0.45\textwidth]{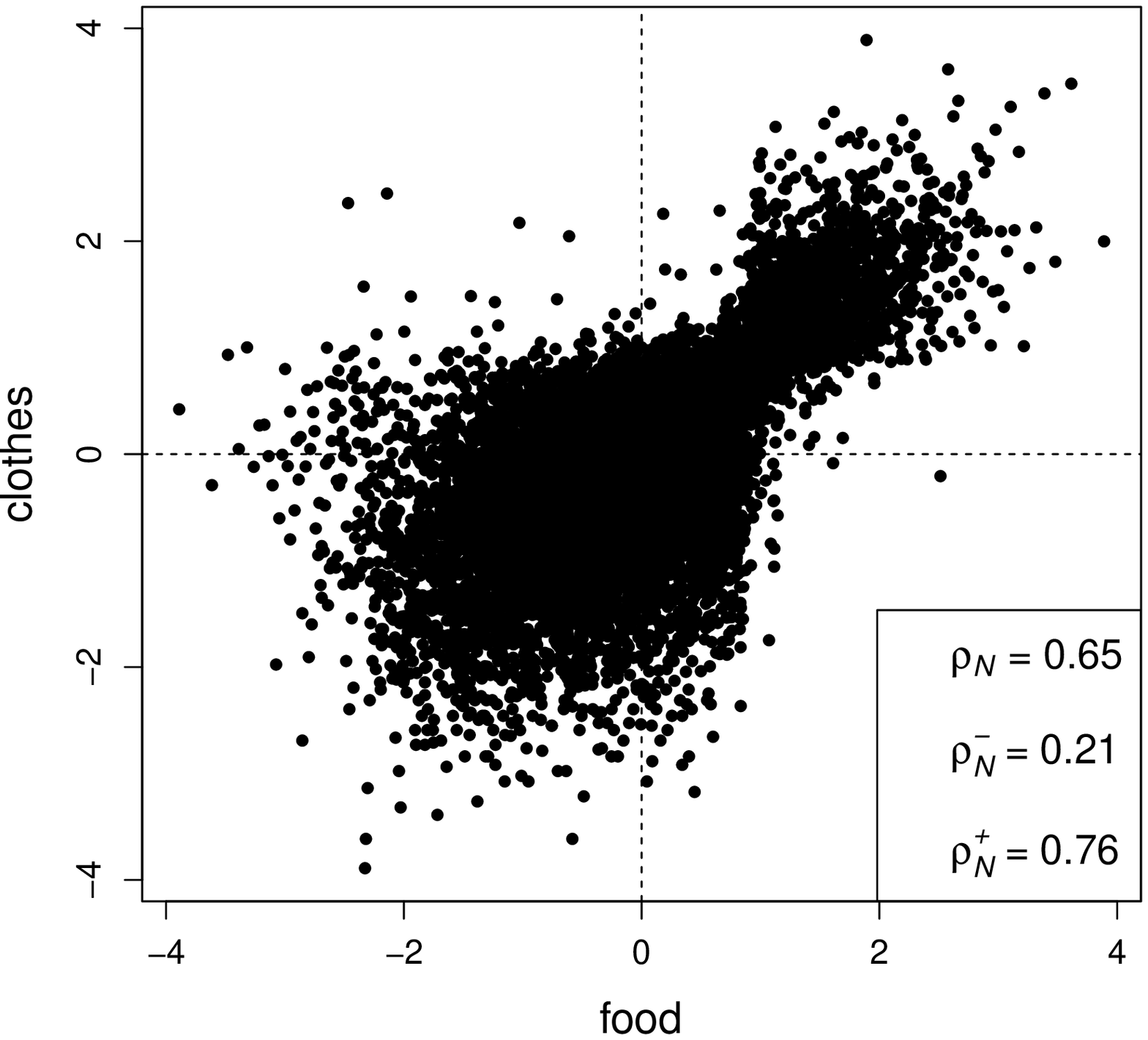}
&
\includegraphics[width=0.45\textwidth]{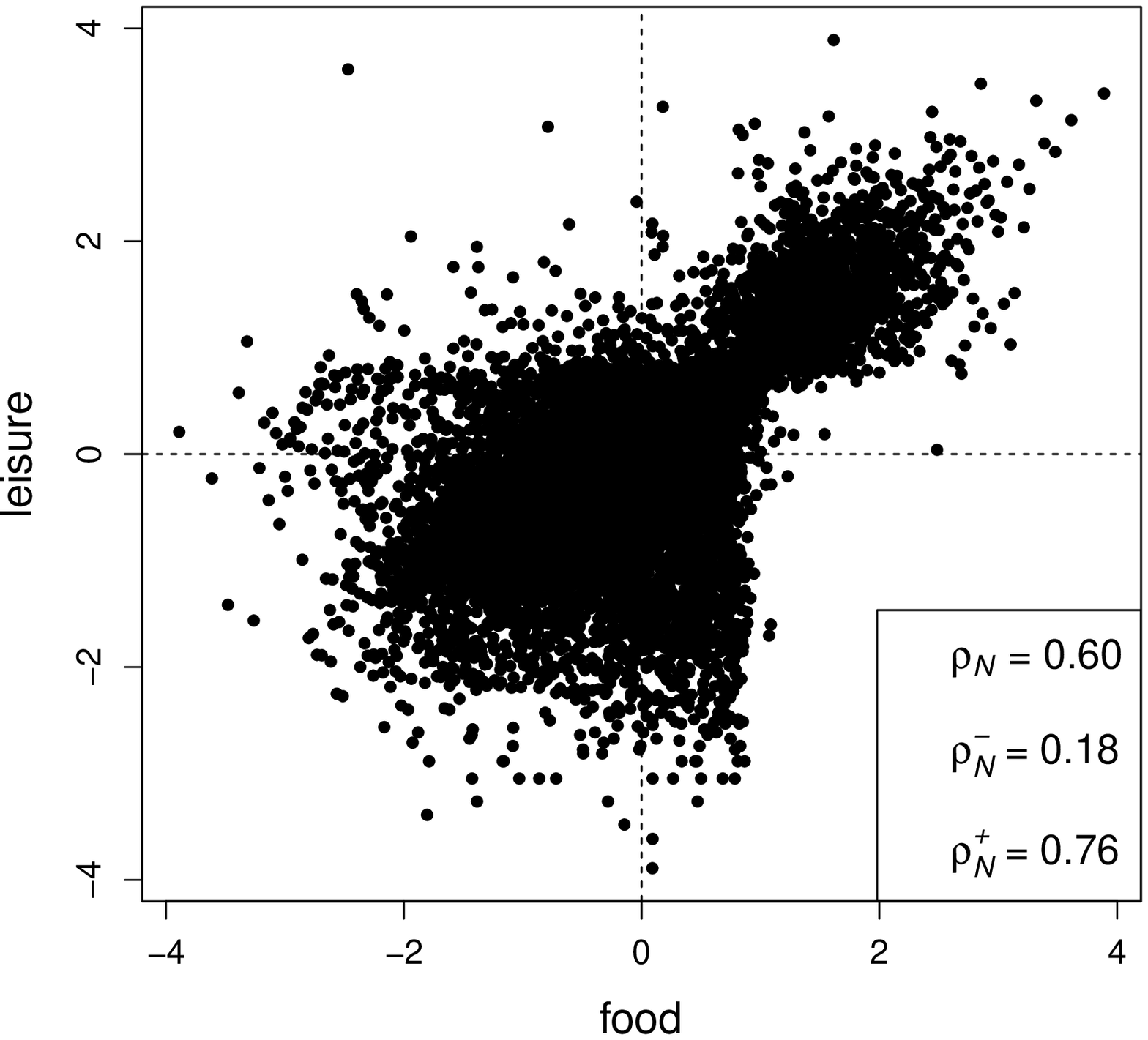}\\
\multicolumn{2}{c}{\includegraphics[width=0.45\textwidth]{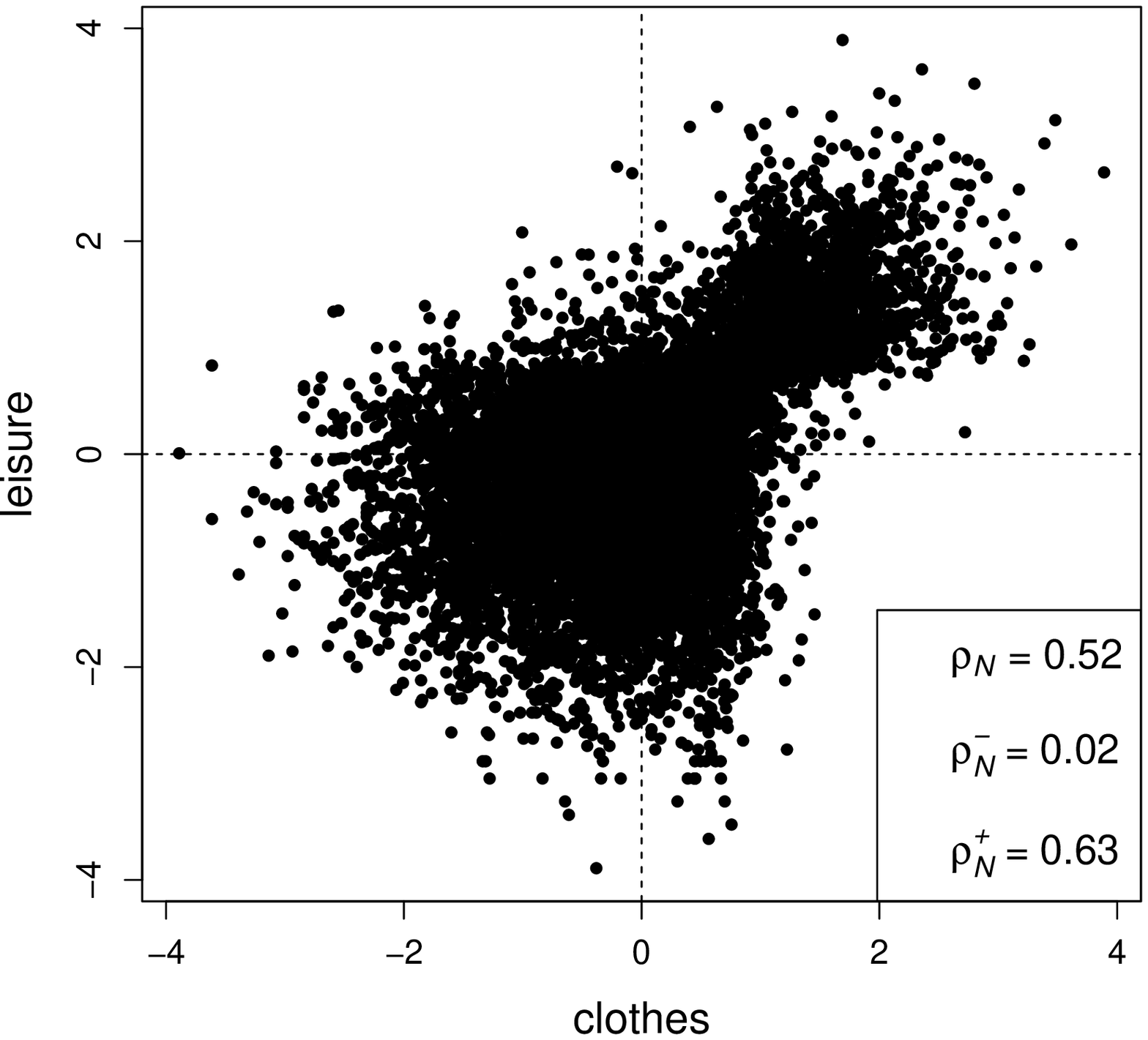}} \\
\end{tabular}
\end{center}
\end{figure}

\cite{Irincheeva&Cantoni&Genton2012-SJS}, with simple descriptive statistics such as scatter  plots of the original data, have shown that these mixed responses  have  reflection asymmetric  dependence, and fitted their latent variable approach  with one and two latent variables. In Figure \ref{normal-scores} we depict the bivariate normal scores plots  for the continuous data along with their correlations and semi-correlations. 
 With a bivariate normal scores plot one can check for deviations from the elliptical shape that would be expected with the BVN copula, and hence assess if tail asymmetry and tail dependence exists on the data. For all the pairs the upper semi-correlation is larger, and interestingly, contrasting the bivariate normal scores plots in Figure \ref{normal-scores} with the contour plots in Figure \ref{contours},  it is apparent that for the continuous 
 variables the linking copulas might be the BB10 copulas.

  \begin{table}[!h]
  \centering
      \caption{\label{tab:correlationSwiss} The   sample correlation $ \rho_N$, lower semi-correlation $\rho_N^{-}$, and upper semi-correlation $\rho_N^{+}$ for each pair of variables, along with  the measures of discrepancy between the sample and the resulting correlation matrix of linear factor analysis with  1, 2 and 3 factors for the Swiss consumption survey dataset.}
         
      \setlength{\tabcolsep}{33pt}  
            
    \begin{tabular}{llccc}
    \toprule
    \multicolumn{2}{c}{pairs of variables}  & $\rho_N$ & $\rho_N^{-}$ & $\rho_N^{+}$ \\
    \midrule
 
    food & clothes & 0.65 & 0.21 & 0.76 \\
    food & leisure  & 0.60 & 0.18 & 0.76 \\
    food & dishwasher  & 0.31 & - & - \\
    food & car  & 0.38 & - & - \\
    food & motorcycle  & 0.11 & - & - \\
    food & bicycles & 0.21 & 0.22 & 0.02 \\
    clothes & leisure  & 0.52 & 0.02 & 0.63 \\
    clothes & dishwasher  & 0.23 & - & - \\
    clothes & car  & 0.25 & - & - \\
    clothes & motorcycle  & 0.07 & - & - \\
    clothes & bicycles & 0.18 & 0.15 & 0.02 \\
    leisure  & dishwasher  & 0.24 & - & - \\
    leisure  & car  & 0.18 & - & - \\
    leisure  & motorcycle  & 0.01 & - & - \\
    leisure  & bicycles & 0.08 & 0.04 & 0.08 \\
    dishwasher  & car  & 0.43 & - & - \\
    dishwasher  & motorcycle  & 0.03 & - & - \\
    dishwasher  & bicycles & 0.24 & - & - \\
    car  & motorcycle  & 0.18 & - & - \\
    car  & bicycles & 0.26 & - & - \\
    motorcycle  & bicycles & 0.21 & - & - \\

    \bottomrule
    \end{tabular}%

       \setlength{\tabcolsep}{48pt}  
    \begin{tabular}{cccc}
  $\#$ factors    & $D_1$& $D_2$ & $D_3$ \\\hline
    1 & 0.27  & 0.06  & 0.26 \\
    2 & 0.12  & 0.02  & 0.06 \\
    3 & 0.03  & 0.01  & 0.01 \\
    \bottomrule
    \end{tabular}%
\end{table}%

Table \ref{tab:correlationSwiss} shows that the sample correlation matrix   of the mixed responses has a 2-factor structure based
on linear factor analysis. The sample semi-correlations  in Table \ref{tab:correlationSwiss}    show that there is more probability in the upper tail and lower tail among the continuous variables and between each of the continuous variables with the count variable, respectively,  suggesting that a factor model with bivariate parametric copulas with asymmetric tail dependence might provide a better fit.
Table \ref{res-swiss} gives the estimated parameters, their standard errors (SE) in Kendall’s tau scale,  the joint log-likelihoods, the $95\%$ CIs of Vuong's tests,  and the $M_2$ statistics for the 1-factor and 2-factor copula models.  The best fitted 1- and  2-factor models result when we use
BB10 copulas with asymmetric quadrant tail independence to  link the  latent variable to each of the continuous observed variables and copulas with lower tail dependence to link the latent variables to the discrete observed variables. 
Once again the  one-factor copula model is not adequate to explain the dependence amongst the mixed responses based on the $M_2$ statistic (Table \ref{res-swiss}, 1-factor).  To apply the $M_2$ statistic,  we transformed the continuous to ordinal variables with 3 categories using the unsupervised strategy in section \ref{gof-section} and the count variable bicycle was treated as ordinal where the 6th category contained all the high counts ($5$ bicycles or more).

\begin{table}[!h]
  \centering
 \caption{ \label{res-swiss}Estimated parameters, their standard errors (SE) in Kendall's $\tau$ scale, joint log-likelihoods, the $95\%$ CIs of Vuong's statistics, and the $M_2$ statistics for the 1- and 2-factor copula models for the Swiss consumption survey dataset.}
 \setlength{\tabcolsep}{7pt}
        \begin{tabular}{lccccccccccccc}
    \toprule
    1-factor  &   & \multicolumn{2}{c}{BVN$^\P$} &   & \multicolumn{2}{c}{$t_5$} &   & \multicolumn{2}{c}{Frank} &   & \multicolumn{1}{l}{Selected} &   &  \\
\cmidrule{1-1}\cmidrule{3-4}\cmidrule{6-7}\cmidrule{9-10}\cmidrule{13-14}      &   & $\hat{\tau}$ & SE &   & $\hat{\tau}$ & SE &   & $\hat{\tau}$ & SE &   & \multicolumn{1}{l}{Copulas} & $\hat{\tau}$ & SE \\
    \midrule
    food &   & 0.69 & 0.01 &   & 0.73 & 0.01 &   & 0.74 & 0.01 &   & \multicolumn{1}{l}{reflected BB10} & 0.79 & 0.00 \\
    clothes &   & 0.53 & 0.01 &   & 0.53 & 0.01 &   & 0.53 & 0.01 &   & \multicolumn{1}{l}{BB10} & 0.38 & 0.00 \\
    leisure &   & 0.47 & 0.01 &   & 0.50 & 0.01 &   & 0.50 & 0.01 &   & \multicolumn{1}{l}{BB10} & 0.39 & 0.00 \\
    dishwasher &   & 0.24 & 0.01 &   & 0.25 & 0.01 &   & 0.23 & 0.01 &   & \multicolumn{1}{l}{reflected Joe} & 0.28 & 0.01 \\
    car &   & 0.27 & 0.01 &   & 0.30 & 0.01 &   & 0.28 & 0.01 &   & \multicolumn{1}{l}{reflected Joe} & 0.23 & 0.01 \\
    motorcycle &   & 0.07 & 0.01 &   & 0.06 & 0.01 &   & 0.08 & 0.01 &   & \multicolumn{1}{l}{reflected Joe} & 0.13 & 0.01 \\
    bicycles &   & 0.15 & 0.01 &   & 0.15 & 0.01 &   & 0.16 & 0.01 &   & \multicolumn{1}{l}{reflected Joe} & 0.17 & 0.01 \\
    \midrule
    AIC &   & \multicolumn{2}{c}{55004.24} &   & \multicolumn{2}{c}{54221.36} &   & \multicolumn{2}{c}{55105.88} &   & \multicolumn{3}{c}{48932.32} \\
    Vuong  $95\%$ CI &   &   &   &   & \multicolumn{2}{c}{(0.032,0.046)} &   & \multicolumn{2}{c}{(-0.015,0.005)} &   & \multicolumn{3}{c}{(0.286,0.324)} \\
    $M_2$ &   & \multicolumn{2}{c}{2775.73} &   & \multicolumn{2}{c}{2734.05} &   & \multicolumn{2}{c}{2808.53} &   & \multicolumn{3}{c}{1626.54} \\
    df &   & \multicolumn{2}{c}{71} &   & \multicolumn{2}{c}{71} &   & \multicolumn{2}{c}{71} &   & \multicolumn{3}{c}{68} \\
    $p-$value  &   & \multicolumn{2}{c}{$<0.001$} &   & \multicolumn{2}{c}{$<0.001$} &   & \multicolumn{2}{c}{$<0.001$} &   & \multicolumn{3}{c}{$<0.001$} \\\midrule
    \end{tabular}%
     \setlength{\tabcolsep}{7pt}
 \begin{tabular}{lcccccccclccc}
    
2-factor          & BVN$^\P$   &       & \multicolumn{2}{c}{$t_7$} &       & \multicolumn{2}{c}{Frank} &             & \multicolumn{3}{l}{Selected} \\
\cmidrule{2-2}\cmidrule{4-5}\cmidrule{7-8}\cmidrule{11-13}    & $\hat{\tau}$  &       & $\hat{\tau}$  & SE    &       & $\hat{\tau}$  & SE    &        & copulas & $\hat{\tau}$ & SE \\
    \midrule
    1st factor          &       &       &       &       &       &       &       &       &       &       &       &         \\
  \hline  
  food	  			& 0.61  &       & 0.34  & 0.03  &       & 0.48  & 0.01  &             & BB10  & 0.38 & 0.00 \\
   clothes  	  		& 0.51  &       & 0.32  & 0.03  &       & 0.42  & 0.01  &            & BB10  & 0.36 & 0.01 \\
   leisure	  		& 0.49  &       & 0.35  & 0.02  &       & 0.42  & 0.01  &              & BB10  & 0.38 & 0.01 \\
   dishwasher    		& 0.14  &       & -0.07 & 0.03  &       & 0.08  & 0.01  &             &reflected  Joe
    & 0.19 & 0.02 \\
   car		  	  & 0.12  &       & -0.13 & 0.03  &       & 0.07  & 0.01  &          & reflected Joe    
   & {0.10} & 0.01 \\
   motorcycle 	 		 & 0.01  &       & -0.10 & 0.02  &       & -0.08 & 0.01  &            & Frank & 0.02 & 0.01 \\
   bicycles				& 0.07  &       & -0.10 & 0.02  &       & -0.05 & 0.01  &          & Frank     & 0.04 & 0.01 \\\hline
    2nd factor                 &       &       &       &       &       &       &       &       &       &       &       &         \\\hline
      
       food 		  		& 0.36  &       & 0.66  & 0.01  &       & 0.66  & 0.01  &            & BB10  &  0.53 & 0.01	 \\
 clothes  	  	& 0.18  &       & 0.46  & 0.02  &       & 0.40  & 0.01  &            & BVN   & 0.28 & 0.01 \\
 leisure 	  		& 0.07  &       & 0.41  & 0.02  &       & 0.36  & 0.01  &              & BB10  & 0.30 & 0.01\\
 dishwasher   & 0.33  &       & 0.37  & 0.01  &       & 0.26  & 0.01  &              & BVN   & 0.42 & 0.01 \\
 car 		  		& 0.48 	 &       & 0.46  & 0.02  &       & 0.36  & 0.01  &            & reflected Joe 
    & 0.35 & 0.01 \\
 motorcycle 	  & 0.19  &       & 0.15  & 0.01  &       & 0.21  & 0.02  &           & reflected  Joe
  & 0.17 & 0.01 \\
 bicycles				& 0.27  &       & 0.27  & 0.01  &       & 0.31  & 0.01  &             & reflected Gumbel
   & 0.27 & 0.01 \\\midrule
    AIC & 54245.91&       & \multicolumn{2}{c}{53482.23} &       & \multicolumn{2}{c}{53514.75} &              & \multicolumn{3}{c}{46233.00} \\
    Vuong $95\%$ CI &       &       & \multicolumn{2}{c}{(0.032,0.045)} &       & \multicolumn{2}{c}{(0.028,0.046)} &             & \multicolumn{3}{c}{(0.386,0.419)} \\
 $M_2$ & 1920.27&       & \multicolumn{2}{c}{1886.66} &       & \multicolumn{2}{c}{1945.07} &              & \multicolumn{3}{c}{450.32} \\
df& 65&       & \multicolumn{2}{c}{64} &       & \multicolumn{2}{c}{64} &              & \multicolumn{3}{c}{59} \\
 $p$-value & $<0.001$&       & \multicolumn{2}{c}{$<0.001$} &       & \multicolumn{2}{c}{$<0.001$} &              & \multicolumn{3}{c}{$<0.001$} \\
    \bottomrule
    \end{tabular}%
    \begin{flushleft}
\begin{footnotesize}
$^\P$: The resulting model is the same as the standard factor model.
\end{footnotesize}  
\end{flushleft}
\end{table}

While it is  revealed  that the selected 2-factor copula model  is the best model (lowest AIC) and there is substantial improvement over the standard 2-factor  model,  it is not apparent  from the $M_2$ statistic that the response patterns are satisfactorily explained by even 2 latent variables. This is not surprising 
since one should expect discrepancies between the postulated parametric model and the population probabilities, when the   sample size is sufficiently large \citep{Maydeu-OlivaresJoe-14-MBR}.    
In  Table \ref{misfit} we list the maximum deviations of observed and expected counts for
each bivariate margin, that is,
$D_{j_1j_2}=n\max_{y_1,y_2}|p_{j_1,j_2,y_1,y_2}-\pi_{j_1,j_2,y_1,y_2}(\hat\thbf)|$. 
From the table, it is revealed, that there is no misfit. The maximum discrepancy  occurs  between the continuous variables food and leisure. For this bivariate margin, 
the discrepancy of 509/9960 maximum occurs in the BVN factor copula model, while this drops to 133/9960 in the selected 2-factor copula model.

\begin{table}[!h]
\setlength{\tabcolsep}{12.5pt}
  \centering
 \caption{\label{misfit} Maximum deviations $D_{j_1j_2}$  of observed and expected counts for
each bivariate margin $(j_1,j_2)$ 
for the 1- and 2-factor copula models for the Swiss consumption survey dataset.}
    \begin{tabular}{cccccccccc}
    \toprule
          & \multicolumn{4}{c}{1-factor model}  &       & \multicolumn{4}{c}{2-factor model} \\
\cmidrule{2-5}\cmidrule{7-10}   $D_{j_1, j_2}$ & BVN   & $t_5$   & Frank   & Selected   &       & BVN   &  $t_7$   & Frank   & Selected  \\
    \midrule
    $D_{1,2}$ & 347   & 317   & 303   & 167  &       & 349   & 311   & 270   		& 40 \\
    $D_{1,3}$ & 511   & 468   & 456   & 183   &       & 509   & 460   & 428   		& 133 \\
    $D_{1,4}$ & 158   & 177   & 163   & 70    &       & 159   & 185   & 161   			& 56 \\
    $D_{1,5}$ & 231   & 189   & 223   & 119   &       & 233   & 181   & 230   		& 60 \\
    $D_{1,6}$ & 87    & 117   & 88    & 60    &       & 87    & 130   & 72    		    	& 12 \\
    $D_{1,7}$ & 78    & 92    & 79    & 88    &       & 78    & 110   & 89    		    	& 81 \\
    $D_{2,3}$ & 442   & 418   & 431   & 69    &       & 433   & 403   & 393  		& 54 \\
    $D_{2,4}$ & 59    & 80    & 84    & 145   &       & 38    & 56    & 64   			& 86 \\
    $D_{2,5}$ & 96    & 107   & 107   & 201   &       & 60    & 47    & 93    			& 36 \\
    $D_{2,6}$ & 18    & 3     & 18    & 27    &       & 19    & 15    & 29    		    	& 39 \\
    $D_{2,7}$ & 51    & 76    & 60    & 83    &       & 49    & 91    & 52    		    	& 61 \\
    $D_{3,4}$ & 182   & 146   & 141   & 196   &       & 253   & 216   & 168   		& 83 \\
    $D_{3,5}$ & 82    & 105   & 106   & 191   &       & 59    & 13    & 83    			& 61 \\
    $D_{3,6}$ & 59    & 58    & 69    & 71    &       & 13    & 23    & 27    		    	& 45 \\
    $D_{3,7}$ & 62    & 54    & 64    & 103   &       & 65    & 67    & 69    			& 59 \\
    $D_{4,5}$ & 289   & 276   & 286   & 223   &       & 66    & 74    & 207   		& 2 \\
    $D_{4,6}$ & 9     & 5     & 11    & 29    &       & 133   & 138   & 100  		    	& 96 \\
    $D_{4,7}$ & 82    & 81    & 81    & 88    &       & 28    & 20    & 46    		    	& 54 \\
    $D_{5,6}$ & 111   & 123   & 111   & 77    &       & 15    & 22    & 19    		    	& 20 \\
    $D_{5,7}$ & 101   & 96    & 95    & 68    &       & 33    & 25    & 40    		    	& 64 \\
    $D_{6,7}$ & 70    & 74    & 70    & 61    &       & 80    & 96    & 87    		    	& 52 \\
\bottomrule
\end{tabular}
\end{table}

For the selected  2-factor model based on the proposed algorithm, note that, without the need for a varimax rotation, the unique loadings show that one factor is loaded only on the discrete variables (dishwasher, car, motorcycle, and bicycles), while both factors are loaded on the continuous variables (food, clothes, and leisure). This reveals that the one latent variable which is only associated with the continuous variables   measures the expenses, while the the other which is associated with all the mixed variables measures the possession.

\section{\label{sec:simulations}Simulations}  
\baselineskip=23.3pt
An extensive simulation study is conducted  to
(a)  examine the performance of the diagnostics  to show that the correlation  matrix of the simulated variables has a factor structure, (b) check the small-sample efficiency  of the sample versions of  $\rho_N,\rho_N^{+},\rho_N^{-}$, (c) gauge the small-sample efficiency of the proposed estimation  
method and investigate the misspecification of the  bivariate pair-copulas, (d)
 examine the reliability of using the heuristic  algorithm to select the correct   bivariate linking copulas, 
and (e)  study the small-sample performance of the  $M_2$ statistic 
after transforming the continuous and count variables to ordinal.

We randomly generated  samples of size $n=\{100,300, 500\}$ from each selected one- and two-factor copula models in  the three application examples in Section \ref{sec-application}. 
We set  the type of the variables, the univariate margins and the  bivariate linking copulas, along with their   univariate  and dependence parameters to mimic the real data. 
The binary variables  don't have  tail asymmetries, hence parametric copulas are  less distinguishable. Therefore instead of binary, we simulated from  ordinal  with 3 equally weighted categories.

Table \ref{tab:simulationDiagnostics} contains the simulated means and   standard deviations (SD) of the discrepancy measures $D_1$, $D_2$ and $D_3$. 
The resultant summaries show that all the discrepancy measures correctly recognize both that the correlation structure has a factor structure and the number of factors. Among the discrepancy measures, $D_2$ has  a good performance even for a small sample size ($n=100$), while this is not the case for $D_1$ and $D_3$ which require larger sample sizes to successively determine the number of adequate factors.

\begin{table}[h!]																					
	\centering		
	\caption{\label{tab:simulationDiagnostics} Small sample of sizes $n=\{100,300, 500\}$ simulations ($10^4$ replications) from the selected factor copula models in Section \ref{sec-application} to assess the measures of discrepancy $D_1$, $D_2$, and $D_3$ between the observed  and the resulting correlation matrix of linear factor analysis for 1, 2 and 3 factors, with resultant means and standard deviations (SD).}

\begin{footnotesize}

	\setlength{\tabcolsep}{13.5pt}	
	\renewcommand{\arraystretch}{.6}

    \begin{tabular}{ccrcccccccc}
    \toprule
    \multicolumn{11}{l}{Political-economic dataset – 1-factor model} \\
    \midrule
      &   &   & \multicolumn{2}{c}{$D_1$} &   & \multicolumn{2}{c}{$D_2$} &   & \multicolumn{2}{c}{$D_3$} \\
\cmidrule{4-5}\cmidrule{7-8}\cmidrule{10-11}    $n$ & $\#$ factors &   & mean & SD &   & mean & SD &   & mean & SD \\
    \midrule
    \multirow{2}[1]{*}{100} & 1 &   & 0.061 & 0.027 &   & 0.016 & 0.006 &   & 0.101 & 0.071 \\
      & 2 &   & 0.022 & 0.016 &   & 0.004 & 0.003 &   & 0.014 & 0.023 \\
      &   &   &   &   &   &   &   &   &   &  \\
    \multirow{2}[0]{*}{300} & 1 &   & 0.038 & 0.017 &   & 0.010 & 0.004 &   & 0.036 & 0.023 \\
      & 2 &   & 0.011 & 0.008 &   & 0.002 & 0.002 &   & 0.004 & 0.005 \\
      &   &   &   &   &   &   &   &   &   &  \\
    \multirow{2}[1]{*}{500} & 1 &   & 0.033 & 0.014 &   & 0.009 & 0.003 &   & 0.024 & 0.015 \\
      & 2 &   & 0.009 & 0.006 &   & 0.002 & 0.001 &   & 0.002 & 0.003 \\
    
    \end{tabular}%

    \begin{tabular}{ccrcccccccc}
    \toprule
    \multicolumn{11}{l}{General social survey – 1-factor model} \\
    \midrule
      &   &   & \multicolumn{2}{c}{$D_1$} &   & \multicolumn{2}{c}{$D_2$} &   & \multicolumn{2}{c}{$D_3$} \\
\cmidrule{4-5}\cmidrule{7-8}\cmidrule{10-11}    $n$ & $\#$ factors &   & mean & SD &   & mean & SD &   & mean & SD \\
    \midrule
    \multirow{3}[1]{*}{100} & 1 &   & 0.178 & 0.048 &   & 0.048 & 0.010 &   & 0.192 & 0.074 \\
      & 2 &   & 0.119 & 0.037 &   & 0.025 & 0.006 &   & 0.077 & 0.039 \\
      & 3 &   & 0.066 & 0.030 &   & 0.010 & 0.004 &   & 0.021 & 0.016 \\
      &   &   &   &   &   &   &   &   &   &  \\
    \multirow{3}[0]{*}{300} & 1 &   & 0.104 & 0.028 &   & 0.028 & 0.006 &   & 0.062 & 0.023 \\
      & 2 &   & 0.068 & 0.021 &   & 0.015 & 0.004 &   & 0.024 & 0.012 \\
      & 3 &   & 0.036 & 0.017 &   & 0.006 & 0.002 &   & 0.006 & 0.005 \\
      &   &   &   &   &   &   &   &   &   &  \\
    \multirow{3}[1]{*}{500} & 1 &   & 0.081 & 0.022 &   & 0.022 & 0.004 &   & 0.038 & 0.014 \\
      & 2 &   & 0.053 & 0.016 &   & 0.012 & 0.003 &   & 0.014 & 0.007 \\
      & 3 &   & 0.028 & 0.013 &   & 0.005 & 0.002 &   & 0.004 & 0.003 \\
   
    \end{tabular}%

    \begin{tabular}{ccrcccccccc}
    \toprule
    \multicolumn{11}{l}{Swiss consumption survey – 1-factor model} \\
    \midrule
      &   &   & \multicolumn{2}{c}{$D_1$} &   & \multicolumn{2}{c}{$D_2$} &   & \multicolumn{2}{c}{$D_3$} \\
\cmidrule{4-5}\cmidrule{7-8}\cmidrule{10-11}    $n$ & $\#$ factors &   & mean & SD &   & mean & SD &   & mean & SD \\
    \midrule
    \multirow{3}[1]{*}{100} & 1 &   & 0.223 & 0.059 &   & 0.059 & 0.011 &   & 0.291 & 0.101 \\
      & 2 &   & 0.144 & 0.046 &   & 0.029 & 0.007 &   & 0.106 & 0.053 \\
      & 3 &   & 0.077 & 0.035 &   & 0.011 & 0.004 &   & 0.028 & 0.022 \\
      &   &   &   &   &   &   &   &   &   &  \\
    \multirow{3}[0]{*}{300} & 1 &   & 0.162 & 0.044 &   & 0.045 & 0.007 &   & 0.156 & 0.044 \\
      & 2 &   & 0.091 & 0.030 &   & 0.018 & 0.005 &   & 0.036 & 0.019 \\
      & 3 &   & 0.044 & 0.021 &   & 0.007 & 0.003 &   & 0.009 & 0.007 \\
      &   &   &   &   &   &   &   &   &   &  \\
    \multirow{3}[1]{*}{500} & 1 &   & 0.150 & 0.039 &   & 0.041 & 0.006 &   & 0.130 & 0.032 \\
      & 2 &   & 0.071 & 0.024 &   & 0.014 & 0.004 &   & 0.022 & 0.011 \\
      & 3 &   & 0.034 & 0.016 &   & 0.005 & 0.002 &   & 0.005 & 0.004 \\
    
    \end{tabular}%

    \begin{tabular}{ccrcccccccc}
    \toprule
    \multicolumn{11}{l}{General social survey – 2-factor model} \\
    \midrule
      &   &   & \multicolumn{2}{c}{$D_1$} &   & \multicolumn{2}{c}{$D_2$} &   & \multicolumn{2}{c}{$D_3$} \\
\cmidrule{4-5}\cmidrule{7-8}\cmidrule{10-11}    $n$ & $\#$ factors &   & mean & SD &   & mean & SD &   & mean & SD \\
    \midrule
    \multirow{3}[1]{*}{100} & 1 &   & 0.360 & 0.066 &   & 0.102 & 0.018 &   & 0.691 & 0.183 \\
      & 2 &   & 0.117 & 0.042 &   & 0.027 & 0.007 &   & 0.118 & 0.059 \\
      & 3 &   & 0.059 & 0.028 &   & 0.010 & 0.004 &   & 0.028 & 0.023 \\
      &   &   &   &   &   &   &   &   &   &  \\
    \multirow{3}[0]{*}{300} & 1 &   & 0.332 & 0.045 &   & 0.101 & 0.012 &   & 0.573 & 0.103 \\
      & 2 &   & 0.066 & 0.023 &   & 0.017 & 0.004 &   & 0.042 & 0.021 \\
      & 3 &   & 0.033 & 0.015 &   & 0.006 & 0.003 &   & 0.009 & 0.008 \\
      &   &   &   &   &   &   &   &   &   &  \\
    \multirow{3}[1]{*}{500} & 1 &   & 0.326 & 0.037 &   & 0.101 & 0.010 &   & 0.552 & 0.078 \\
      & 2 &   & 0.052 & 0.017 &   & 0.014 & 0.004 &   & 0.027 & 0.014 \\
      & 3 &   & 0.026 & 0.012 &   & 0.005 & 0.002 &   & 0.006 & 0.005 \\
  
    \end{tabular}%

    \begin{tabular}{ccrcccccccc}
    \toprule
    \multicolumn{11}{l}{Swiss consumption survey – 2-factor model} \\
    \midrule
      &   &   & \multicolumn{2}{c}{$D_1$} &   & \multicolumn{2}{c}{$D_2$} &   & \multicolumn{2}{c}{$D_3$} \\
\cmidrule{4-5}\cmidrule{7-8}\cmidrule{10-11}    $n$ & $\#$ factors &   & mean & SD &   & mean & SD &   & mean & SD \\
    \midrule
    \multirow{3}[1]{*}{100} & 1 &   & 0.249 & 0.070 &   & 0.060 & 0.013 &   & 0.343 & 0.129 \\
      & 2 &   & 0.130 & 0.047 &   & 0.026 & 0.007 &   & 0.111 & 0.056 \\
      & 3 &   & 0.065 & 0.031 &   & 0.010 & 0.004 &   & 0.028 & 0.023 \\
      &   &   &   &   &   &   &   &   &   &  \\
    \multirow{3}[0]{*}{300} & 1 &   & 0.200 & 0.047 &   & 0.048 & 0.009 &   & 0.198 & 0.061 \\
      & 2 &   & 0.075 & 0.028 &   & 0.017 & 0.004 &   & 0.040 & 0.020 \\
      & 3 &   & 0.036 & 0.017 &   & 0.006 & 0.003 &   & 0.009 & 0.007 \\
      &   &   &   &   &   &   &   &   &   &  \\
    \multirow{3}[1]{*}{500} & 1 &   & 0.191 & 0.038 &   & 0.046 & 0.007 &   & 0.171 & 0.045 \\
      & 2 &   & 0.059 & 0.021 &   & 0.014 & 0.004 &   & 0.026 & 0.013 \\
      & 3 &   & 0.027 & 0.013 &   & 0.005 & 0.002 &   & 0.006 & 0.005 \\
    \bottomrule
    \end{tabular}%

\end{footnotesize}
																	
\end{table}%

To check the small-sample efficiency of the sample versions of  $\rho_N$, $\rho_N^{+}$ and $\rho_N^{-}$ we have generated 
$10^4$ random  samples of size $n=\{100,300,500\}$ from all the aforementioned bivariate copulas that join the distributions of two continuous variables, two ordinal variables, one continuous and one ordinal variable, one continuous and one count variable, one ordinal and one count, and two count variables with small ($\tau=0.3$), moderate ($\tau=0.5$) and strong dependence ($\tau=0.7$).  
Representative results are shown in Table \ref{tab:simulationRhoGumbel} for the Gumbel copula.
Note that the count variable was treated as ordinal  with 5 categories where the 5th category contained all the  counts greater than 3.  The resultant biases, root mean square errors (RMSE), and standard deviations (SD), scaled by $n$, show the estimation of the correlations and semi-correlations is highly efficient.    
Note in passing that because only part of the data are used in computing sample semi-correlations, their variability is larger  than the correlations. However, if there is a consistent direction to the tail asymmetry based on semi-correlations, this is useful information for choosing potential bivariate parametric  copulas. 

\begin{sidewaystable}[htbp]
  \centering
	\caption{\label{tab:simulationRhoGumbel} Small sample of sizes $n=\{100, 300, 500\}$ simulations ($10^4$ replications)  from the Gumbel copula with Kendall's  $\tau = \{0.3, 0.5, 0.7\}$ for mixed continuous, ordinal, and count data with resultant
biases, root mean square errors (RMSE) and standard deviations (SD),  scaled by $n$, for the  estimated correlation $\rho_N$, lower semi-correlation $\rho_N^{-}$,  and upper semi-correlation $\rho_N^{+}$.	
}

  \begin{footnotesize}

	\setlength{\tabcolsep}{4pt}	
	\renewcommand{\arraystretch}{1}			
     \begin{tabular}{cclccccccccccccccccccccccc}
    \toprule
      &   &   & \multicolumn{3}{c}{(continuous, continuous)} &   & \multicolumn{3}{c}{(continuous, ordinal)} &   & \multicolumn{3}{c}{(continuous, count)} &   & \multicolumn{3}{c}{(ordinal, ordinal)} &   & \multicolumn{3}{c}{(ordinal, count)} &   & \multicolumn{3}{c}{(count, count)} \\
\cmidrule{4-6}\cmidrule{8-10}\cmidrule{12-14}\cmidrule{16-18}\cmidrule{20-22}\cmidrule{24-26}    $n$ & $\tau$ &   & $\rho_N$ & $\rho_N^{-}$ & $\rho_N^{+}$ &   & $\rho_N$ & $\rho_N^{-}$ & $\rho_N^{+}$ &   & $\rho_N$ & $\rho_N^{-}$ & $\rho_N^{+}$ &   & $\rho_N$ & $\rho_N^{-}$ & $\rho_N^{+}$ &   & $\rho_N$ & $\rho_N^{-}$ & $\rho_N^{+}$ &   & $\rho_N$ & $\rho_N^{-}$ & $\rho_N^{+}$ \\
    \midrule
    \multirow{12}[6]{*}{100} & \multirow{4}[2]{*}{0.3} & True values & 0.46 & 0.16 & 0.46 &   & 0.46 & 0.16 & 0.46 &   & 0.46 & 0.16 & 0.46 &   & 0.46 & 0.16 & 0.46 &   & 0.46 & 0.16 & 0.46 &   & 0.46 & 0.16 & 0.46 \\
      &   & $n$Bias & -1.09 & -0.79 & -4.18 &   & -1.05 & 0.16 & -9.25 &   & 1.62 & 1.40 & -4.41 &   & -0.55 & 2.35 & -9.86 &   & 0.62 & 4.88 & -8.25 &   & 2.03 & 9.54 & -5.58 \\
      &   & $n$SD & 8.57 & 18.10 & 16.83 &   & 9.03 & 18.64 & 16.71 &   & 9.23 & 16.54 & 20.46 &   & 9.31 & 18.95 & 18.03 &   & 9.37 & 17.08 & 21.78 &   & 9.55 & 14.73 & 24.24 \\
      &   & $n$RMSE & 8.64 & 18.12 & 17.34 &   & 9.09 & 18.64 & 19.10 &   & 9.37 & 16.60 & 20.93 &   & 9.32 & 19.10 & 20.55 &   & 9.39 & 17.76 & 23.29 &   & 9.76 & 17.55 & 24.87 \\
\cmidrule{2-26}      & \multirow{4}[2]{*}{0.5} & True values & 0.70 & 0.36 & 0.67 &   & 0.70 & 0.36 & 0.67 &   & 0.70 & 0.36 & 0.67 &   & 0.70 & 0.36 & 0.67 &   & 0.70 & 0.36 & 0.67 &   & 0.70 & 0.36 & 0.67 \\
      &   & $n$Bias & -0.99 & -1.65 & -3.98 &   & -0.16 & -0.38 & -10.21 &   & 2.43 & 0.10 & -3.78 &   & 0.26 & 3.93 & -8.58 &   & 1.41 & 7.42 & -8.18 &   & 2.80 & 14.88 & -5.57 \\
      &   & $n$SD & 5.77 & 15.73 & 11.72 &   & 6.26 & 15.67 & 12.41 &   & 6.19 & 14.49 & 14.93 &   & 6.30 & 15.91 & 13.52 &   & 6.35 & 14.71 & 16.73 &   & 6.34 & 12.18 & 17.35 \\
      &   & $n$RMSE & 5.85 & 15.82 & 12.37 &   & 6.26 & 15.68 & 16.07 &   & 6.65 & 14.49 & 15.40 &   & 6.31 & 16.39 & 16.01 &   & 6.51 & 16.48 & 18.62 &   & 6.94 & 19.23 & 18.22 \\
\cmidrule{2-26}      & \multirow{4}[2]{*}{0.7} & True values & 0.88 & 0.64 & 0.85 &   & 0.88 & 0.64 & 0.85 &   & 0.88 & 0.64 & 0.85 &   & 0.88 & 0.64 & 0.85 &   & 0.88 & 0.64 & 0.85 &   & 0.88 & 0.64 & 0.85 \\
      &   & $n$Bias & -0.71 & -2.12 & -2.74 &   & 0.78 & -2.21 & -8.77 &   & 2.16 & -5.21 & -0.28 &   & 0.55 & 4.34 & -4.46 &   & 1.23 & 6.29 & -4.75 &   & 2.12 & 13.76 & -2.19 \\
      &   & $n$SD & 2.71 & 10.76 & 5.99 &   & 3.02 & 10.84 & 7.29 &   & 2.80 & 10.74 & 8.40 &   & 3.07 & 10.48 & 7.77 &   & 3.03 & 10.24 & 10.91 &   & 2.94 & 7.26 & 9.42 \\
      &   & $n$RMSE & 2.80 & 10.96 & 6.59 &   & 3.12 & 11.06 & 11.40 &   & 3.53 & 11.94 & 8.40 &   & 3.12 & 11.35 & 8.96 &   & 3.27 & 12.02 & 11.90 &   & 3.63 & 15.56 & 9.67 \\
    \midrule
    \multirow{12}[6]{*}{300} & \multirow{4}[2]{*}{0.3} & True values & 0.46 & 0.16 & 0.46 &   & 0.46 & 0.16 & 0.46 &   & 0.46 & 0.16 & 0.46 &   & 0.46 & 0.16 & 0.46 &   & 0.46 & 0.16 & 0.46 &   & 0.46 & 0.16 & 0.46 \\
      &   & $n$Bias & -1.44 & -1.48 & -5.96 &   & -2.88 & 1.14 & -26.54 &   & 5.52 & 4.43 & -12.35 &   & -1.56 & 7.52 & -28.59 &   & 2.04 & 14.61 & -25.56 &   & 6.36 & 28.76 & -16.44 \\
      &   & $n$SD & 15.04 & 30.94 & 28.32 &   & 15.75 & 31.26 & 27.83 &   & 16.11 & 28.02 & 33.34 &   & 16.32 & 32.42 & 30.34 &   & 16.45 & 28.98 & 36.06 &   & 16.65 & 25.39 & 40.28 \\
      &   & $n$RMSE & 15.11 & 30.97 & 28.94 &   & 16.01 & 31.28 & 38.46 &   & 17.03 & 28.37 & 35.55 &   & 16.39 & 33.29 & 41.69 &   & 16.58 & 32.46 & 44.20 &   & 17.82 & 38.36 & 43.50 \\
\cmidrule{2-26}      & \multirow{4}[2]{*}{0.5} & True values & 0.70 & 0.36 & 0.67 &   & 0.70 & 0.36 & 0.67 &   & 0.70 & 0.36 & 0.67 &   & 0.70 & 0.36 & 0.67 &   & 0.70 & 0.36 & 0.67 &   & 0.70 & 0.36 & 0.67 \\
      &   & $n$Bias & -1.23 & -2.48 & -5.34 &   & -0.77 & -1.16 & -30.78 &   & 7.39 & -0.39 & -11.03 &   & 0.64 & 11.81 & -25.37 &   & 4.11 & 21.74 & -25.71 &   & 8.39 & 44.61 & -16.40 \\
      &   & $n$SD & 9.99 & 26.98 & 19.09 &   & 10.87 & 26.60 & 20.40 &   & 10.62 & 24.72 & 24.25 &   & 11.08 & 27.48 & 22.59 &   & 11.09 & 25.22 & 27.56 &   & 10.96 & 20.84 & 28.88 \\
      &   & $n$RMSE & 10.06 & 27.09 & 19.82 &   & 10.90 & 26.63 & 36.93 &   & 12.94 & 24.72 & 26.64 &   & 11.10 & 29.91 & 33.97 &   & 11.82 & 33.30 & 37.69 &   & 13.80 & 49.24 & 33.22 \\
\cmidrule{2-26}      & \multirow{4}[2]{*}{0.7} & True values & 0.88 & 0.64 & 0.85 &   & 0.88 & 0.64 & 0.85 &   & 0.88 & 0.64 & 0.85 &   & 0.88 & 0.64 & 0.85 &   & 0.88 & 0.64 & 0.85 &   & 0.88 & 0.64 & 0.85 \\
      &   & $n$Bias & -0.83 & -2.93 & -3.43 &   & 1.42 & -7.89 & -28.35 &   & 5.84 & -18.52 & -1.43 &   & 1.31 & 12.56 & -13.92 &   & 3.27 & 17.32 & -16.87 &   & 5.97 & 40.56 & -7.09 \\
      &   & $n$SD & 4.60 & 18.37 & 9.35 &   & 5.16 & 18.37 & 11.94 &   & 4.71 & 18.05 & 13.58 &   & 5.34 & 18.16 & 13.05 &   & 5.26 & 17.59 & 18.02 &   & 5.04 & 12.35 & 15.54 \\
      &   & $n$RMSE & 4.68 & 18.61 & 9.96 &   & 5.35 & 19.99 & 30.76 &   & 7.50 & 25.86 & 13.66 &   & 5.50 & 22.08 & 19.08 &   & 6.20 & 24.68 & 24.69 &   & 7.81 & 42.40 & 17.08 \\
    \midrule
    \multirow{12}[6]{*}{500} & \multirow{4}[2]{*}{0.3} & True values & 0.46 & 0.16 & 0.46 &   & 0.46 & 0.16 & 0.46 &   & 0.46 & 0.16 & 0.46 &   & 0.46 & 0.16 & 0.46 &   & 0.46 & 0.16 & 0.46 &   & 0.46 & 0.16 & 0.46 \\
      &   & $n$Bias & -1.37 & -1.08 & -7.04 &   & -4.45 & 2.42 & -44.04 &   & 9.65 & 7.93 & -20.25 &   & -2.25 & 12.60 & -47.33 &   & 3.75 & 24.71 & -42.32 &   & 10.96 & 48.14 & -27.35 \\
      &   & $n$SD & 19.06 & 39.98 & 36.96 &   & 19.95 & 39.89 & 35.47 &   & 20.49 & 35.93 & 42.97 &   & 20.75 & 41.68 & 39.18 &   & 21.00 & 37.65 & 46.91 &   & 21.35 & 32.71 & 52.79 \\
      &   & $n$RMSE & 19.11 & 40.00 & 37.63 &   & 20.44 & 39.97 & 56.55 &   & 22.64 & 36.79 & 47.51 &   & 20.87 & 43.54 & 61.45 &   & 21.33 & 45.04 & 63.17 &   & 24.00 & 58.20 & 59.46 \\
\cmidrule{2-26}      & \multirow{4}[2]{*}{0.5} & True values & 0.70 & 0.36 & 0.67 &   & 0.70 & 0.36 & 0.67 &   & 0.70 & 0.36 & 0.67 &   & 0.70 & 0.36 & 0.67 &   & 0.70 & 0.36 & 0.67 &   & 0.70 & 0.36 & 0.67 \\
      &   & $n$Bias & -1.11 & -2.31 & -6.27 &   & -1.16 & -1.38 & -51.40 &   & 12.58 & -0.39 & -18.48 &   & 1.34 & 19.78 & -41.78 &   & 7.16 & 36.42 & -42.79 &   & 14.31 & 74.64 & -27.23 \\
      &   & $n$SD & 12.58 & 35.08 & 24.56 &   & 13.67 & 34.07 & 26.18 &   & 13.31 & 31.87 & 31.24 &   & 14.04 & 35.55 & 29.22 &   & 13.99 & 32.68 & 36.00 &   & 13.91 & 26.57 & 37.57 \\
      &   & $n$RMSE & 12.63 & 35.16 & 25.35 &   & 13.72 & 34.10 & 57.68 &   & 18.31 & 31.88 & 36.29 &   & 14.10 & 40.69 & 50.99 &   & 15.71 & 48.93 & 55.92 &   & 19.95 & 79.23 & 46.40 \\
\cmidrule{2-26}      & \multirow{4}[2]{*}{0.7} & True values & 0.88 & 0.64 & 0.85 &   & 0.88 & 0.64 & 0.85 &   & 0.88 & 0.64 & 0.85 &   & 0.88 & 0.64 & 0.85 &   & 0.88 & 0.64 & 0.85 &   & 0.88 & 0.64 & 0.85 \\
      &   & $n$Bias & -0.76 & -2.83 & -3.63 &   & 2.03 & -13.45 & -47.97 &   & 9.47 & -31.71 & -2.93 &   & 2.21 & 21.01 & -22.91 &   & 5.43 & 28.29 & -28.60 &   & 9.95 & 67.47 & -11.82 \\
      &   & $n$SD & 5.81 & 23.66 & 11.69 &   & 6.48 & 23.48 & 15.30 &   & 5.95 & 23.22 & 17.23 &   & 6.75 & 23.36 & 16.73 &   & 6.68 & 22.72 & 23.21 &   & 6.42 & 15.65 & 20.03 \\
      &   & $n$RMSE & 5.86 & 23.82 & 12.24 &   & 6.79 & 27.06 & 50.35 &   & 11.18 & 39.30 & 17.48 &   & 7.10 & 31.42 & 28.37 &   & 8.61 & 36.29 & 36.84 &   & 11.84 & 69.26 & 23.26 \\
    \bottomrule
    \end{tabular}%
 \end{footnotesize}
      
\end{sidewaystable}%

Table \ref{simest-res} contains the resultant biases, RMSEs, and SDs, scaled by $n$, for the estimates obtained using the estimation approach in Section \ref{Est-section}. The results show that the proposed estimation approach   is highly efficient according to the simulated biases, SDs and RMSEs. We further investigated the misspesification of the bivariate pair-copulas by deriving the same statistics but from 1-factor model with BVN pair copulas, i.e. the standard 1-factor model.  Once again,  the simulated data are based on the selected 1-factor copula models in Section \ref{sec-application}. Table \ref{simestBVN-res} contains the resultant biases, RMSEs, and SDs, scaled by $n$. The results show that the Kendall's tau estimates are not robust to pair-copula misspesification if the true (simulated) factor copula model has different dependence in the middle of the data, e.g.  when the BB10 copulas that can provide a non convex shape of dependence (see e.g., Fugure \ref{contours}) are used to specify the true  factor  copula model (Table \ref{simestBVN-res}, Swiss consumption survey). As we have already mentioned the Kendall's $\tau$ only accounts for the dependence dominated by the middle of the data, and it is expected to be similar among parametric families of copulas that provide a convex shape of dependence (Table \ref{simestBVN-res}, Political-economic dataset and general social survey).

Table \ref{simres-M2} contains   four common nominal levels of  the $M_2$ statistic under the factor copula models for mixed data. We transformed the continuous  and count variables  to ordinal  with $K=\{3, 4, 5\}$  and $K=\{3, 4\}$ categories, respectively, using the unsupervised strategies proposed in Section \ref{m2gof}.  We also transformed the count variables to ordinal with $K=5$ categories by treating them as ordinal where the 5th category contained all the  counts greater than 3.  As  the observed levels are close to nominal levels,  it is demonstrated that the $M_2$ statistic  remains  reliable for mixed data and  that the information loss under transformation to ordinal is minimal. 

\begin{landscape}
\begin{table}[!h]
  \centering
\caption{\label{simest-res}Small sample of sizes $n=\{100,300, 500\}$ simulations ($10^4$ replications) from the selected factor copula models in Section \ref{sec-application} with resultant
biases, root mean square errors (RMSE) and standard deviations (SD),  scaled by $n$, for the estimated parameters.}

\begin{footnotesize}

  \setlength{\tabcolsep}{10pt}
      \begin{tabular}{lccccccccccccccccccc}
    \toprule
            \multicolumn{20}{l}{Political-economic dataset -- 1-factor model} \\
    \midrule
    $\tau$ & \multicolumn{3}{c}{0.51} &       & \multicolumn{3}{c}{0.58} &       & \multicolumn{3}{c}{0.80} &       & \multicolumn{3}{c}{0.68} &       & \multicolumn{3}{c}{0.74} \\
\cmidrule{2-4}\cmidrule{6-8}\cmidrule{10-12}\cmidrule{14-16}\cmidrule{18-20}    $n$     & 100   & 300   & 500   &       & 100   & 300   & 500   &       & 100   & 300   & 500   &       & 100   & 300   & 500   &       & 100   & 300   & 500 \\\hline
    $n$Bias & 0.88  & 2.30  & 3.17  &       & -1.36 & -3.39 & -4.87 &       & 0.75  & -0.55 & 0.64  &       & 0.21  & -0.27 & 0.19  &       & 0.29  & 2.57  & 0.73 \\
    $n$SD & 4.28  & 7.60  & 9.63  &       & 4.19  & 7.50  & 9.08  &       & 5.41  & 10.91 & 11.98 &       & 4.58  & 8.43  & 9.84  &       & 4.46  & 14.92 & 11.78 \\
    $n$RMSE & 4.37  & 7.95  & 10.13 &       & 4.40  & 8.23  & 10.31 &       & 5.47  & 10.92 & 12.00 &       & 4.59  & 8.44  & 9.84  &       & 4.47  & 15.13 & 11.80 \\
   
    \end{tabular}%

\setlength{\tabcolsep}{4.5pt}
    \begin{tabular}{lccccccccccccccccccccccccccc}
    \toprule
        \multicolumn{28}{l}{General social survey  -- 1-factor model} \\
    \midrule
    $\tau$ & \multicolumn{3}{c}{0.30} &       & \multicolumn{3}{c}{-0.14} &       & \multicolumn{3}{c}{0.46} &       & \multicolumn{3}{c}{0.33} &       & \multicolumn{3}{c}{0.55} &       & \multicolumn{3}{c}{-0.14} &       & \multicolumn{3}{c}{-0.27} \\
\cmidrule{2-4}\cmidrule{6-8}\cmidrule{10-12}\cmidrule{14-16}\cmidrule{18-20}\cmidrule{22-24}\cmidrule{26-28}    $n$ & 100   & 300   & 500   &       & 100   & 300   & 500   &       & 100   & 300   & 500   &       & 100   & 300   & 500   &       & 100   & 300   & 500   &       & 100   & 300   & 500   &       & 100   & 300   & 500 \\\hline
    $n$Bias & -0.11 & -0.86 & -1.66 &       & -0.06 & -0.10 & -0.19 &       & 0.29  & 0.17  & 0.53  &       & 0.19  & 0.26  & 0.27  &       & 0.72  & 0.89  & 0.94  &       & -0.18 & -0.37 & -0.30 &       & -0.09 & -0.03 & -0.03 \\
    $n$SD & 7.46  & 12.41 & 16.01 &       & 6.55  & 11.12 & 14.00 &       & 8.53  & 13.76 & 17.97 &       & 8.33  & 14.07 & 17.75 &       & 9.45  & 14.63 & 18.92 &       & 6.89  & 11.89 & 15.05 &       & 7.75  & 12.90 & 16.54 \\
    $n$RMSE & 7.46  & 12.44 & 16.10 &       & 6.55  & 11.12 & 14.00 &       & 8.53  & 13.76 & 17.98 &       & 8.33  & 14.07 & 17.75 &       & 9.47  & 14.65 & 18.94 &       & 6.89  & 11.90 & 15.05 &       & 7.75  & 12.90 & 16.54 \\
   
    \end{tabular}%

\setlength{\tabcolsep}{4.6pt}
    \begin{tabular}{lccccccccccccccccccccccccccc}
    \toprule
    \multicolumn{28}{l}{Swiss consumption survey -- 1-factor model} \\
    \midrule
    $\tau$ & \multicolumn{3}{c}{0.69} &       & \multicolumn{3}{c}{0.38} &       & \multicolumn{3}{c}{0.39} &       & \multicolumn{3}{c}{0.28} &       & \multicolumn{3}{c}{0.23} &       & \multicolumn{3}{c}{0.13} &       & \multicolumn{3}{c}{0.17} \\
\cmidrule{2-4}\cmidrule{6-8}\cmidrule{10-12}\cmidrule{14-16}\cmidrule{18-20}\cmidrule{22-24}\cmidrule{26-28}    $n$ & 100   & 300   & 500   &       & 100   & 300   & 500   &       & 100   & 300   & 500   &       & 100   & 300   & 500   &       & 100   & 300   & 500   &       & 100   & 300   & 500   &       & 100   & 300   & 500 \\\hline
    $n$Bias & -15.95 & -0.78 & -0.04 &       & -7.85 & -1.57 & -3.22 &       & -8.03 & -1.62 & -3.22 &       & 0.08  & 0.09  & 0.26  &       & 0.10  & 0.06  & -0.11 &       & 0.23  & 0.12  & 0.40  &       & 0.13  & 0.06  & 0.20 \\
    $n$SD & 8.81  & 9.93  & 13.16 &       & 9.58  & 6.24  & 7.98  &       & 9.54  & 6.52  & 8.11  &       & 7.69  & 13.02 & 16.80 &       & 7.72  & 13.02 & 17.02 &       & 7.46  & 13.01 & 16.90 &       & 7.51  & 12.79 & 16.67 \\
    $n$RMSE & 18.23 & 9.96  & 13.16 &       & 12.38 & 6.43  & 8.60  &       & 12.47 & 6.72  & 8.73  &       & 7.69  & 13.02 & 16.81 &       & 7.72  & 13.02 & 17.02 &       & 7.46  & 13.01 & 16.91 &       & 7.51  & 12.79 & 16.67 \\
    
    \end{tabular}%

\setlength{\tabcolsep}{7pt}
    \begin{tabular}{lccccccccccccccccccccccccccc}
    \toprule
   \multicolumn{8}{l}{General social survey -- 2-factor model}   &       & \multicolumn{1}{l}{ } &       &       &       & \multicolumn{3}{c}{$n=500$} &       &       &       &       &       &       &       &       &       &       &       &  \\
    \midrule
          & \multicolumn{13}{c}{1st factor}                                                                       &       & \multicolumn{13}{c}{2nd factor} \\
\cmidrule{2-14}\cmidrule{16-28}    $\tau$  & 0.34  &       & 0.49  &       & 0.18  &       & -0.13 &       & -0.13 &       & 0.44  &       & 0.11  &       & 0.40  &       & -0.14 &       & 0.65  &       & 0.29  &       & 0.49  &       & -0.24 &       & -0.26 \\\hline
    $n$Bias & 1.18  &       & -7.19 &       & 1.40  &       & 0.31  &       & 0.19  &       & 1.45  &       & -0.44 &       & -0.96 &       & 0.19  &       & -0.05 &       & 0.22  &       & 2.59  &       & -2.47 &       & 0.00 \\
    $n$SD & 16.17 &       & 17.21 &       & 19.25 &       & 18.83 &       & 18.63 &       & 19.05 &       & 17.66 &       & 18.52 &       & 18.32 &       & 22.72 &       & 17.68 &       & 26.90 &       & 21.77 &       & 16.33 \\
    $n$RMSE & 16.21 &       & 18.65 &       & 19.30 &       & 18.84 &       & 18.63 &       & 19.11 &       & 17.67 &       & 18.54 &       & 18.32 &       & 22.72 &       & 17.68 &       & 27.03 &       & 21.91 &       & 16.33 \\
   
    \end{tabular}%

    \begin{tabular}{lccccccccccccccccccccccccccc}
    \toprule
    \multicolumn{13}{l}{Swiss consumption survey -- 2-factor model}                                       & \multicolumn{3}{c}{$n=500$} &       &       &       &       &       &       &       &       &       &       &       &  \\
    \midrule
          & \multicolumn{13}{c}{1st factor}                                                                       &       & \multicolumn{13}{c}{2nd factor} \\
\cmidrule{2-14}\cmidrule{16-28}    $\tau$ & 0.34  &       & 0.36  &       & 0.38  &       & 0.19  &       & 0.09  &       & 0.02  &       & 0.04  &       & 0.53  &       & 0.28  &       & 0.30  &       & 0.42  &       & 0.35  &       & 0.17  &       & 0.27 \\\hline
    $n$Bias & -2.31 &       & -1.60 &       & -0.69 &       & -3.01 &       & -1.04 &       & -0.54 &       & 2.89  &       & -4.27 &       & 0.64  &       & 1.00  &       & 3.41  &       & 1.27  &       & 0.59  &       & -4.15 \\
    $n$SD & 7.43  &       & 13.67 &       & 16.12 &       & 27.11 &       & 25.31 &       & 21.27 &       & 21.31 &       & 20.37 &       & 17.98 &       & 19.05 &       & 21.20 &       & 20.89 &       & 19.41 &       & 21.55 \\
    $n$RMSE & 7.78  &       & 13.77 &       & 16.14 &       & 27.27 &       & 25.33 &       & 21.28 &       & 21.51 &       & 20.82 &       & 17.99 &       & 19.08 &       & 21.47 &       & 20.93 &       & 19.42 &       & 21.95 \\
    \bottomrule
    \end{tabular}%
   \end{footnotesize}
\end{table}%
\end{landscape}

\begin{landscape}
\begin{table}[!h]

\vspace{3cm}

  \centering
\caption{\label{simestBVN-res}Small sample of sizes $n=\{100,300, 500\}$ simulations ($10^4$ replications) from the selected 1-factor copula models in Section \ref{sec-application} with resultant
biases, root mean square errors (RMSE) and standard deviations (SD),  scaled by $n$, for the estimated parameters under an 1-factor copula model with BVN copulas, i.e. the standard factor model.}

\begin{small}

  \setlength{\tabcolsep}{10pt}																					
	\begin{tabular}{lccccccccccccccccccc}																																			
	\toprule																																			
	\multicolumn{20}{l}{Political-economic	dataset	--	1-factor	model}	\\																														
	\midrule																																			
	$\tau$	&	\multicolumn{3}{c}{0.51}	&	&	\multicolumn{3}{c}{0.58}	&	&	\multicolumn{3}{c}{0.80}	&	&	\multicolumn{3}{c}{0.68}	&	&	\multicolumn{3}{c}{0.74}	\\																				
\cmidrule{2-4}\cmidrule{6-8}\cmidrule{10-12}\cmidrule{14-16}\cmidrule{18-20}	$n$	&	100	&	300	&	500	&	&	100	&	300	&	500	&	&	100	&	300	&	500	&	&	100	&	300	&	500	&	&	100	&	300	&	500	\\
	\midrule																																			
	$n$Bias	&	-0.35	&	-0.96	&	-1.56	&	&	-1.40	&	-3.90	&	-6.41	&	&	-1.29	&	-6.19	&	-10.54	&	&	0.51	&	0.89	&	1.40	&	&	-0.18	&	-1.08	&	-2.15	\\
	$n$SD	&	5.24	&	9.16	&	11.57	&	&	4.95	&	8.57	&	11.13	&	&	6.03	&	9.52	&	12.14	&	&	4.60	&	7.91	&	10.01	&	&	4.42	&	7.49	&	9.69	\\
	$n$RMSE	&	5.25	&	9.21	&	11.68	&	&	5.15	&	9.42	&	12.85	&	&	6.17	&	11.35	&	16.07	&	&	4.63	&	7.96	&	10.11	&	&	4.42	&	7.57	&	9.92	\\																																
	\end{tabular}%

  \setlength{\tabcolsep}{4.4pt}																													
	\begin{tabular}{lccccccccccccccccccccccccccc}																																							
	\toprule																																																	
	\multicolumn{28}{l}{General	social	survey	--	1-factor	model	}	\\																												
	\midrule																																																	
	$\tau$	&	\multicolumn{3}{c}{0.30}	&	&	\multicolumn{3}{c}{-0.14}	&	&	\multicolumn{3}{c}{0.46}	&	&	\multicolumn{3}{c}{0.33}	&	&	\multicolumn{3}{c}{0.55}	&	&	\multicolumn{3}{c}{-0.14}	&	&	\multicolumn{3}{c}{-0.27}	\\																												
\cmidrule{2-4}\cmidrule{6-8}\cmidrule{10-12}\cmidrule{14-16}\cmidrule{18-20}\cmidrule{22-24}\cmidrule{26-28}	$n$	&	100	&	300	&	500	&	&	100	&	300	&	500	&	&	100	&	300	&	500	&	&	100	&	300	&	500	&	&	100	&	300	&	500	&	&	100	&	300	&	500	&	&	100	&	300	&	500	\\
	\midrule																																																	
	$n$Bias	&	-1.68	&	-5.15	&	-8.75	&	&	-0.93	&	-3.17	&	-5.45	&	&	-0.09	&	-1.34	&	-2.11	&	&	-0.16	&	-0.92	&	-1.86	&	&	0.65	&	-0.07	&	-0.75	&	&	-2.20	&	-6.94	&	-11.32	&	&	-0.99	&	-2.53	&	-4.30	\\
	$n$SD	&	7.66	&	12.91	&	16.57	&	&	8.14	&	13.62	&	17.45	&	&	9.08	&	14.45	&	18.82	&	&	8.56	&	14.28	&	18.05	&	&	10.46	&	15.79	&	20.38	&	&	8.67	&	14.79	&	18.91	&	&	8.51	&	13.60	&	17.53	\\
	$n$RMSE	&	7.84	&	13.90	&	18.73	&	&	8.19	&	13.99	&	18.28	&	&	9.08	&	14.52	&	18.94	&	&	8.57	&	14.31	&	18.15	&	&	10.48	&	15.79	&	20.39	&	&	8.95	&	16.34	&	22.03	&	&	8.57	&	13.83	&	18.05	\\																																															
	\end{tabular}%

	\setlength{\tabcolsep}{4.25pt}																																					
	\begin{tabular}{lccccccccccccccccccccccccccc}																																							
	\toprule																																																	
	\multicolumn{28}{l}{Swiss	consumption	survey	--	1-factor	model	}	\\																										
	\midrule																																																	
	$\tau$	&	\multicolumn{3}{c}{0.69}	&	&	\multicolumn{3}{c}{0.38}	&	&	\multicolumn{3}{c}{0.39}	&	&	\multicolumn{3}{c}{0.28}	&	&	\multicolumn{3}{c}{0.23}	&	&	\multicolumn{3}{c}{0.13}	&	&	\multicolumn{3}{c}{0.17}	\\																												
\cmidrule{2-4}\cmidrule{6-8}\cmidrule{10-12}\cmidrule{14-16}\cmidrule{18-20}\cmidrule{22-24}\cmidrule{26-28}	$n$	&	100	&	300	&	500	&	&	100	&	300	&	500	&	&	100	&	300	&	500	&	&	100	&	300	&	500	&	&	100	&	300	&	500	&	&	100	&	300	&	500	&	&	100	&	300	&	500	\\
	\midrule																																																	
	$n$Bias	&	-16.40	&	-53.51	&	-90.99	&	&	3.02	&	8.67	&	14.58	&	&	3.02	&	8.30	&	13.91	&	&	-2.90	&	-8.03	&	-13.20	&	&	-2.26	&	-6.53	&	-11.09	&	&	-1.33	&	-4.02	&	-6.20	&	&	-3.02	&	-8.83	&	-14.50	\\
	$n$SD	&	12.86	&	21.36	&	26.83	&	&	10.08	&	17.97	&	23.59	&	&	10.36	&	18.07	&	23.67	&	&	9.36	&	16.40	&	21.17	&	&	9.17	&	15.68	&	20.71	&	&	8.77	&	15.06	&	19.80	&	&	8.36	&	14.33	&	18.63	\\
	$n$RMSE	&	20.84	&	57.62	&	94.87	&	&	10.53	&	19.95	&	27.73	&	&	10.79	&	19.88	&	27.45	&	&	9.80	&	18.27	&	24.94	&	&	9.45	&	16.99	&	23.49	&	&	8.87	&	15.58	&	20.75	&	&	8.88	&	16.83	&	23.61	\\	
	  \bottomrule																																														
\end{tabular}	

    \end{small}		
\end{table}%

\end{landscape}

\begin{table}[!h]
  \centering
    \setlength{\tabcolsep}{7pt}
\caption{\label{simres-M2}
Small sample of sizes $n=\{100,300, 500\}$ distribution for $M_2$ ($10^4$ replications). 
Empirical rejection levels at $\alpha=\{0.20,0.10,0.05,0.01\}$, degrees of freedom (df), and  mean under the factor copula models. Continuous and count variables  are transformed   to ordinal  with $K=\{3, 4, 5\}$  and $K=\{3, 4\}$ categories, respectively, using the general strategies proposed in Section \ref{m2gof}.  Count variables area also transformed to ordinal with $K=5$ categories by treating them as ordinal where the 5th category contained all the  counts greater than 3.}
    \begin{tabular}{lccccccccccc}
    \toprule
          & \multicolumn{3}{c}{$n=100$} &       & \multicolumn{3}{c}{$n=300$} &       & \multicolumn{3}{c}{$n=500$} \\
\cmidrule{2-4}\cmidrule{6-8}\cmidrule{10-12}    & $K=3$   & $K=4$   & $K=5$      &       &  $K=3$   & $K=4$   & $K=5$     &       &  $K=3$   & $K=4$   & $K=5$   \\
    \midrule
  \multicolumn{12}{l}{   Political-economic dataset -- 1-factor model} \\
    \midrule
    df    & 92    & 121   & 152   &       & 92    & 121   & 152   &       & 92    & 121   & 152 \\
    mean  & 89.3  & {118.3} & {148.4} &       & {91.0 }& {119.7} & {152.6} &       & {91.0} & {119.6} & {152.3} \\
    $\alpha=0.20$ & 0.183 & 0.192 & 0.197 &       & 0.196 & 0.194 & 0.195 &       & 0.196 & 0.189 & 0.190 \\
    $\alpha=0.10$ & {0.121} & 0.125 & 0.134 &       & 0.122 & 0.121 & 0.119 &       & {0.114} & 0.109 & 0.109 \\
    $\alpha=0.05$ & {0.083} & 0.089 & 0.098 &       & 0.076 & 0.077 & 0.077 &       & {0.072} & {0.070} & 0.067 \\
    $\alpha=0.01$ & 0.044 & 0.046 & 0.055 &       & 0.036 & 0.034 & 0.037 &       & {0.027} & {0.030} & 0.026 \\

   \bottomrule
    \multicolumn{12}{l}{General social survey -- 1-factor model} \\
    \midrule
    df    & 161   & 239   & 329   &       & 161   & 239   & 329   &       & 161   & 239   & 329 \\
    mean  & 161.5 & 240.0 & 333.0 &       & 160.7 & {239.4} & 329.7 &       & 161.3 & 240.2 & 329.6 \\
    $\alpha=0.20$ & 0.213 & 0.220 & 0.240 &       & 0.202 & 0.216 & 0.203 &       & 0.211 & 0.228 & 0.212 \\
    $\alpha=0.10$ & 0.110 & 0.121 & 0.122 &       & 0.106 & 0.118 & 0.102 &       & 0.118 & 0.127 & 0.108 \\
    $\alpha=0.05$ & 0.058 & 0.070 & 0.061 &       & 0.054 & 0.067 & 0.051 &       & 0.065 & 0.073 & 0.056 \\
    $\alpha=0.01$ & 0.013 & 0.018 & 0.014 &       & 0.014 & 0.019 & 0.012 &       & 0.016 & 0.023 & 0.011 \\

    \bottomrule
 \multicolumn{12}{l}{Swiss consumption survey -- 1-factor model}\\
     \midrule
    df    & 74    & 128   & 194   &       & 74    & 128   & 194   &       & 74    & 128   & 194 \\
    mean  & 75.4  & 130.1 & {197.8} &       & 74.6  & 128.5 & 195.1 &       & 74.5  & 128.0 & 194.4 \\
    $\alpha=0.20$ & 0.229 & 0.239 & 0.254 &       & 0.214 & 0.209 & 0.221 &       & 0.210 & 0.202 & 0.207 \\
    $\alpha=0.10$ & 0.121 & 0.135 & 0.147 &       & 0.111 & 0.104 & 0.113 &       & 0.105 & 0.099 & 0.103 \\
    $\alpha=0.05$ & 0.067 & 0.076 & 0.086 &       & 0.056 & 0.055 & 0.060 &       & 0.051 & 0.053 & 0.053 \\
    $\alpha=0.01$ & 0.016 & 0.024 & 0.030 &       & 0.011 & 0.013 & 0.013 &       & 0.012 & 0.011 & 0.012 \\

    \end{tabular}%
    
  \setlength{\tabcolsep}{9.5pt}
    \begin{tabular}{lccccccc}
    \toprule
     & \multicolumn{3}{c}{General social survey -- 2-factor model} &       & \multicolumn{3}{c}{Swiss consumption survey -- 2-factor model} \\
\cmidrule{2-4}\cmidrule{6-8}  
 &\multicolumn{3}{c}{$n=500$}&&\multicolumn{3}{c}{$n=500$}\\
 \cmidrule{2-4}\cmidrule{6-8} 
       & $K=3$ & $K=4$ & $K=5$ &       & $K=3$ & $K=4$ & $K=5$ \\

    \midrule
    df    & 154   & 232   & 322   &       &  65  & 119   & 185 \\
    mean  & 154.8 & 234.0 & 323.3 &       & 65.6 & 119.7 &  185.5\\
    $\alpha=0.20$ & 0.217 & 0.234 & 0.214 &       & 0.217 & 0.215 & 0.217 \\
    $\alpha=0.10$ & 0.113 & 0.131 & 0.116 &       & 0.114 & 0.111 & 0.113 \\
    $\alpha=0.05$ & 0.065 & 0.075 & 0.059 &       & 0.060 & 0.057 & 0.060 \\
    $\alpha=0.01$ & 0.018 & 0.022 & 0.018 &       & 0.013 & 0.013 & 0.017 \\

    \bottomrule
    \end{tabular}%

\end{table}%

Table \ref{simres-alg} presents the number of times that the true bivariate parametric copulas  are chosen over 100 simulation runs.  If the true copula has distinct dependence properties with medium to strong dependence, then the algorithm performs extremely well as the sample size increases. Low selection rates occur if the true copulas have low dependence or  similar tail dependence properties, since for that case it is difficult to distinguish amongst parametric families of copulas \citep{Nikoloulopoulos&Karlis2008-CSDA}. For example,

\begin{table}[h!]
  \centering
  \caption{\label{simres-alg}Frequencies of the true bivariate copula  identified using the model selection algorithm  from 100 simulation runs. }

 \begin{small}

\setlength{\tabcolsep}{21pt}
    \begin{tabular}{lcccccc}
    
    \toprule
    \multicolumn{7}{l}{Political-economic dataset -- 1-factor model} \\
    \midrule
          & \multicolumn{2}{c}{Continuous} &       & \multicolumn{3}{c}{Ordinal} \\
\cmidrule{2-3} \cmidrule{5-7}
    $n$ & 1-reflected Joe
     & Joe   &       & reflected Joe
     & Joe   & Gumbel \\
   \midrule
    100   & 88    & 81    &       & 45    & 82    & 34 \\
    300   & 88    & 93    &       & 54    & 83    & 60 \\
    500   & 91    & 100   &       & 66    & 100   & 79 \\
    \bottomrule
    \end{tabular}
\setlength{\tabcolsep}{8.5pt}    
    \begin{tabular}{lccccccccc}
  
    \multicolumn{10}{l}{General social survey -- 1-factor model} \\
    \midrule
          & \multicolumn{2}{c}{Continuous} & \multicolumn{1}{c}{} & \multicolumn{3}{c}{Ordinal} & \multicolumn{1}{c}{} & \multicolumn{2}{c}{Count} \\
   
 \cmidrule{2-3} \cmidrule{5-7} \cmidrule{9-10}
 
    $n$ & Joe   & 2-reflected Joe 
    &       & $t_5$ & $t_5$ & reflected Gumbel
    &       & 2-reflected Joe
    &2-reflected  Gumbel
    \\
    \midrule
    100   & 68    & 63    &       & 27    & 19    & 27    &       & 56    & 28 \\
    300   & 89    & 79    &       & 41    & 43    & 49    &       & 65    & 55 \\
    500   & 91    & 85    &       & 61    & 65    & 74    &       & 73    & 68 \\
    \bottomrule
    \end{tabular}%
 
 \setlength{\tabcolsep}{8pt}  
 
    \begin{tabular}{lccccccccc}
   
    \multicolumn{10}{l}{Swiss consumption survey -- 1-factor model} \\
    \midrule
          & \multicolumn{3}{c}{Continuous} &       & \multicolumn{3}{c}{Ordinal} &       & Count  \\
  
 \cmidrule{2-4} \cmidrule{6-8} \cmidrule{10-10} 
 
    $n$ & reflected BB10
    & BB10  & BB10  &       & reflected Joe
    & reflected Joe 
    & reflected Joe
    &       & reflected Joe
    \\
    \midrule
    100   & 27    & 94    & 91    &       & 61    & 60    & 41    &       & 56 \\
    300   & 50    & 99    & 98    &       & 64    & 71    & 63    &       & 68 \\
    500   & 70    & 98    & 98    &       & 68    & 74    & 71    &       & 72 \\
    \bottomrule
    \end{tabular}%
 \setlength{\tabcolsep}{1pt}  
 \begin{tabular}{lccccccccc}
  
    \multicolumn{10}{l}{General social survey -- 2-factor model} \\
    \midrule
    1st Factor & \multicolumn{2}{c}{Continuous} & \multicolumn{1}{c}{} & \multicolumn{3}{c}{Ordinal} & \multicolumn{1}{c}{} & \multicolumn{2}{c}{Count} \\
   \cmidrule{2-3} \cmidrule{5-7}\cmidrule{9-10}

    $n$ & reflected Gumbel
    & reflected Joe
    &       & BVN   & 1-reflected Joe
     & 1-reflected Joe 
     &       & reflected Joe
     & Gumbel \\
    \midrule
    100   & 22    & 40    &       & 10    & 19    & 19    &       & 50    & 6 \\
    300   & 26    & 52    &       & 11    & 42    & 36    &       & 79    & 16 \\
    500   & 19    & 67    &       & 13    & 52    & 53    &       & 83    & 39 \\
    \bottomrule
  
    2nd Factor & \multicolumn{2}{c}{Continuous} &       & \multicolumn{3}{c}{Ordinal} &       & \multicolumn{2}{c}{Count} \\
    \cmidrule{2-3} \cmidrule{5-7}\cmidrule{9-10}
    $n$ & Gumbel &2-reflected  Joe
     &       &reflected  Joe
     & Gumbel & $t_5$ &       & BVN   & 2-reflected Gumbel
     \\\midrule
    100   & 13    & 28    &       & 28    & 7     & 14    &       & 21    & 17 \\
    300   & 26    & 39    &       & 56    & 30    & 45    &       & 28    & 47 \\
    500   & 32    & 67    &       & 65    & 53    & 59    &       & 33    & 70 \\
    \bottomrule
    \end{tabular}%
 \setlength{\tabcolsep}{7.5pt}  
    \begin{tabular}{lccccccccc}
   
       \multicolumn{10}{l}{Swiss consumption survey -- 2-factor model}
\\\midrule    1st Factor & \multicolumn{3}{c}{Continuous} &       & \multicolumn{3}{c}{Ordinal} &       & Count \\
   \cmidrule{2-4} \cmidrule{6-8} \cmidrule{10-10} 
    $n$ & BB10  & BB10  & BB10  &       & reflected Joe
    & reflected Joe
     & Frank &       & Frank \\\midrule
    100   & 57    & 77    & 55    &       & 31    & 28    & 23    &       & 34 \\
    300   & 81    & 94    & 82    &       & 51    & 40    & 19    &       & 21 \\
    500   & 88    & 94    & 87    &       & 49    & 50    & 21    &       & 16 \\\midrule
    2nd Factor & \multicolumn{3}{c}{Continuous} &       & \multicolumn{3}{c}{Ordinal} &       & Count \\  \cmidrule{2-4} \cmidrule{6-8} \cmidrule{10-10} 
    $n$ & BB10  & BVN   & BB10  &       & BVN   & reflected Joe 
    & reflected Joe 
    &       & reflected Gumbel 
     \\\midrule
    100   & 5     & 14    & 28    &       & 10    & 29    & 31    &       & 10 \\
    300   & 27    & 29    & 43    &       & 22    & 49    & 40    &       & 16 \\
    500   & 39    & 39    & 60    &       & 31    & 55    & 63    &       & 31 \\
    \bottomrule
    \end{tabular}%
   \end{small}
   
\end{table}

\begin{itemize}

\item  in the results from the 2-factor model for the general social survey, the true copula for the first continuous variable (1st factor) is the reflected Gumbel 
with $\tau=0.34$ and is only selected  a considerable small number of times. The algorithm instead selected with a high probability  the  reflected Joe 
(results not shown here due to space constraints), because both reflected Joe 
and Gumbel 
copulas provide similar dependence properties, i.e., lower tail dependence. 
\item  in the results from the 2-factor model for the Swiss consumption survey, the variables with Frank copulas have the lowest selection rates. This is due to the fact that their true Kendall's $\tau$'s parameters are close to 0  (independence).
\end{itemize} 

\baselineskip=24.5pt

\section{\label{sec-discussion}Discussion}
We have extended the factor copula model proposed in \cite{Krupskii&Joe-2013-JMVA} and \cite{Nikoloulopoulos2015-PKA} to the case of mixed continuous and discrete responses.
It is the most general factor model as (a) it has the  standard factor model with 
an additive latent structure as a special case  when the BVN copulas are used, (b) it can have a latent structure that is not additive if other than BVN copulas are called, (c)
the  parameters of the univariate distributions are separated from the copula (dependence) parameters which are interpretable as
dependence of an observed variable with a latent variable, or conditional dependence of an observed variable with a latent variable given preceding latent variables.
Other non-linear  (e.g., \citealt{Rizopoulos&Moustaki-2008-bjmsp}), semi- (e.g., \citealt{gruhl-etal-2013}) or non-parametric  models  (e.g., \citealt{Kelava-etal-2017}) with latent variables  have either  an additive latent structure  or  allow polynomial  and interaction terms to be added in the linear predictor, hence are not as general. 
Another mixed-variable model in the  literature that is called factor copula  model  \citep{Murray-etal-2013} is restricted to the MVN copula as the model proposed by  \cite{gruhl-etal-2013}, hence has an additive latent structure.

We have shown that the factor copula models provide a substantial  improvement over the standard factor model  on the basis of the log-likelihood principle, Vuong's and $M_2$ statistics.  Hence,  superior statistical inference for the loading parameters of interest can be achieved. This improvement relies on the
fact that the latent variable distribution is expressed via factor copulas instead of the 
 MVN distribution. The latter   is restricted to linear and reflection symmetric  dependence. \cite{Rizopoulos&Moustaki-2008-bjmsp} stressed that the inadequacy of normally distributed latent variables can be caused by the non-linear dependence on the latent variables. 
The factor copula can provide flexible  reflection asymmetric tail and non-linear dependence  as it is a truncated canonical vine copula \citep{Brechmann-Czado-Aas-2012} rooted at the latent variables.
 \cite{Joe&Li&Nikoloulopoulos2010} show that in order for a vine copula to have (tail) dependence for all bivariate margins, it is only necessary for the bivariate copulas in level 1  to have (tail) dependence and it is not necessary for the conditional bivariate copulas in levels $2, \ldots, d-1$   to have tail dependence. 
The 1-factor copula has bivariate copulas with tail dependence in the 1st level and independence copulas in all the remaining levels of the vine (truncated after the 1st level). 
The 2-factor copula has bivariate copulas with tail dependence in the 1st and 2nd level and independence copulas in all the remaining levels (truncated after the 2nd level). Hence, the tail dependence among the latent variables and each of the observed variables is inherited to the tail dependence among the observed variables.

\begin{figure}[!h]
\begin{center}
\caption{\label{rankings} Comparison of the political-economic risk rankings obtained via our selected model, the standard factor model, and the mixed-data factor analysis of \cite{Quinn2004}.}
\begin{tabular}{cc}
\includegraphics[width=0.45\textwidth]
{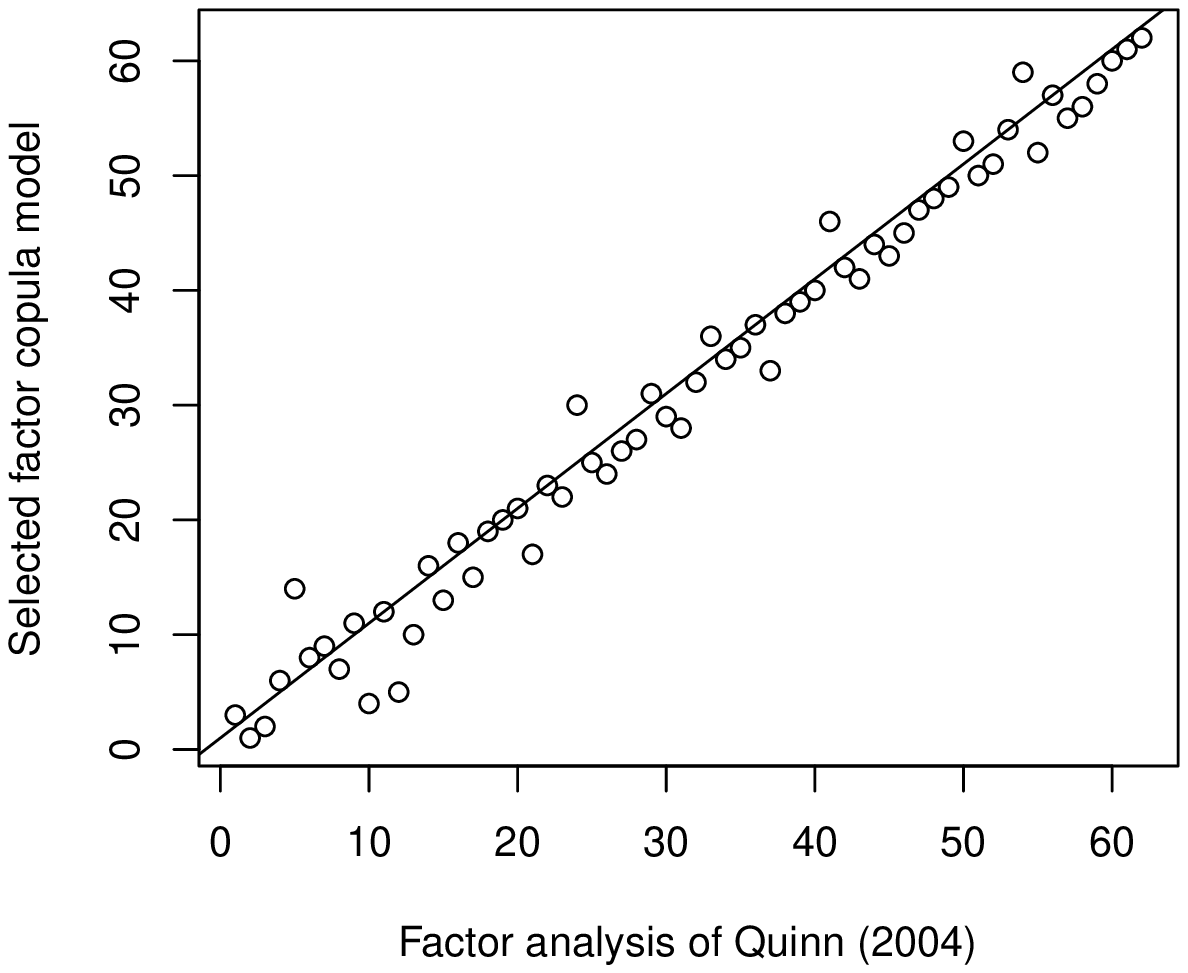}
&
\includegraphics[width=0.45\textwidth]
{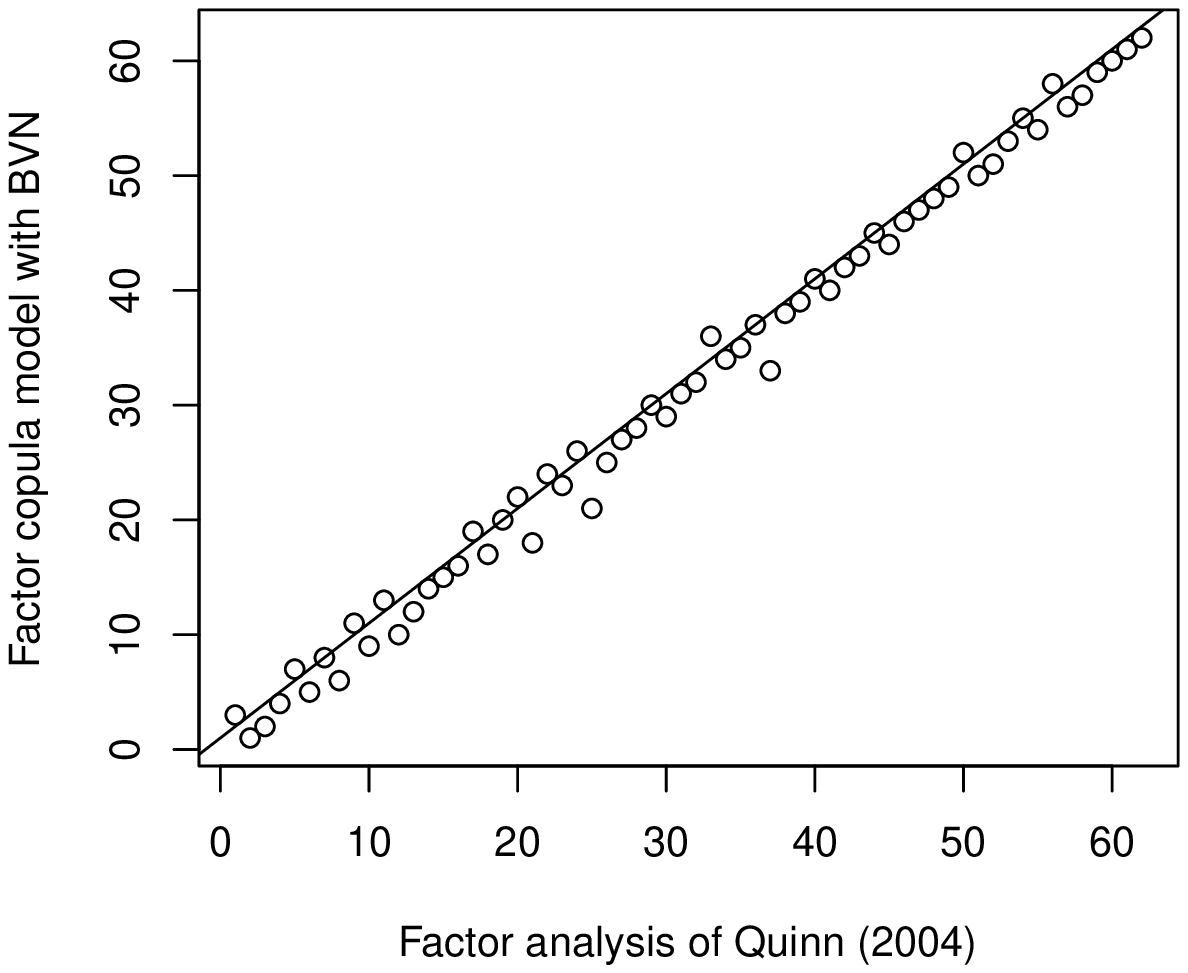}\\
\multicolumn{2}{c}{\includegraphics[width=0.45\textwidth]
{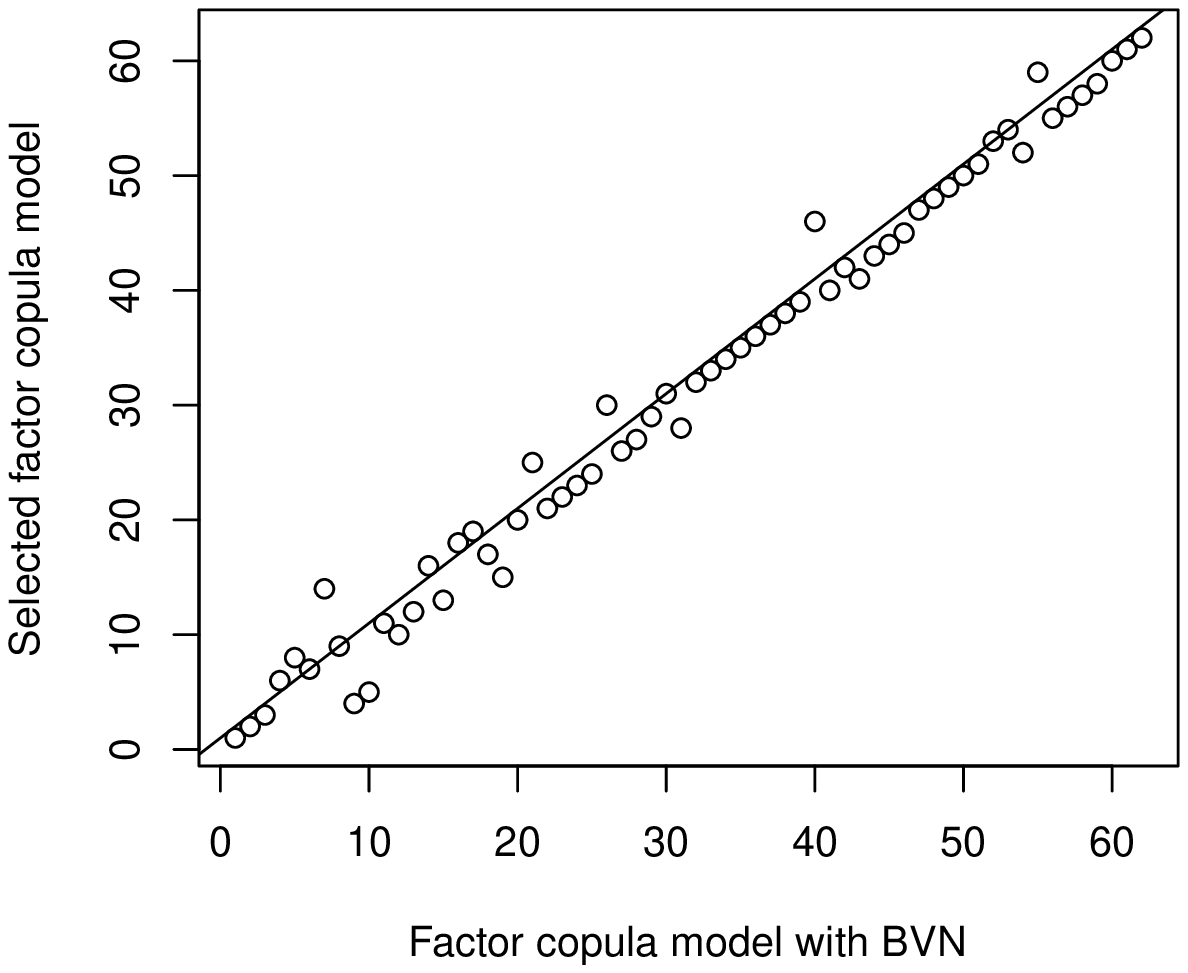}}\\
\end{tabular}
\end{center}
\end{figure}

Even in the cases, where the effect of misspecifying the bivariate linking copula choice to build the factor copula models can be seen  as minimal for the  Kendall's $\tau$ (loading) parameters,  
the tail dependence varies, as explained in Section \ref{bivcop}, and is a property to consider when choosing amongst different families of copulas and hence affects prediction. \cite{Rabe-Hesketh-etal-2003-StatMod} highlighted the importance of the correct distributional assumptions for the prediction of latent scores. The latent scores will essentially show the effect of different model  assumptions,  because it is an inference that depends on the joint distribution. 
The factor copula models have bivariate copulas  that  link the latent variables to each of the observed variables. If these bivariate copulas have upper or lower tail dependence, then this type of dependence is inherited to the dependence  between the factor scores and each of the observed variables. Hence,  
factor scores are  fairly different than the ones for the standard factor model if the sample size is sufficient.
 Figure \ref{rankings} demonstrates these differences by revisiting the political-economic dataset in Section \ref{PE-sec} and comparing the political-economic risk ranking obtained via our selected model, the factor copula model with BVN copulas (standard factor model), and the mixed-data factor analysis of \cite{Quinn2004}. It is revealed that  even for a small sample size ($n=62$) there are differences.  Between  the factor copula model with BVN copulas and the factor analysis model of \cite{Quinn2004}, there are small to moderate differences, because while these models share the same latent variables distribution, the former  model does not assume  the observed variables to be normally distributed, but rather   uses the empirical distribution of the continuous observed variables, i.e. allows the  margins to be quite free and not restricted by  normal distribution. The differences at the lower panel graph are solely due the miss-specification the latent variable distribution.  
  
As stated by many researchers (e.g.,  \citealt{Rabe-Hesketh&Skrondal-2001-Biometrics,Rabe-Hesketh&Skrondal-2004}), the major difficulty of all the models with latent variables is identifiability. 
 For example, for the standard factor model or the more flexible model in \cite{Irincheeva&Cantoni&Genton2012-SJS} 
one of  
loadings in the second factor has to be set to zero, 
because the model with $2d$ loadings is not identifiable.
The standard factor model  arises as special case of our model if we use  as bivariate linking copulas the BVN copulas. Hence, for the 2-factor copula model with BVN copulas, one of the BVN copulas in the second factor has to be set as an independence copula. 
However, using other than BVN copulas, the 2-factor copula model   is near-identifiable with  $2d$ bivariate linking copulas  as it has been demonstrated  by  \cite{Krupskii&Joe-2013-JMVA} and \cite{Nikoloulopoulos2015-PKA}.

\section*{Software}
Our modelling framework is implemented in the package \pkg{FactorCopula} \citep{blinded} within the open source statistical environment \proglang{R} \citep{CRAN}. All the analyses presented in Sections \ref{PE-sec} and \ref{GSS-section} are given as code examples in the package.

\section*{Acknowledgements}
We would like to thank the referees and Professor Harry Joe (University of British Columbia) for their careful reading and comments that have led to an improved presentation and Dr  Irina Irincheeva (University of Bern) and Professor Marc Genton (King Abdullah University of Science and Technology) for sharing the Swiss consumption survey dataset. The simulations presented in this paper were carried out on the High Performance Computing Cluster supported by the Research and Specialist Computing Support service at the University of East Anglia.

\end{document}